\documentclass[a4paper,fleqn,usenatbib]{mn2e}
\usepackage{newtxtext}
\usepackage[T1]{fontenc}
\usepackage{ae,aecompl}

\usepackage{mathrsfs,amsmath,graphics,color,epsfig,amssymb,mathtools}
\usepackage{mathptmx}
\usepackage{epsfig}
\usepackage{booktabs}
\usepackage{rotating,rotfloat}
\usepackage{longtable}
\usepackage{graphicx}
\usepackage{url} 
\usepackage{listings}
\usepackage{color, soul}
\usepackage{pslatex} 
\usepackage{booktabs}
\usepackage{dcolumn}
\usepackage{multicol}
\usepackage{subcaption}
\usepackage{multirow}
\usepackage{ctable}

\usepackage{hyperref} 

\newcommand\tsys{T_{\mathrm{sys}}}
\newcommand\psd{\mathrm{PSD}}
\newcommand\snr{\mathrm{SNR}}
\newcommand\cest{\hat{C}_\ell}
\newcommand\vt{\mathrm{v}_\mathrm{t}}

\title[Simulations of $1/f$ Noise for IM]{Impact of Simulated $1/f$ Noise for HI Intensity Mapping Experiments}

\author[S. \,Harper et al.]
  {S.~Harper,$^1$\thanks{E-mail: stuart.harper@manchester.ac.uk}
  C.~Dickinson,$^1$\thanks{E-mail: clive.dickinson@manchester.ac.uk} R.A.~Battye,$^1$ S.\,Roychowdhury,$^1$ I.W.A.~Browne,$^1$ 
 \newauthor   
 Y.-Z.~Ma,$^{2}$ 
L.~C.~Olivari,$^1$ 
T.~Chen$^1$\\
$^1$Jodrell Bank Centre for Astrophysics, Alan Turing Building, School of Physics \& Astronomy, The University of Manchester, Oxford Road, \\
Manchester M13 9PL, UK \\
$^2$School of Chemistry and Physics, University of KwaZulu-Natal, Westville Campus, Private Bag X54001, Durban, 4000, South Africa;
 }

\date{Accepted XXX. Received YYY; in original form ZZZ}

\pubyear{2015}

\begin{document}
\label{firstpage}
\pagerange{\pageref{firstpage}--\pageref{lastpage}}
\maketitle

\begin{abstract}
Cosmology has entered an era where the experimental limitations are not due to instrumental sensitivity but instead due to inherent systematic uncertainties in the instrumentation and data analysis methods. The field of HI intensity mapping (IM) is still maturing, however early attempts are already systematics limited. One such systematic limitation is $1/f$ noise, which largely originates within the instrumentation and manifests as multiplicative gain fluctuations. To date there has been little discussion about the possible impact of $1/f$ noise on upcoming single-dish HI IM experiments such as BINGO, FAST or SKA. Presented in this work are Monte-Carlo end-to-end simulations of a 30\,day HI IM survey using the SKA-MID array covering a bandwidth of 950 and 1410\,MHz. These simulations extend $1/f$ noise models to include not just temporal fluctuations but also correlated gain fluctuations across the receiver bandpass. The power spectral density of the spectral gain fluctuations are modelled as a power-law, and characterised by a parameter $\beta$. It is found that the degree of $1/f$ noise frequency correlation will be critical to the success of HI IM experiments. Small values of $\beta$ ($\beta$ < 0.25) or high correlation is preferred as this is more easily removed using current component separation techniques. Spectral index of temporal fluctuations ($\alpha$) is also found to have a large impact on signal-to-noise. Telescope slew speed has a smaller impact, and a scan speed of 1\,deg\,s$^{-1}$ should be sufficient for a HI IM survey with the SKA.
\end{abstract}
\begin{keywords}
cosmology: observations $-$ large-scale structure of Universe $-$ radio lines: galaxies $-$ methods: statistical $-$ data analysis $-$ instrumentation: spectrographs
\end{keywords}


\section{Introduction}\label{sec:intro}

Cosmology has entered an era where most cosmological parameters are known to the one per\,cent level using a combination of probes including the cosmic microwave background (CMB) \citep{PlanckX2016}, galaxy redshift surveys \citep[e.g.,][]{Eisenstein2005,Alam2015,Aubourg2015,Hinton2017}, and Type Ia supernovae \citep[e.g.,][]{Riess1998,Betoule2014}. Another possible cosmological probe that is yet to be exploited is the hyperfine 21\,cm spin-flip transition of neutral hydrogen (HI), which can be used to, for example, trace large-scale structure and the ionization history of the Universe \citep{Pritchard2012}. Given the abundance of hydrogen within the Universe one might expect HI emission to be a bright signal. However, given the expected lifetime of $10^7$\,years for the upper transition state of the 21\,cm line, HI emission in reality is extremely faint. To date the detection of individual galaxies up to modest redshifts of $z<0.4$ has proved extremely challenging using the current generation telescopes \citep{Zwaan2001,Zwaan2004,Giovanelli2005,Verheijen2010,Catinella2015,Fernandez2016}. By the stacking of HI emission from many galaxies \citep{Lah2009,Rhee2016},  exploiting strongly lensed systems \citep[e.g.,][]{Hunt2016}, and measurements of damped Lyman-$\alpha$ systems \citep[e.g.,][]{Noterdaeme2012,Neeleman2016} it is known that there is a vast quantity of neutral hydrogen present throughout the cosmic history of the Universe.  It is this extremely faint HI signal and the potential cosmological information within that was the original science driver behind the Square Kilometer Array (SKA) proposal \citep{Wilkinson1991}.

HI emission as a probe of cosmological structure is considered extremely promising as it is expected to have far more information content than the cosmic microwave background (CMB). This is because tracing HI emission allows for the Universe to be divided in to redshift slices and the evolution of structure to be studied. HI cosmology is expected to yield both vastly improved constraints on parameters that describe the evolution of dark energy, neutrino masses, HI bias, HI mass density, and the astrophysical processes occurring within the Epoch-of-Reionization (EoR) \citep{Loeb2004,Santos2005,Sethi2005,Fan2006,McQuinn2008,Loeb2008,Pritchard2008}. However, contemporary HI instrumentation lacks the sensitivity to conduct a traditional galaxy survey using the 21\,cm line. Such a survey today would suffer from high levels of incompleteness, uncertainties in redshifts and long survey times. For these reasons there has been a drive to find an alternative method of measuring the cosmic HI signal. 

The proposed alternative to HI galaxy surveys is known as the HI intensity mapping (IM) method. HI IM foregoes the detection of individual galaxies in favour of the total integral HI flux density over large cosmological volumes. The benefits of HI IM are no bias in galaxy selection, intrinsic knowledge of redshift, and a higher surface brightness relative to galaxy surveys. HI IM has the potential to be one of the most important tools available to understanding the evolution of large-scale structure, as it contains several orders-of-magnitude more information than the CMB \citep{Battye2004,Wyithe2008b,Masui2010,Battye2013,Santos2014,Santos2015,Padmanabhan2015,Bull2015,Ma2016}. 

There have been several attempts at probing the post-reionization HI signal using the intensity mapping technique with existing telescopes at the Parkes \citep{Pen2009,Anderson2017} and Green Bank \citep{Chang2010,Masui2010,Masui2013,Shaw2014,Wolz2015} observatories. However, due to systematic limitations from the instrumentation and radio-frequency interference (RFI) the HI power spectrum was only detected in cross-correlation with optical surveys. There are a number of bespoke post-reionization epoch experiments planned such as the Baryonic Acoustic Oscillations from Integrated Neutral Gas Observations \citep[BINGO][]{Battye2013},  Canadian Hydrogen Intensity Mapping Experiment \citep[CHIME][]{Bandura2014}, Tianlai Telescope \citep{Chen2012}, Five hundred metre Aperture Spherical Telescope \citep[FAST][]{Smoot2014,Bigot2016}, Hydrogen Intensity and Real-Time Analysis experiment \citep[HIRAX][]{Newburgh2016},  and a potential survey with the SKA-MID array \citep{Santos2015}. There has also been a recent proposal to perform a HI IM survey with MeerKAT \citep{Santos2017}.  At lower frequencies there are numerous phased array telescopes planning to probe the EoR such as  the Precision Array for Probing the Epoch of Reionzation \citep[PAPER][]{Parsons2010},  Low Frequency Array \citep[LOFAR][]{vanHaarlem2013},  the Giant Metrewave Radio Telescope  \citep[GMRT][]{Paciga2011},  Murchison Widefield Array  \citep[MWA][]{Bowman2013}, and  Hydrogen Epoch of Reionization Array \citep[HERA][]{DeBoer2017}. Similarly there are plans to perform intensity mapping using other emission lines such as CO \citep{Lidz2011,Keating2015,Li2016,Keating2016}, Lyman-$\alpha$ \citep{Peterson2012,Croft2016}, and CII \citep{Crites2014}. 

The focus of this paper will be on the HI IM survey discussed in \citet{Bull2015} that proposes the use of the SKA-MID array as a collection of single-dish radio telescopes  as opposed to a  single interferometer.  This is sometimes referred to as autocorrelation or total-power observations. The forecast for such a survey as described in \citet{Bull2015} suggests that the cosmological constraints on dark energy parameters at redshifts less than $z < 1$ will be comparable to upcoming Stage IV  dark energy experiments such as \textit{Euclid} \citep{Albrecht2006}. The claims in \citet{Bull2015} were further supported by simulations of such an SKA survey by \citet{Alonso2015}, where it was demonstrated that existing methods for component separation are sufficient to remove astrophysical foregrounds in intensity. However, for real HI IM data there will be challenges due to instrumental systematics such as standing waves, frequency dependent asymmetric beams, or polarisations leakage. 

The current work will investigate the impact on the recovery of the HI angular power spectrum in the presence of a single instrumental systematic that is commonly referred to as $1/f$ noise. As many instrumental systematics are time dependent, determining the exact impact each has on the underlying HI signal requires end-to-end simulations of signal on the sky, through a receiver model and resulting observed sky map. This work presents such end-to-end simulations with the intent of determining the impact of $1/f$ noise on the recovery of the HI angular power spectrum.

The paper proceeds as follows: In Section \ref{sec:overview} an overview of single-dish HI IM will be given as well as a brief review of $1/f$ noise; Section \ref{sec:pipeline} will describe the simulation pipeline that was used to generate the mock SKA HI IM data and the data analysis methods; Section \ref{sec:results} will describe how $1/f$ noise will impact SKA observations; and Section \ref{sec:discussion} will derive an empirical model of the impact of $1/f$ noise and potential avenues of future investigation into the impact of $1/f$ noise on HI IM. Finally, in Section \ref{sec:conclusion} the key results and insights of the paper are summarised. 

\section{Single-Dish Intensity Mapping and 1/f Noise}\label{sec:overview}

The SKA array will be a revolutionary interferometric radio observatory. It has been proposed in \citet{Santos2015} and \citet{Bull2015} that the approximately 200 15\,m dishes in the SKA-MID array could each be used as independent single-dish (SD) radio telescopes (i.e. autocorrelation or total-power observations). The advantage of SD HI IM over interferometric surveys is increased HI surface brightness and sensitivity to HI on large angular scales. The above papers suggest that a HI IM survey using the SKA could yield cosmological constraints that are competitive with Stage IV dark-energy surveys on comparable time scales.

There are disadvantages to using the SKA as an array of SD instruments. The first is the poor resolution of SD observations, which limits an autocorrelation SKA HI IM survey to studying late-time cosmology of $z < 1$ (assuming approximately $\sim$1\,degree resolution). This is especially important if the objective is to measure the characteristic \textit{wiggles} of baryonic acoustic oscillations  within the HI power spectrum on scales of $~0.1$\,Mpc$^{-1}$. This was discussed extensively in \citet{Bull2015}, and is easily summarised by looking at the extent of the beam resolution in $k$-space as 
\begin{equation}\label{eqn:kmin}
	k_{\mathrm{beam}} \approx \frac{2 \pi D_{\mathrm{dish}}}{ r \lambda} ,
\end{equation}
where $D_{\mathrm{dish}}$ is the diameter of the telescope dish (accounting for under-illumination), $\lambda$ is the wavelength of observation and $r$ is the co-moving distance at a given redshift. For even the largest SD radio telescopes, such as the 500\,m diameter FAST telescope, the maximum redshift at which BAOs could still be measured is $z < 4$.

Another disadvantage of using the SD approach is the challenge of separating spurious contaminants from the HI signal. When observing the sky with an interferometric array there are several ways to utilise the combined information of all the telescopes to suppress external sources of RFI \citep[e.g.][]{Hellbourg2014}, or instrumental effects such as standing waves and $1/f$ noise that are specific to individual receivers within the array. However, this is not possible with SD radio telescopes and suppressing systematics or spurious signals requires a combination of careful modelling and a well designed observing strategy.

This work will focus on the contamination of SD observations by $1/f$ noise, which is a form of correlated noise that is ubiquitous to radio receiver systems and manifests as small gain fluctuations. When binned into a map of the sky $1/f$ noise manifests as large-scale spatial fluctuations that are not trivial to separate from the true underlying sky signal. Exactly how the the fluctuations manifest in sky maps depends on the noise properties (for example: the Gaussianity of the noise, the characteristic time scale of the fluctuations or whether it is stationary) and the details of the observing strategy. For these reasons $1/f$ noise has for a long time has been of interest to SD observers in both radio \citep[e.g.,][]{Seiffert2002} and sub-millimetre \citep[e.g.,][]{Emerson1988,Traficante2011}. The removal of $1/f$ noise has also become an area of research that has resulted in several advanced map-making methods \citep[e.g.,][]{Natoli2001,Sutton2010}.

Before discussing $1/f$ noise it should be made clear that it is a phenomenon separate to the well understood concept of thermal noise. Thermal noise is caused by the thermal motion of charge carriers and has an r.m.s. related to the noise temperature of the receiver through the  well known radiometer equation
\begin{equation}\label{eqn:radiometer}
	\sigma = A \tsys \sqrt{ \frac{ f_{\mathrm{sr}}}{\delta\nu } } ,
\end{equation}
where $A$ is constant that is dependent on the receiver system of order unity \citep{Wilson2009}, $T_{\mathrm{sys}}$ is the system temperature, $f_{\mathrm{sr}}$ is the sample rate, and $\delta\nu$ is the system channel width. Thermal noise can be accurately modelled as a Gaussian white noise distribution \citep{Wilson2009}.

There have been several proposed physical explanations for $1/f$ noise within electronic circuits stemming from its initial discovery \citep{johnson1925,Nyquist1928}. However, even with its prevalence in a range of physical systems including cognition \citep{gilden1995}, biomechanics \citep{kobayashi1982}, geological records  \citep{Mandelbrot1969}, music \citep{voss1978}, and others \citep[e.g.,][]{Press1978}, a fundamental description of $1/f$ noise has yet to be found. Fortunately $1/f$ noise can still be modelled and phenomenologically described by a small number of statistical properties. 

It is common in astronomy to define the $\psd$ of a receiver contaminated with thermal and $1/f$ noise as \citep[e.g.,][]{Seiffert2002,BigotSazy2015}
\begin{equation}\label{eqn:astro_fnoise}
	\psd(f) = \frac{\tsys^2}{\delta\nu} \left[1 + \left(\frac{f_k}{f}\right)^\alpha \right] ,
\end{equation}
where $\tsys$ is the system temperature of the receiver, $\delta \nu$ is the channel bandwidth, $f_k$ is known as the knee frequency, and $\alpha$ is spectral index of the noise. The unity term inside the brackets of Eqn.~\ref{eqn:astro_fnoise} describes the thermal noise contribution and the $1/f$ noise is described by the reciprocal power-law on the \textit{right}. The reciprocal power-law, from which $1/f$ noise derives its name, is its key property and implies that for $\alpha > 0$ long time-scale fluctuations will have more power than short time scale fluctuations. It is common to find named types of $1/f$ noise that describe specific spectral indices, such as pink noise ($\alpha = 1$), and brown noise ($\alpha = 2$), or sometimes generally referred to as red noise for any $\alpha > 0$. Another property of $1/f$ noise is that the fluctuations need not be Gaussian distributed, which could impact how it averages down over time. However, in this work it is assumed that $1/f$ noise does have Gaussian properties, this should be confirmed through measurements from real receiver systems. A third important property of $1/f$ noise is that Eqn.~\ref{eqn:astro_fnoise} only defines it for a finite period, as the $1/f$ noise term in Eqn.~\ref{eqn:astro_fnoise} is unbounded and tends to infinity at the zeroth frequency. This of course cannot be true because it would be in violation of both Parseval's theorem and conservation of energy. One simple solution to this paradox would be to suppose that on some sufficiently long time-scale the $1/f$ PSD flattens. This must be true, but intriguingly, for semi-conductors at least, no such turn off has been observed even after months of continuous observation \citep{Caloyannides1974,Mandal2009}. For real astronomical observations this is also not true as no observation is taken over infinite time, and data calibration effectively acts as a high-pass filter on the lowest frequency $1/f$ noise modes.

For many past CMB and other SD experiments each receiver has a single output that is integrated over a wide bandwidth. Therefore the $1/f$ noise fluctuations in each receiver are entirely independent and as such can be sufficiently characterised by the parameters $f_k$ and $\alpha$ from Eqn.~\ref{eqn:astro_fnoise}. For the spectroscopic receivers used in IM experiments each receiver will have many outputs. In this instance the $1/f$ noise fluctuations in two output channels from one receiver are likely to be highly correlated. Similarly, the width of each channel is arbitrary with wider channels having lower thermal noise levels. However, if the $1/f$ noise is highly correlated then it will not average down with wider channel widths and so will effectively increase the $f_k$ defined in Eqn.~\ref{eqn:astro_fnoise}.  For this reason the use of $f_k$ to characterise $1/f$ noise properties of a spectroscopic system should be used with care, as it depends on both the channel bandwidth and the correlations of the $1/f$ noise in frequency . 

At present there is very little known about the frequency correlations of $1/f$ noise in spectroscopic receivers. This work will explore the impact of frequency correlated $1/f$ noise on a simulated SKA HI IM survey. Due to lack of knowledge as to what the true functional form for the PSD of frequency correlated $1/f$ noise may be, a simple power-law model is adopted. Modifying Eqn.~\ref{eqn:astro_fnoise} to include frequency correlations results in
\begin{equation}\label{eqn:astro_fnoise2}
	\psd(f, \omega) = \frac{\tsys^2}{\delta\nu} \left[1 + C(\beta, N_\nu) \left(\frac{f_k}{f}\right)^\alpha \left( \frac{1}{\omega \Delta \nu } \right)^{\frac{1-\beta}{\beta}} \right] ,
\end{equation}
where $\omega$ is the inverse spectroscopic frequency wavenumber, $\Delta \nu$ is the total receiver bandwidth, $\beta$ is used to parametrise the spectral index of the PSD, and $C(\beta, N_\nu)$ is a constant that is discussed in more detail in Section~\ref{sec:1f}. Eqn.~\ref{eqn:astro_fnoise2} describes a 2-dimensional PSD for which the temporal and spectroscopic $1/f$ noise fluctuations are separable. The spectral index of the frequency correlations is defined in the limits $0 < \beta < 1$, where $\beta = 0$ describes $1/f$ noise fluctuations that are identical in every frequency channel, whereas $\beta = 1$ would describe $1/f$ noise that is independent in every channel. In Section~\ref{sec:1f} the $1/f$ noise model and simulations will be discussed in detail.

\begin{figure}
\centering
\includegraphics[width=0.46\textwidth]{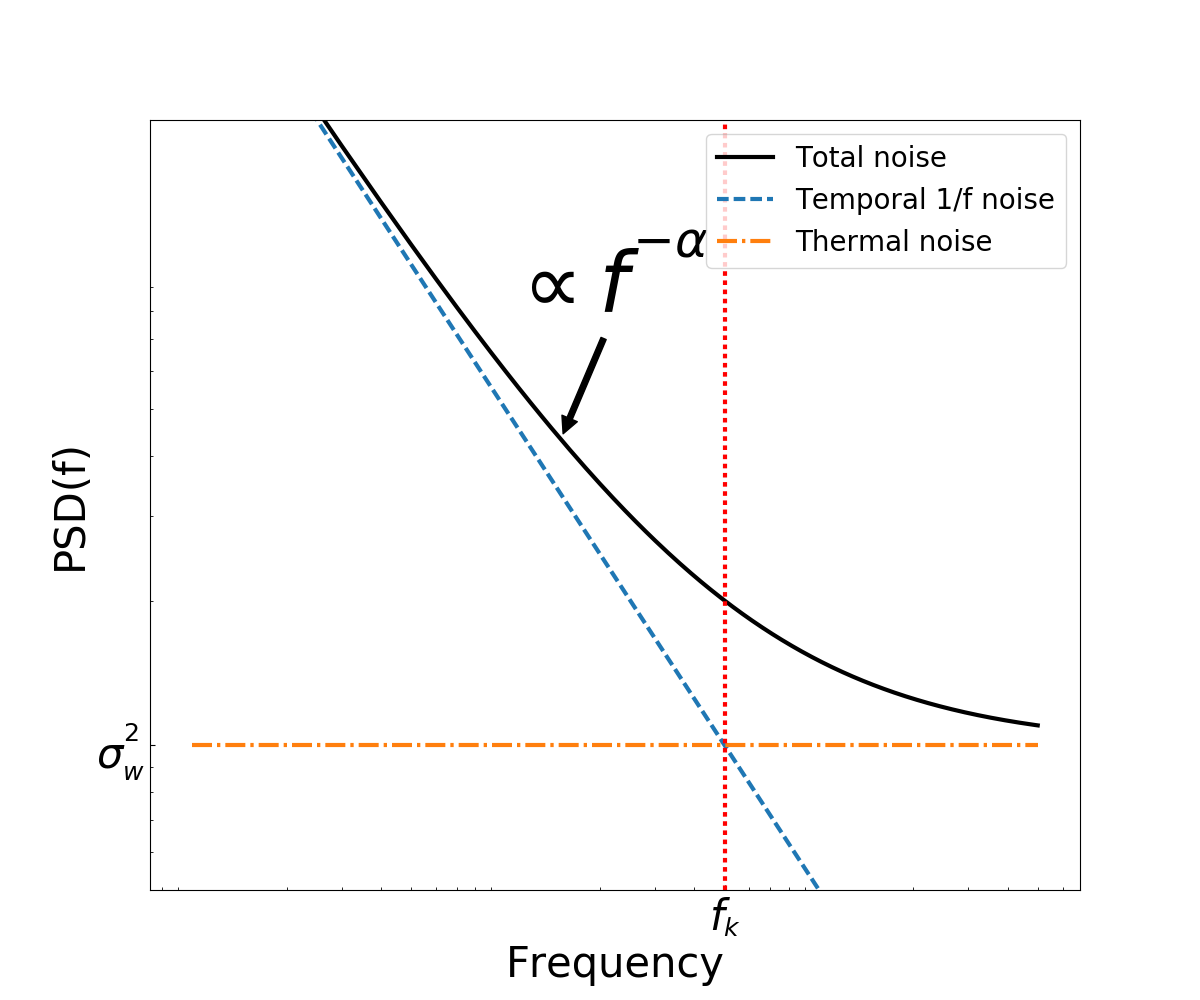}
\includegraphics[width=0.46\textwidth]{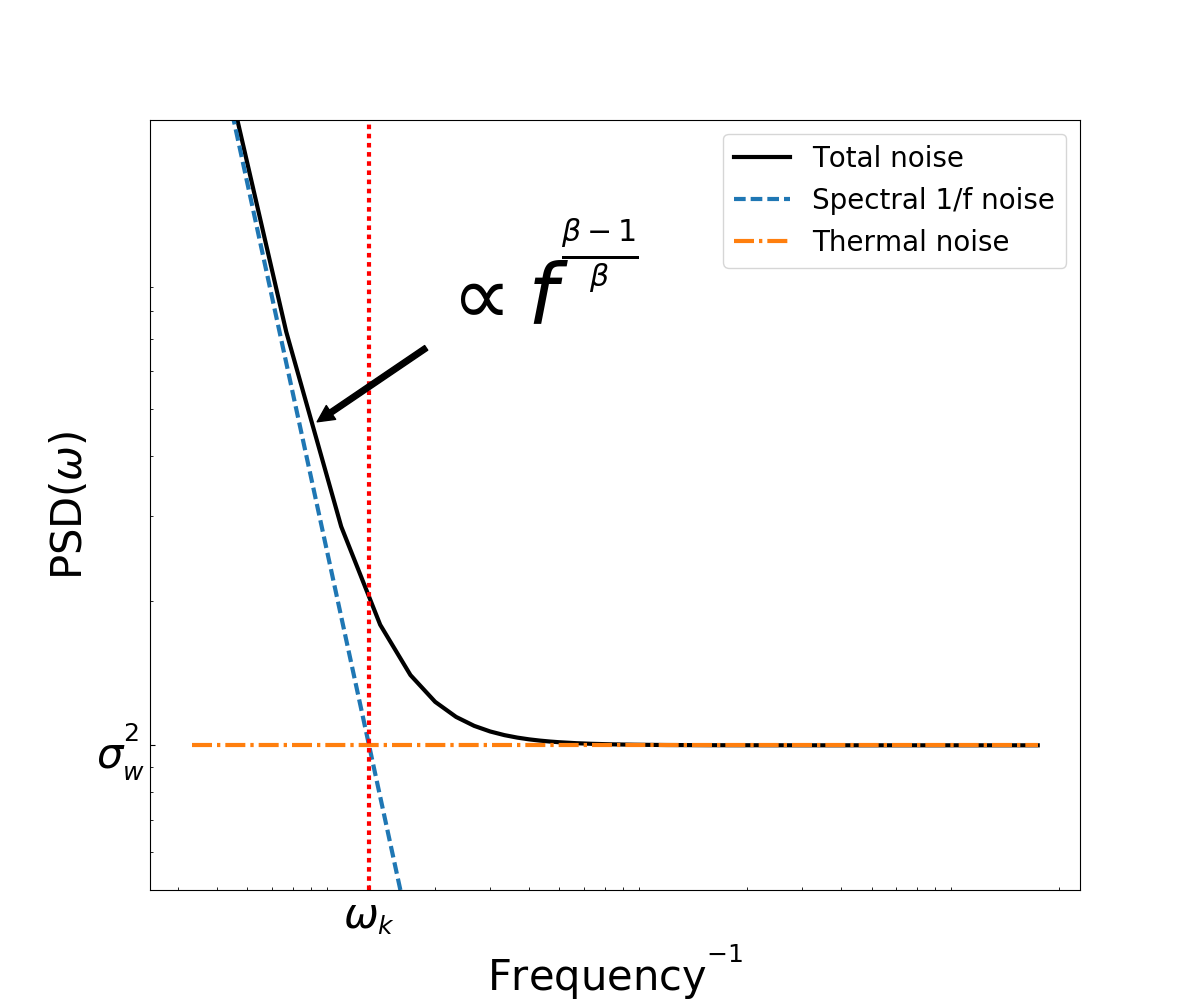}
\caption{The power spectral density of the temporal (\textit{top}) and spectroscopic (\textit{bottom}) $1/f$ noise as described by Eqn.~\ref{eqn:astro_fnoise2}. Both plots have the same thermal noise level ($\omega_w^2$), and an $\alpha = 1$ and a $\beta = 0.25$ for the temporal and spectroscopic PSD respectively. The \textit{black line} is the combined spectral power of the $1/f$ noise slope and the thermal noise. Both the thermal noise and the $1/f$ noise have equal spectral power at $f_k$ and $\omega_k$ marked by the \textit{red dotted-line} in the \textit{top} and \textit{bottom} plots respectively.}\label{fig:fnoisepsd}
\end{figure}

Fig.~\ref{fig:fnoisepsd} shows a toy example of Eqn.~\ref{eqn:astro_fnoise2}, taking slices of the PSD along the smallest wavenumber mode in time and frequency for an arbitrary receiver system with a $\beta = 0.25$ and $\alpha = 1$. It is expected that $1/f$ noise will be highly correlated, and as such the $\beta$ value will be small. Fig.~\ref{fig:fnoisepsd} shows that in this case the spectroscopic $1/f$ noise slope will be very steep, and may only dominate on the very largest spectroscopic scales making measurements of $\beta$ challenging. Fig.~\ref{fig:fnoisepsd} also defines a \textit{knee} for the spectroscopic $1/f$ noise fluctuations ($\omega_k$) that is not defined in Eqn.~\ref{eqn:astro_fnoise2}. This knee is intrinsically linked to the temporal knee frequency $f_k$ by
\begin{equation}
	\omega_k = \left(f_k T_\mathrm{obs}\right)^{\frac{\alpha \beta}{1-\beta}} \Delta \nu^{-1} ,
\end{equation}
where $T_\mathrm{obs}$ is the total observing time per receiver and $\Delta \nu$ is the total bandwidth of the receiver.

\section{Pipeline and Simulations}\label{sec:pipeline}

Mock datasets are a crucial component of any precision cosmology experiment. For systematics in radiometers that are time dependent, the only way to study how these contaminants interact with the signal of interest is through end-to-end simulations. Within this section a description of an end-to-end simulation pipeline for SD HI IM experiments is given. The pipeline simulates the expected total power outputs from each radiometer within an arbitrary array. This includes modelling the telescope array and scanning strategy in subsection \ref{sec:design}, the expected emission from sky (including the cosmological HI signal) in subsection \ref{sec:sky}, and the generation of the $1/f$ noise signal in subsection \ref{sec:1f}. The pipeline has been designed to be modular such that in the future it can be expanded to include other systematics expected to impact HI IM experiments such as RFI, standing waves, calibration errors and more. The pipeline is written in a combination of \texttt{Python} and \texttt{FORTRAN}, and makes use of MPI functionality though the \texttt{MPI4Py} module \citep{Dalcin2011}.

\subsection{Instrument Design}\label{sec:design}

The simulated SKA array described in this Section is based on the design outlined in the SKA 2016 baseline document \citet{SKABaseline2016}, and several recent SKA HI IM forecast and modelling papers \citep{Santos2015,Bull2015,Alonso2015,Villaescusa2017}. The simulated survey will use a subset of the Band 2 frequencies spanning the frequency range $950 < \nu < 1410$\,MHz with $N_{\nu}=23$ frequency channels, each with a channel width of $\delta \nu = 20$\,MHz. The choice of channel width was motivated to optimise between the effectiveness of the component separation, which performs better with more channels \citep{Olivari2016}, and computational efficiency. The corresponding upper redshift limit will be $z \approx 0.5$. The choice of Band 2 over Band 1 is based on the limited resolution of the SKA1-MID dishes at redshifts greater than $z > 0.5$ resulting in an insensitivity to the BAO signal as discussed in Section \ref{sec:overview}. The number of dishes used is $N_{\mathrm{dish}} = 200$, which is several more than the current baseline SKA description, but parametrisation of the results presented later will allow this number to be scaled to that of the resulting final array. The precise telescope positions do not have a significant impact on the final results. Table \ref{table:survey} provides a description of all the other parameters.

\ctable[
  caption = Input parameters describing the simulated SKA1-MID array and instrumentation.,
  label = table:survey,
  width = 0.45\textwidth
]{l|c|c}{
\tnote[a]{The slightly smaller dish sizes used in the MeerKAT array ($D_\mathrm{dish} = 13.5$\,m) is ignored for simplicity.}
\tnote[b]{The total system temperature also includes a frequency and position dependent contribution from the sky, however the receiver temperature ($T_\mathrm{rx}$) and CMB temperature ($T_\mathrm{CMB}=2.73$\,K) are assumed to be constant.}
	}{\FL Description & Parameter & Value \ML
  Dish Diameter & $D_{\mathrm{dish}}$ &  15\,m\tmark[a] \NN
  No. Dishes & $N_{\mathrm{dish}}$ & 200  \NN
 Receiver + CMB & $T_\mathrm{CMB} + T_{\mathrm{rx}}$ & 20\,K\tmark[b] \NN
 No. Polarimeters & $N_{\mathrm{pol}}$ & 2   \NN
 No. Channels & $N_{\nu}$ & 23 \NN 
 Bandwidth & $\Delta \nu$ & $950 < \nu < 1410$\,MHz  \NN
 Channel width & $\delta \nu$ & 20\,MHz \NN
 Sample Rate & $f_\mathrm{sr}$ & 4\,Hz \NN
 Integration Time & $T_\mathrm{obs}$ & 30\,days \NN
 Elevation & E & 55\,deg \NN
 Slew Speed & $\vt$ & $0.5  <  \vt < 2.0$\,deg\,s$^{-1}$ \LL}

The beam of each telescope is assumed to be Gaussian and the full-width half-maximum (FWHM) of the beam scales with wavelength ($\lambda$) as
\begin{equation}
\theta_{\mathrm{FWHM}} = 1.1 \frac{\lambda}{D_{\mathrm{dish}}} .
\end{equation}
The scaling factor of 1.1 comes from measurements of the SKA-MID primary beam models (Robert Lehmensiek, \textit{priv. comm.}). For these simulations it is assumed that all observations are taken at the same resolution equal to the resolution at 950\,MHz, which corresponds to $\theta_\mathrm{FWHM} = 1.33$\,deg. This approximation was made to simplify the noise estimates in data analysis stage of the simulations.

The chosen observing strategy is to continuously slew the SKA dishes 360\,deg at a constant elevation of 55\,deg. Observing at a constant elevation is commonly used by ground based SD survey instruments such as \citep[CBASS,][]{Irfan2015} and \citep[QUIJOTE,][]{Rubino2010}. The advantage of scanning at a constant elevation is to control systematics such as the local horizon or ground pick-up. In practice several elevations would be used to more evenly distribute observing time over the entire observed field, however to simplify the data analysis of the simulations only one elevation is used here. The choice of 55\,deg elevation results in approximately a survey area of 20500\,sq.\,deg, which was chosen to be comparable with the survey areas proposed in \citet{Santos2015} and \citet{Bull2015}. Simulations were generated for surveys with slews speeds of $\vt = 0.5$, 1, and 2\,deg\,s$^{-1}$. These speeds were chosen as the SKA-MID dishes are designed to slew at speeds $< 1$\,deg\,s$^{-1}$, while maintaining sufficient pointing accuracy \citep{SKABaseline2016}. The simulations at 2\,deg\,s$^{-1}$ are provided to determine whether there is sufficient motivation to push to higher slew speeds. The chosen receiver sample rate of 4\,Hz is sufficient to Nyquist sample the sky up to 2\,deg\,s$^{-1}$ at an elevation of 55\,deg.

\subsection{Sky Model}\label{sec:sky}

\subsubsection{Synchrotron}\label{sec:sync}

Synchrotron radiation originates from the interaction of relativistic cosmic ray electrons (CRE) interacting within the Galactic magnetic field. The dominant source of CRE are generated in supernovae and have a lifetime of approximately $10^5 - 10^6$\,years, this means that synchrotron emission is a tracer for relatively recent star formation within the Galaxy \citep{Condon1992}. At low radio frequencies synchrotron emission is vastly brighter than any other emission from the sky. The best measurement of the all sky-distribution of Galactic synchrotron intensity comes from a reprocessing of the \citet{Haslam1982} 408\,MHz all-sky map by \citet{Remazeilles2015}.

It is well known that Galactic synchrotron emission varies smoothly with frequency and the ensemble average of the emission along a given line-of-sight can be approximated by a power-law \citep[e.g.,][]{Scheuer1968,OliveiraCosta2008,PlanckXXV2016}. The synchrotron model used in this work is
\begin{equation}
	T(\nu, \hat{n}) = T_{408\mathrm{\,MHz}}(\nu, \hat{n}) \left(\frac{\nu}{408\mathrm{\,MHz}}\right)^{\alpha_s(\nu, \hat{n})},
\end{equation}
where $\nu$ is frequency, $\hat{n}$ is a line-of-sight on the sky, $T_{408\mathrm{\,MHz}}$ is the pixel amplitude of the 408\,MHz map, and $\alpha_s$ is the spectral index for a given line-of-sight. The spectral index from pixel-to-pixel was estimated using the all-sky spectral index map by \citet{Platania2003}. The spectral indices do not include spectral curvature. The \textit{top} image of Fig.~\ref{fig:skymaps} shows the simulated Galactic synchrotron emission at 1190\,MHz for the simulated SKA strip of sky. The bright features to the \textit{left} and \textit{right} of the image are saturated slices through the Galactic plane emission, which were masked before performing the component separation analysis. Looking at the total simulated emission map, which combines all foregrounds and the HI signal, it is clear that the synchrotron emission is the dominant foreground by at least an order-of-magnitude. It is also clear that, even after masking out the brightest low Galactic latitude emission, the residual foreground components are still far brighter than the underlying HI signal.

\subsubsection{Free-Free}\label{sec:freefree}

\begin{figure*}
\centering
\includegraphics[width=0.9\textwidth]{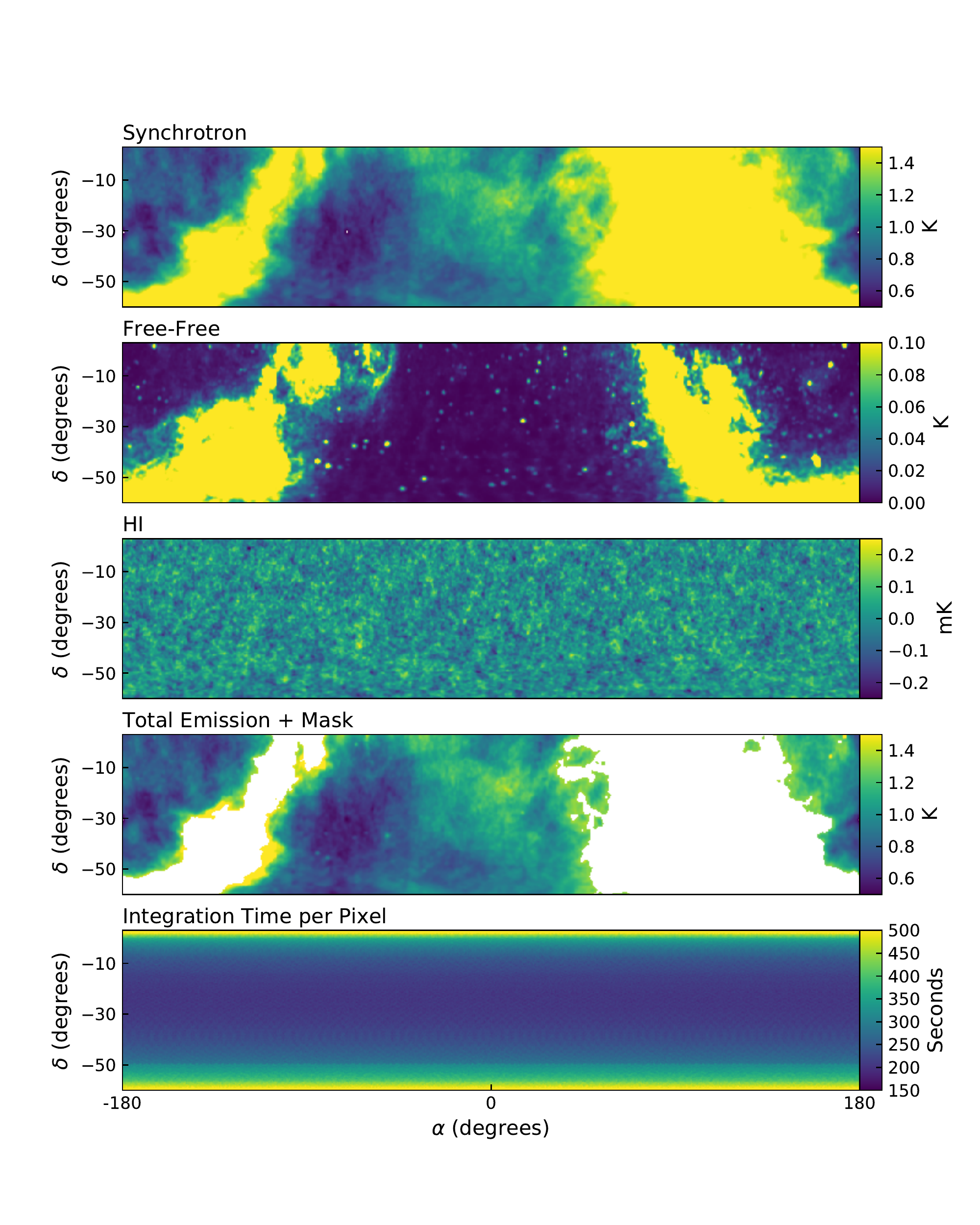}
\caption{Cartesian projections in celestial coordinates of the simulated SKA HI IM survey strip for (from \textit{top} to \textit{bottom}): Galactic synchrotron emission (K), Galactic free-free emission (K), the cosmic HI intensity field (mK), the combined sky emission (K), and the integration time distribution per pixel (seconds) for the adopted scanning strategy. All images use a \texttt{HEALPix} grid with $N_\mathrm{side}=512$ and centred at Right Ascension $\alpha = 0$ and Declination $\delta = -28.5$\,deg. All colour scales are linear with high brightness regions saturated to highlight the low brightness emission. Due to the projection of spherical data onto a Cartesian plane low declination region will look stretched in the Right Ascension direction. }\label{fig:skymaps}
\end{figure*}

Free-free radiation is generated from unbound interactions within regions of the interstellar medium that have been ionized by nearby OB stars. Free-free emission is a very well understood Galactic foreground component and for diffuse, optically thin HII regions have a spectral index of $\alpha_s = -2.1$ at frequencies around $1$\,GHz \citep[e.g.,][]{Draine2011}. For low Galactic latitudes the radio free-free emission was simulated using estimates of emission measure from the \textit{Planck} Galactic component maps \citep{PlanckX2016}. However, for these particular simulations such low Galactic latitudes are masked, and the remaining intermediate-to-high Galactic latitude free-free emission was modelled using the combined all-sky H$\alpha$ emission map and H$\alpha$-to-radio relations described in \citet{Dickinson2003}. For both models the mean electron temperature was fixed at 7000\,K for all lines-of-sight \citep{Alves2012}. Although free-free emission is the sub-dominant foreground as shown in Fig.~\ref{fig:skymaps}, it will still act to flatten the overall foreground spectrum at higher frequencies. Therefore free-free emission effectively adds spectral curvature to the foreground components, making the foreground signal more challenging to characterise and remove.

\subsubsection{HI}\label{sec:HI}

The HI emission at low redshifts of $z < 0.5$ is assumed to be confined to galaxies. The HI signal was simulated using the model for the mean HI brightness described in \citet{Battye2013}
\begin{equation}
\bar{T}_{\mathrm{obs}}(z) = 44\,\mu\mathrm{K} \left(\frac{\Omega_{\mathrm{HI}}(z)h }{2.45 \times 10^{-4}} \right) \frac{(1 + z)^2}{E(z)} ,
\end{equation}
where $\bar{T}_\mathrm{obs}$ is the mean observed brightness temperature of the HI at redshift $z$, $\Omega_{\mathrm{HI}}$ is the neutral HI fraction, and $E(z)$ describes the Hubble expansion. It is assumed that there is no evolution in the neutral HI fraction with redshift and a value of $\Omega_{\mathrm{HI}} = 6.2 \times 10^{-4}$ is adopted \citep{Switzer2013}. 

The underlying cold dark matter power spectrum, $P_{\mathrm{cdm}}$, was generated using  the \texttt{CAMB} software package \citep{Lewis2002}. To project the matter spectrum into the angular power spectrum at each redshift  the Limber approximation was assumed \citep{Datta2007},
\begin{equation}
C_{\ell} = \frac{H_0 b^2}{c} \int \mathrm{d}z E(z) \left[ \frac{\bar{T}_{\mathrm{obs}}(z) D(z)}{r(z)} \right]^2 P_{\mathrm{cdm}}\left(\frac{\ell + \frac{1}{2}}{r} \right)
\end{equation}
where $r(z)$ is the comoving distance out to redshift $z$, $D(z)$ is the growth factor and $b$ is the HI bias (assumed here to be constant and unity). Each map realisation was generated using the \texttt{synfast} module within the \texttt{HEALPix} package \citep{Gorski2005}. Fig.~\ref{fig:skymaps} shows the HI emission for the simulated SKA strip at 1190\,MHz. The figure demonstrates the foreground subtraction challenge faced by HI IM experiments as the typical fluctuations of the HI signal in figure are of the order $\sim 10$\,$\mu$K, while the largely Galactic synchrotron foreground fluctuations are several orders-of-magnitude higher around $\sim 100$\,mK. 

\subsection{Modelling 1/f Noise}\label{sec:1f}

The $1/f$ noise model used in these simulations assumes that the source of the $1/f$ fluctuations in the time-ordered data (TOD) originate as correlated fluctuations in the gain of the receiver amplifiers. The $1/f$ noise fluctuations are therefore small multiplicative variations around the system temperature. The $1/f$ noise can be represented in the TOD as 
\begin{equation}
	\Delta T(t, \nu) = \delta G(t, \nu) \tsys (t, \nu) ,
\end{equation}
where $\Delta T(t, \nu)$ is the power of the $1/f$ fluctuations for a given interval, which is the combination of the instantaneous fluctuation in the gain $\delta G$ and system temperature $\tsys$. It is the fluctuations in the gain, $\delta G$, that this section will describe.

The power spectral density ($\psd$) of $\delta G$ can be described by four parts. The first is 
\begin{equation}\label{eqn:timecorr}
	F(f) = \frac{1}{\delta \nu}\left( \frac{f_k}{f} \right)^\alpha ,
\end{equation}
which is the power spectrum of the temporal fluctuations as in Eqn.~\ref{eqn:astro_fnoise} but without the white noise component. $\delta \nu$ describes the frequency channel bandwidth, $f_k$ is the knee frequency, and $\alpha$ is the spectral index of the noise.

The second component of the $1/f$ power spectrum describes the correlations of the noise in frequency. At present there is no literature on the possible functional form of the $1/f$ noise frequency-space PSD, and so a conservative power-law model was adopted (similar to the temporal fluctuations)
\begin{equation}\label{eqn:freqcorr}
	H(\omega) = \left( \frac{\omega_0}{\omega} \right)^{\frac{1-\beta}{\beta}}
\end{equation}
where $\omega$ is the Fourier mode of the spectral frequency (i.e. the wave number), $\omega_0$ is the smallest wave number ($1/N_\nu\Delta\nu$), and $0 \leq \beta \leq 1$ describes the correlations in frequency. For example, $\beta = 1$ describes $1/f$ noise that is entirely uncorrelated in frequency and $\beta=0$ would result in the perfect correlation of the $1/f$ noise across channels. This model assumes no physical interpretation for the origin of the $1/f$ noise frequency correlations.

\begin{figure*}
\centering
	\begin{subfigure}[b]{0.45\textwidth}
    	\includegraphics[width=\textwidth]{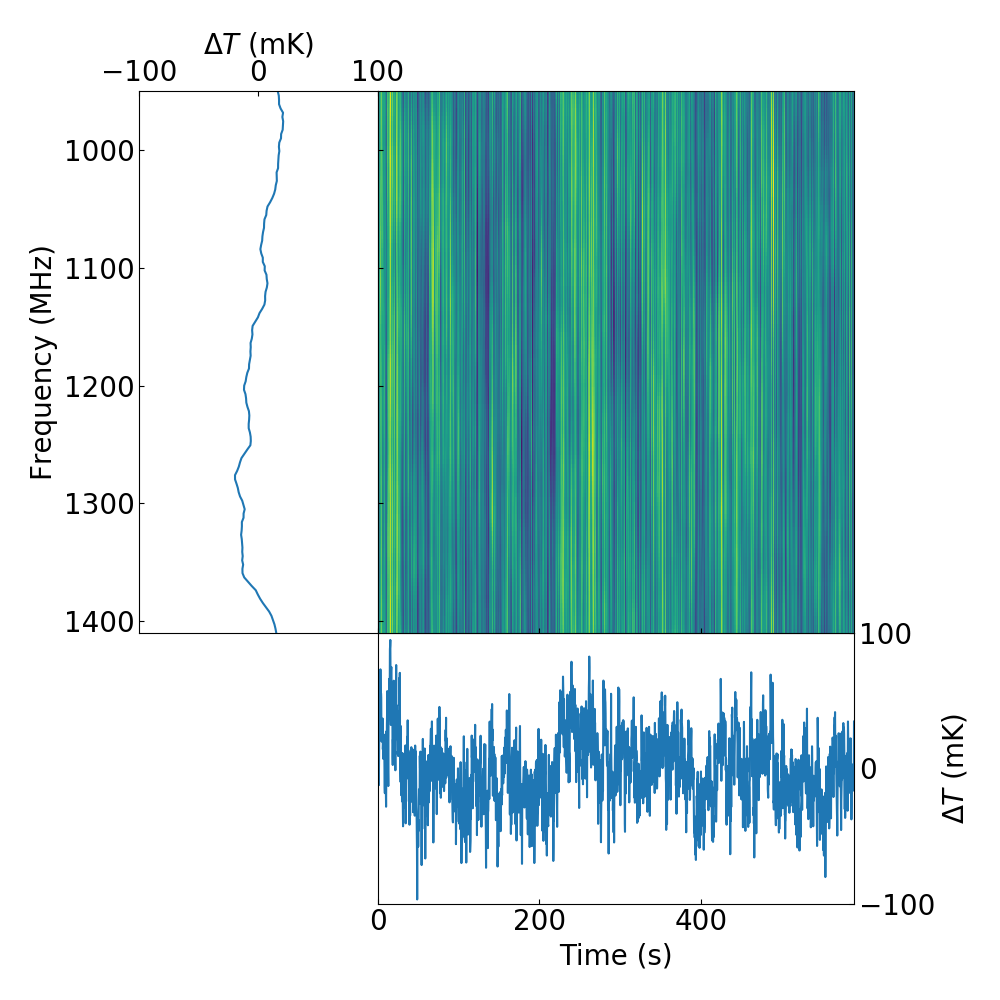}
        \caption{$\beta = 0.25$, $\alpha = 1$, $f_{\mathrm{k}} = 1$\,Hz}\label{fig:TODGridA}
	\end{subfigure}
	\begin{subfigure}[b]{0.45\textwidth}
    	\includegraphics[width=\textwidth]{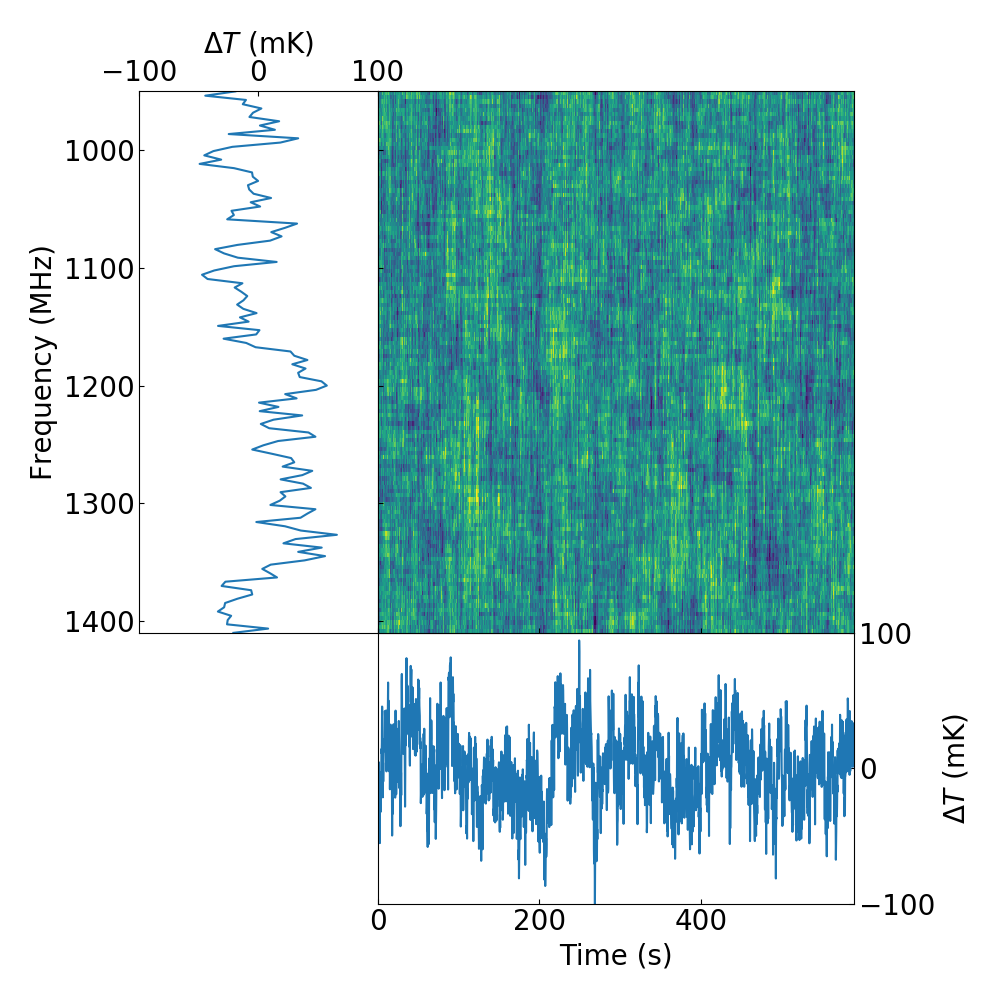}
        \caption{$\beta = 0.5$, $\alpha = 1$, $f_{\mathrm{k}} = 1$\,Hz}\label{fig:TODGridB}
	\end{subfigure}
	\begin{subfigure}[b]{0.45\textwidth}
    	\includegraphics[width=\textwidth]{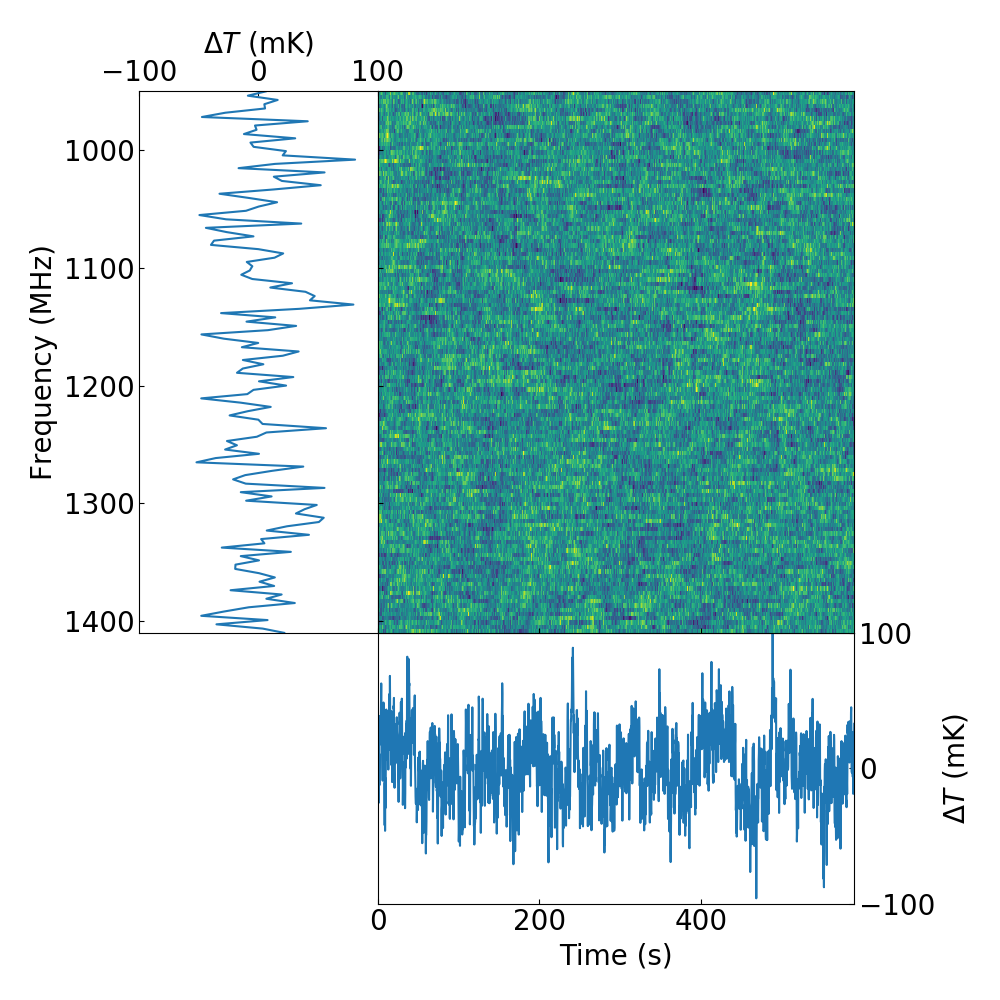}
        \caption{$\beta = 0.75$, $\alpha = 1$, $f_{\mathrm{k}} = 1$\,Hz}\label{fig:TODGridC}
	\end{subfigure}
	\begin{subfigure}[b]{0.45\textwidth}
    	\includegraphics[width=\textwidth]{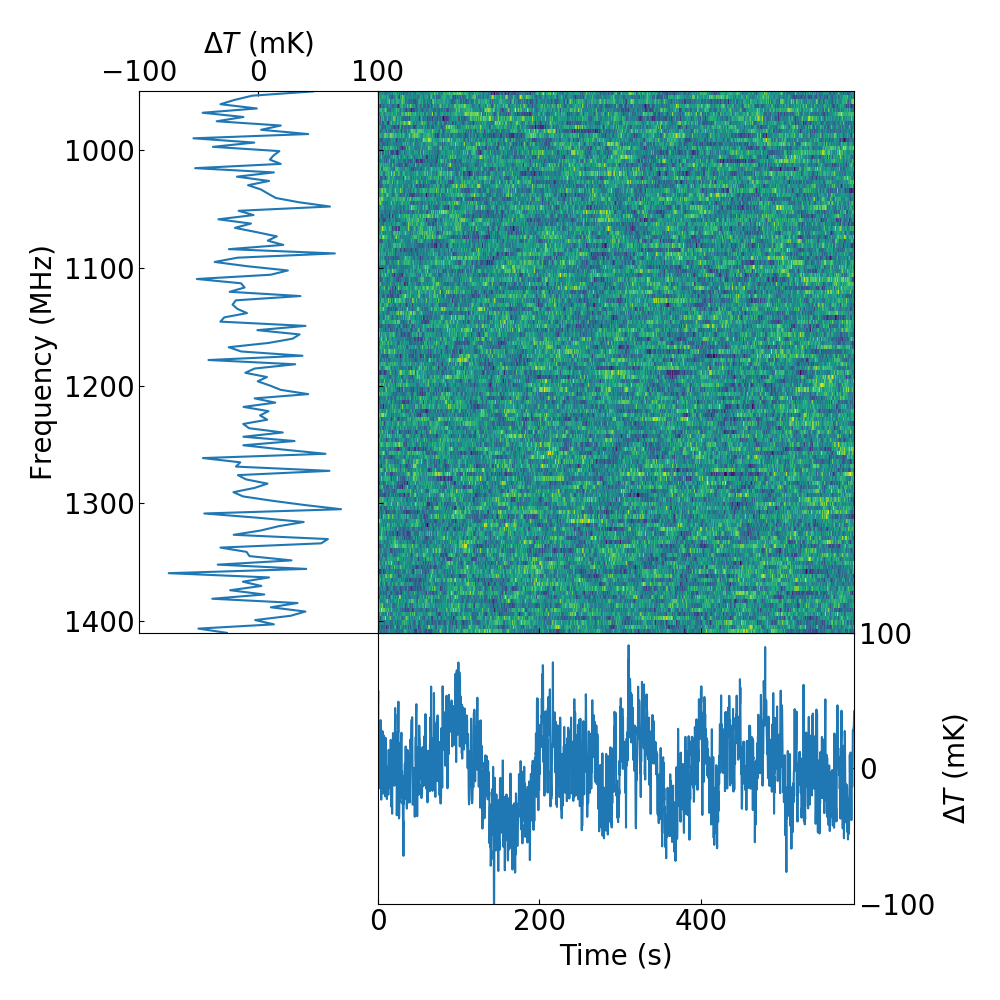}
        \caption{$\beta = 1.0$, $\alpha = 1$, $f_{\mathrm{k}} = 1$\,Hz}\label{fig:TODGridD}
	\end{subfigure}

\caption{Time-frequency plots of $1/f$ noise generated using the simulation pipeline. Fig.~\ref{fig:TODGridA} is example of highly correlated $1/f$ noise ($\beta = 0.25$), while Fig.~\ref{fig:TODGridB} ($\beta = 0.5$) and Fig.~\ref{fig:TODGridC} ($\beta = 0.75$) are increasingly uncorrelated in frequency, and finally Fig.~\ref{fig:TODGridD} ($\beta = 1$) is an example of completely uncorrelated 1/f noise where each channel has a unique noise realisation. The images represent waterfall plots, with time along the $x$-axis and frequency along the $y$-axis. The parallel and vertical plots associated with each image are slices along either a single channel or time interval respectively. These figures do not include any white noise contribution.}\label{fig:TODGrid}
\end{figure*}

Fig.~\ref{fig:TODGrid} shows examples of $1/f$ noise simulated with different values of $\beta$ with stationary temporal noise properties (e.g., the total r.m.s., $f_k$ and $\alpha$ are all constant with time). The figure shows the Fourier transform of Eqn.~\ref{eqn:timecorr} and Eqn.~\ref{eqn:freqcorr} combined with randomised phases. The mean frequency spectrum and the mean temporal fluctuations are both zero. The simulated gain fluctuations should be interpreted as ripples across a 2D surface. Strictly there is no white noise component in these simulations, however when $\beta = 1$ the frequency spectrum is entirely uncorrelated giving an effective appearance of white noise. This means that the number of modes needed to describe the $\beta = 1$ $1/f$ noise is equal to the number of channels, therefore removing the $1/f$ noise will be very challenging for typical component separation methods. Conversely, fewer modes are required for $1/f$ noise with $\beta < 1$, implying that component separation will find it easier to remove the $1/f$ noise contamination. 

\begin{figure*}
\centering
	\begin{subfigure}[b]{0.48\textwidth}
    	\includegraphics[width=\textwidth]{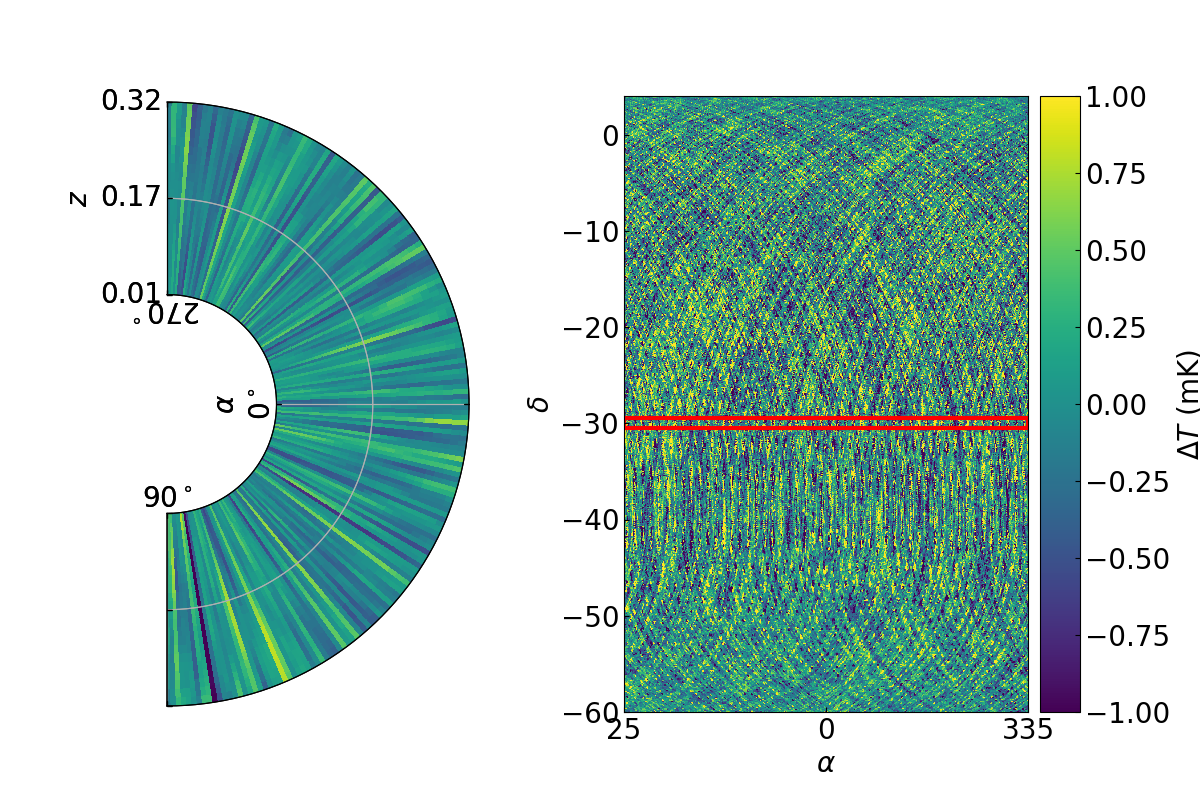}
        \caption{$\beta = 0.25$, $\alpha = 1$, $f_{\mathrm{k}} = 1$\,Hz}\label{fig:RadialGridA}
	\end{subfigure}
	\begin{subfigure}[b]{0.48\textwidth}
    	\includegraphics[width=\textwidth]{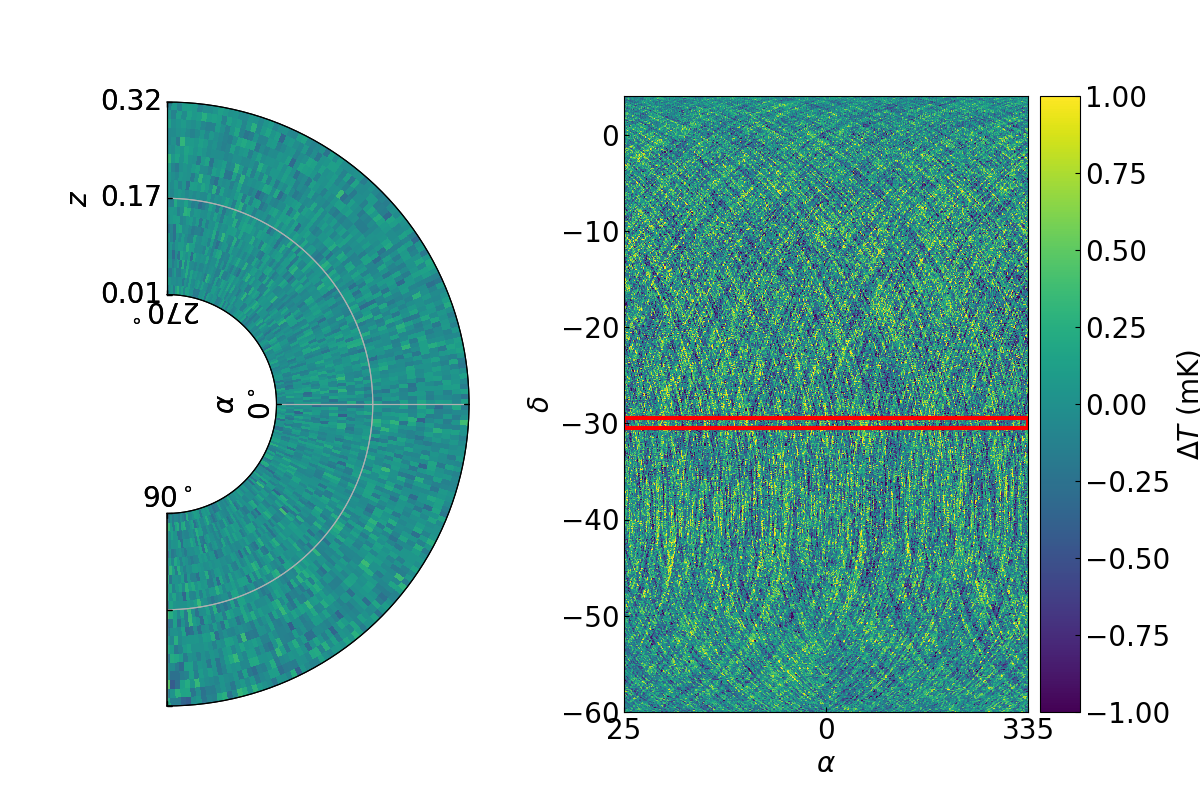}
        \caption{$\beta = 1.0$, $\alpha = 1$, $f_{\mathrm{k}} = 1$\,Hz}\label{fig:RadialGridB}
	\end{subfigure}
\caption{Spatial and radial slices of $1/f$ noise with $\beta = 0.25$ (Fig.~\ref{fig:RadialGridA}), and $\beta = 1.0$ (Fig.~\ref{fig:RadialGridB}). The radial slice through redshift in the \textit{left-hand} plots is taken from the declination highlighted by the red stripe in the \textit{right-hand} spatial image. The radial slices are presented as arcs to illustrate that these are cuts through a sphere. Due to the different frequency correlation parameters the $1/f$ noise averages down far slower in Fig.~\ref{fig:RadialGridA} than in Fig.~\ref{fig:RadialGridB}. }\label{fig:RadialGrid}
\end{figure*}

For each example in Fig.~\ref{fig:TODGrid} the r.m.s. of the temporal fluctuations is fixed, however the amplitude of the fluctuations in frequency depend on the $\beta$ correlation parameter, with the uncorrelated noise having far larger fluctuations than the correlated fluctuations. Over long integrations the uncorrelated $1/f$ noise will average as $\sqrt{1/T}$ (where $T$ is integration time), whereas the correlated $1/f$ noise will average down slower depending on the $\beta$ parameter. This can result in the reversal of what is observed in Fig.~\ref{fig:TODGrid}, as demonstrated by Fig.~\ref{fig:RadialGrid}. The figure shows $1/f$ noise within the simulated SKA field after 30\,days of integration, and that the resulting $1/f$ noise fluctuations in frequency are higher for $\beta = 0.25$ than $\beta = 1$. For the analysis in this work the higher amplitude of the correlated fluctuations after integration is not problematic as component separation is very effective at subtracting smoothly varying correlated signals. However, such an effect could be a concern for experiments interested in the cosmological HI signals frequency-space correlations such as redshift space distortions. Further, the assumption that the $1/f$ noise integrates down as $\sqrt{1/T}$ is dependent on both the Gaussianity and stationarity of the $1/f$ noise, either of which may not be true for real instruments. 

For small numbers of channels, $\beta$ and the amplitude of the $1/f$ noise frequency fluctuations become highly dependent when using Eqn.~\ref{eqn:freqcorr} due to aliasing. For example, in the extreme case of just one channel, the choice of $\beta$ no longer matters and the r.m.s. of the fluctuations must be zero. However, for these simulations a sufficient number of frequency channels have been used such that the dependence of $\beta$ and  amplitude of the fluctuations on the number of frequency channels is expected to be minimal.

A high-pass cosine filter is applied to the $\psd$ to remove correlations on very long time-scales. The filter is used in these simulations to avoid any additional complexity in the interpretation of the results that may be due using an advanced map-making method. The filter has the form
\begin{equation}\label{eqn:window}
	W(f) = \begin{cases}
    	1, & \text{if}\quad |f| > f_c + \frac{f_w}{2} \\
        0, & \text{if}\quad |f| < f_c - \frac{f_w}{2} \\
        \frac{1}{2}\left[ 1 + \cos\left(\pi \frac{|f| - f_c - f_w/2}{f_w}\right) \right], & \text{otherwise}
    \end{cases}
\end{equation}
where $f_c$ is the temporal frequency for which half the signal power is filtered, and $f_w$ is the width of the filter. The choice of the filter cutoff scale was chosen to synchronise with the period of a single 360\,deg slew in azimuth as this is both easy to achieve in reality and suppresses spikes in the power spectrum caused by the interaction of the $1/f$ noise and the scan strategy at the pixel scale. Though scan speed is one variable investigated in this paper, the actual details of the scan strategy are not. The overall power of the $1/f$ noise angular power spectrum is only weakly dependent on filter scale used as long as the filter period is greater than the slew time.

The final component needed to characterise the $1/f$ noise $\psd$ is a normalisation factor that forces the temporal variance of $\delta G$ to be constant regardless of the $\beta$ chosen for the correlations in frequency. The correction factor is
\begin{equation}\label{eqn:fudge}
	C(\beta, N_{\nu}) = \frac{N_{\nu}}{1 + \frac{\sum\limits_{i=1}^{N_\nu/2}H(\omega_i, \beta)\Delta\omega}{\int H (\omega, 0)\text{d}\omega } \left(N_\nu - 1 \right) } ,
\end{equation}
where $N_\nu$ is the number of channels. For a small number of channels it is important to use a discrete sum and not the continuous integral over Eqn.~\ref{eqn:freqcorr}.

The resulting $1/f$ $\psd$ can now be described by
\begin{equation}
	\psd(f,\omega) = F(f) H(\omega) W(f) C(\beta, N_\nu) ,
\end{equation}
where $F(f)$ is from Eqn.~\ref{eqn:timecorr}, $H(\omega)$ is Eqn.~\ref{eqn:freqcorr}, $W(f)$ is Eqn.~\ref{eqn:window} and $C(\beta, N_\nu)$ is Eqn.~\ref{eqn:fudge}. Each $1/f$ noise TOD realisation is generated using Gaussian distributed phases with means equal to the square-root of the power as
\begin{equation}
	Z(f, \omega) = \sqrt{\psd(f,\omega)}\left( x + iy \right)
\end{equation}
where $x$ and $y$ are Gaussian random variates with zero mean and unit variance. The Fourier transform of $Z$ results in the $1/f$ gain fluctuations 
\begin{equation}
	\delta G (t, \nu) = \int \int \text{d}f \text{d}\omega \, Z(f, \omega) e^{2\pi i  (t f + \nu \omega ) } .
\end{equation}
It is an assumption of these simulations that the $1/f$ noise is Gaussian distributed. In real data that may not be true. Also, as the $1/f$ noise is multiplicative with system temperature, which can be time variable, it is non-stationary in terms of its variance (e.g., the r.m.s. of the $1/f$ noise increases when observing bright sources on the sky), but the $f_k$, $\alpha$ and $\beta$ of the $1/f$ noise are all assumed stationary, which also may not be true in real observations. Alone or in combination these effects will influence how the $1/f$ noise averages down on long time scales.

\subsection{Sky Mask}\label{sec:mask}

Before performing the component separation and power spectrum estimation discussed in the proceeding subsections, the output of the sky maps from the simulation pipeline were multiplied with a mask. The mask  is based on a threshold cut of 40\,K in the Haslam 408\,MHz sky map \citep{Haslam1982,Remazeilles2015}, which was chosen to agree with other SKA simulations  by  \citet{Alonso2015}. It removes just the brightest regions of the Galactic plane. The mask also removes \textit{edge-pixels} that, due to the arrangement of the SKA dishes, have very few hits in the highest and lowest observed declinations relative to the rest of the observed sky. Fig.~\ref{fig:skymaps} shows the regions that have been masked, the resulting total observed sky fraction after masking is $f_\mathrm{sky} = 0.3$.

\subsection{Component Separation}\label{sec:pca}

There are currently a wide range of parametric and non-parametric  methods for component separation that have been developed during the past decade for CMB experiments, but are now being applied to HI IM. Examples include FASTICA \citep{Chapman2012}, correlated component analysis \citep{Bonaldi2015}, GMCA \citep{Chapman2012}, and GNILC \citep{Olivari2016}. For the purposes of these simulations the method of Principle Component Analysis (PCA) was chosen. PCA is not as sophisticated a method as those mentioned above, and as such is not as robust at removing foregrounds. However, PCA has the advantage of being computationally fast, making it highly suited to Monte-Carlo simulations. Also, PCA relies upon eigenvalue decomposition, which is at the heart of most non-parametric component separation methods.

The method of PCA  works by assuming that a map that is composed of $N_{\mathrm{pix}}$ pixels and $N_{\mathrm{\nu}}$ frequencies can be decomposed into a set of $N_{\mathrm{\nu}}$ independent components via a mixing matrix  as
\begin{equation}
\textbf{m} = \textbf{U} \textbf{s} ,
\end{equation}
where $\textbf{m}$ is the set of $N_{\mathrm{\nu}}$ maps, $\textbf{U}$ is the mixing matrix and $\textbf{s}$ is the set of $N_\nu$ independent components. The formalism adopted here is that vectors are lower-case, bold characters, matrices are upper-case, bold characters, and rank-0 variates are lower-case and not bold. 

The  covariance  matrix of the maps is defined as 
\begin{equation}
\textbf{m} \textbf{m}^{\mathrm{T}} =  \textbf{C} = \textbf{U} \textbf{s} \textbf{s}^{\mathrm{T}} \textbf{U}^{\mathrm{T}} = \textbf{U} \Lambda \textbf{U}^{\mathrm{T}} ,
\end{equation}
where $\textbf{C}$ is the covariance matrix, and it has been assumed that since each component of $\textbf{s}$ is independent,  the outer product yields a  diagonal matrix $\Lambda$. To determine the mixing matrix $\textbf{U}$ and  the diagonal matrix $\Lambda$ an eigenvalue decomposition can be performed on $\textbf{C}$.  This means that diagonal elements of $\Lambda$ are the eigenvalues of $\textbf{C}$ and the columns  of the mixing matrix $\textbf{U}$ are the corresponding eigenvectors. The mixing matrix $\textbf{U}$ also obeys the condition that 
\begin{equation}
\textbf{U}^{\mathrm{T}} \textbf{U} = \textbf{U} \textbf{U}^{\mathrm{T}} = \textbf{I} ,
\end{equation}
where $\textbf{I}$ is the identity matrix. 

The eigenvalues of $\textbf{C}$ describe the relative variance of each of the components in $\textbf{m}$. As the Galactic foregrounds are bright relative to the HI signal, and highly correlated it is assumed that they can be described by just a few bright eigenvalues \citep[e.g.,][]{BigotSazy2015}. Therefore, to remove the foregrounds, a subspace of the mixing matrix $\tilde{\textbf{U}}$ is derived by removing the eigenvectors associated with the largest eigenvalues.  The  foreground subtracted map is then defined as
\begin{equation}
	\tilde{\textbf{U}} \textbf{U}^{\mathrm{T}} \textbf{m}  =  \tilde{\textbf{m}}
\end{equation}
where $\tilde{\textbf{m}}$ is the estimated foreground subtracted sky.

After applying the PCA method to the simulated datasets with no $1/f$ noise it was found that the optimal number of eigenmodes to subtract were $N_\lambda = 6$. This minimises the recovered variance of the HI power spectrum due to foregrounds, but at the cost of the two lowest frequency channels being over subtracted. For the rest of these simulations these two channels are neglected.

Non-parametric methods of foreground subtraction are limited by the available number of frequency channels, $N_{\nu}$, and cannot remove a foreground component that can be decomposed into $N_{\mathrm{\lambda}} > N_{\nu}$. Similarly if the sky has large variations in the frequency response of the foregrounds then  PCA may work better along some lines-of-sight more than others. One very important property of PCA is that it will always remove components that are completely correlated in frequency. This means that $1/f$ noise with $\beta = 0$ will always be perfectly removed. This has been verified and therefore is not included in the results as it is equivalent to the white noise only case. A simple proof of this assertion is presented in Appendix \ref{app:A}.

\subsection{Power Spectrum Estimation}\label{sec:power}

For each realisation, frequency and eigenmode the angular power spectrum was measured. Each map was decomposed into spherical harmonic coefficients ($a_{lm}$) \citep{Peebles1973,Wandelt2001},
\begin{equation}
	a_{\ell m} = \int\limits_{\Omega}  m(\Omega) Y_{\ell m}(\Omega) \mathrm{d}\Omega,
\end{equation}
where $m$ is the observed sky map, $Y_{\ell m}$ are spherical harmonics and $\Omega$ describes a given line-of-sight. The angular power spectrum is then
\begin{equation}
	C_\ell^\text{obs} =  \frac{1}{2 \ell + 1}\sum\limits_m | a_{\ell m}|^2  .
\end{equation}
Here, $C^\text{obs}_\ell$ describes the observed HI angular power spectrum after foreground subtraction. The observed HI angular power spectrum, in the presence of Gaussian white noise only, is related to the underlying HI spectrum realisation by
\begin{equation}
	C_\ell = \left< C_\ell^\text{obs} - N_\ell \right> = \left< \cest \right>,
\end{equation}
where the brackets indicate an average over many noise realisations of the white noise power spectrum, and $\cest$ is the estimate of the $C_\ell$. Note that $C_\ell$ in this notation is not the underlying theoretical HI angular power spectrum, these are related by
\begin{equation}
	\left< C_\ell \right> = C_\ell^\text{theory}B_\ell^2,
\end{equation}
where the average now is over many HI sky realisations and $B_\ell$ is a beam window function.

The spherical harmonic transform and power spectrum estimation of the simulated maps were performed using the publicly available \texttt{PolSpice} software \citep{Chon2004}. As only a partial fraction of the sky was observed, neighbouring $C_\ell$ components will be correlated. The correlations can be damped by binning components in $\ell$ with an approximate analytical bin width of \citep{Hauser1973}
\begin{equation}
	\Delta \ell \approx \frac{\pi}{\Delta\theta_{0}},
\end{equation}
where $\Delta \theta_0$ is the span of the polar angle over the observed sky area. For these simulations a binning width of $\Delta \ell = 5$ was chosen. To further reduce the impact of the masked sky the correlation functions were apodized by 110\,deg, with a maximum angular size set to 110\,deg. These values were determined empirically by finding the values that minimized high-$\ell$ ringing while not biasing the recovered angular power spectrum at small-$\ell$.

Before power spectrum estimation each map is smoothed to a common FWHM resolution of 1.3\,deg (corresponding to the resolution in the 950\,MHz channel). As smoothing is performed after masking, power will be lost from pixels near the mask boundaries, which will result in a biased recovery of the HI power spectrum. To avoid this \texttt{PolSpice} was provided an apodized weight map. The weights were derived by calculating the distance of each unmasked pixel from the nearest masked region using the \texttt{HEALPix} routine \texttt{process\_mask}, and passing these distances through a cosine filter function with a characteristic width of 2\,deg.

Estimated uncertainties in the angular power spectrum were measured from the standard deviation of all realisations in a given $\Delta \ell$ bin using
\begin{equation}\label{eqn:fmodel}
 \Delta F_\ell = \sqrt{ \left< \hat{C}_\ell^2 \right> - \left< \hat{C}_\ell \right>^2} ,
\end{equation}
where $\Delta F_\ell$ here denotes the uncertainty in the power spectrum due to a combination of white noise, $1/f$ noise, residual foregrounds and any other contaminant within the power spectrum. Note that in most cases discussed later only the additional uncertainty over the expected white noise is of interest, therefore an estimate of the white noise uncertainty is required. The mean angular power spectrum for white noise that is uniformly distributed is well known as \citep{Knox1995}
\begin{equation}\label{eqn:meanwhite}
	N_\ell = \sigma_\text{pix}^2 \Omega_\text{pix} ,
\end{equation}
where $\sigma_\text{pix}$ is the standard deviation of pixels in the map domain, and $\Omega_\text{pix}$ is the solid angle of each pixel. However, for specific realisations, the mean white noise angular power will be different from  Eqn.~\ref{eqn:meanwhite} and a transfer function that describes the non-uniform noise distribution and the influence of data processing (e.g., in this case foreground subtraction from Section~\ref{sec:pca}) should also be included. An alternative method therefore, which was adopted here, was to simply measure the mean white noise power from simulations of white noise plus signal and foregrounds only, post-foreground subtraction.

To calculate the underlying white noise power spectrum uncertainty it was assumed that the underlying sky signal is non-varying. This means that the additional statistical uncertainty due the white noise is not
\begin{equation}\label{eqn:knox}
	\Delta N_\ell = \sqrt{\frac{2}{(2 \ell + 1)}}   \left( C_\ell + N_\ell \right) ,
\end{equation}
as derived in \citet{Knox1995}, but instead is given by
\begin{equation}\label{eqn:knoxnew}
	\Delta N_\ell = \sqrt{\frac{2}{(2 \ell + 1)}} N_\ell \sqrt{1 + 2 \frac{C_\ell}{N_\ell}} ,
\end{equation}
which is derived in Appendix~\ref{app:B}.

\subsection{Summary}

To conclude this Section the following is a complete summary of each step taken in the process of the simulations.
\begin{enumerate}
 \item Generate sky foreground and signal maps from models described in Sections \ref{sec:sync}, \ref{sec:freefree}, and \ref{sec:HI}.
 \item Simulate TODs for the observing strategy using the model in Section \ref{sec:design}.
 \item For each arbitrary block of TOD produce a correlated $1/f$ noise signal following the method outlined in Section \ref{sec:1f}.
 \item Sample from model sky using observing strategy to produce a noise-free data stream, and add $1/f$ noise plus white noise. Combine the TOD into a realisation map of the sky.
 \item Separate each sky map realisation into eigenmode components using PCA as described in Section \ref{sec:pca}.
 \item Calculate the angular power spectrum of every map for each frequency and eigenmode.
\end{enumerate}

\section{Results}\label{sec:results}

The previous Section describes the key components of the HI IM simulation pipeline. The simulation pipeline was used to generate several thousand mock 30\,day SKA HI IM surveys with unique $1/f$ noise properties but identical astrophysical foreground and HI signals. Using these mocks, HI angular power spectra were recovered. The results presented in this Section are predominantly concerned with the impact of $1/f$ noise correlations in the frequency direction on the statistical and systematic errors of the recovered HI power spectra. 

Subsection \ref{sec:results1} will simply describe the range of $1/f$ noise parameters that are explored. Subsection \ref{sec:results2} will describe the properties of the mean $1/f$ noise angular power spectra (e.g., how it varies in frequency and harmonic space). Then Subsections \ref{sec:results3}, \ref{sec:results4}, and \ref{sec:results5} will describe the impact of $1/f$ noise on the recovery of the HI angular power spectrum in terms of first the increased statistical uncertainty, second through a bias term, and finally the combined impact of these two terms respectively.

\subsection{Parameter Space}\label{sec:results1}

The discussions of $1/f$ noise in Section \ref{sec:1f} imply that there are four key parameters that must be considered: the spectral index of the temporal $1/f$ fluctuations ($\alpha$), the proxy for the spectral index of the $1/f$ frequency fluctuations ($\beta$), the telescope scanning speed ($\vt$), and the $1/f$ knee frequency ($f_k$) at 20\,MHz channel width. The full range of parameters explored by these simulations are listed in Table \ref{table:ParameterSpace}. For each input parameter there are 100 Monte-Carlo realisations.

For most of the discussions in this Section a baseline reference simulation will be used with fixed parameters of $\alpha = 1$ and $\vt = 1$deg\,s$^{-1}$.

\begin{table*}
\centering
\caption{The full range of the parameter space explored by this work. Each column represents the input parameters used to generate a given set of 100 Monte-Carlo simulations of the 30 day SKA HI IM survey. For each set of realisations with a given $\beta$ only one $f_k$ was simulated, and other knee frequencies were generated by scaling the noise appropriately before component separation or power spectrum estimation.}
\label{table:ParameterSpace}
\begin{tabular}{l|cccccccccccc}
\hline
$\alpha$                            & \multicolumn{12}{c}{1}                                                                                                      \\
$\vt$ (deg\,s$^{-1}$) & \multicolumn{4}{c}{0.5}                 & \multicolumn{4}{c}{1}                   & \multicolumn{4}{c}{2}                   \\
$\beta$                             &  0.25      & 0.5     & 0.75  & 1    & 0.25      & 0.5     & 0.75  & 1   &  0.25      & 0.5     & 0.75  & 1    \\
$f_k$ (Hz)                          & \multicolumn{4}{c}{10, 1, 0.1, 0.01} & \multicolumn{4}{c}{10, 1, 0.1, 0.01} & \multicolumn{4}{c}{10, 1, 0.1, 0.01} \\ \hline
$\alpha$                            & \multicolumn{12}{c}{2}                                                                                                      \\
$\vt$ (deg\,s$^{-1}$) & \multicolumn{4}{c}{0.5}                 & \multicolumn{4}{c}{1}                   & \multicolumn{4}{c}{2}                   \\
$\beta$                             &  0.25      & 0.5     & 0.75  & 1    & 0.25      & 0.5     & 0.75  & 1   &  0.25      & 0.5     & 0.75  & 1    \\
$f_k$ (Hz)                          & \multicolumn{4}{c}{10, 1, 0.1, 0.01} & \multicolumn{4}{c}{10, 1, 0.1, 0.01} & \multicolumn{4}{c}{10, 1, 0.1, 0.01} \\ \hline
\end{tabular}
\end{table*}

\subsection{Simulated 1/f Power Spectra}\label{sec:results2}

\begin{figure}
\centering 
\includegraphics[width=0.5\textwidth]{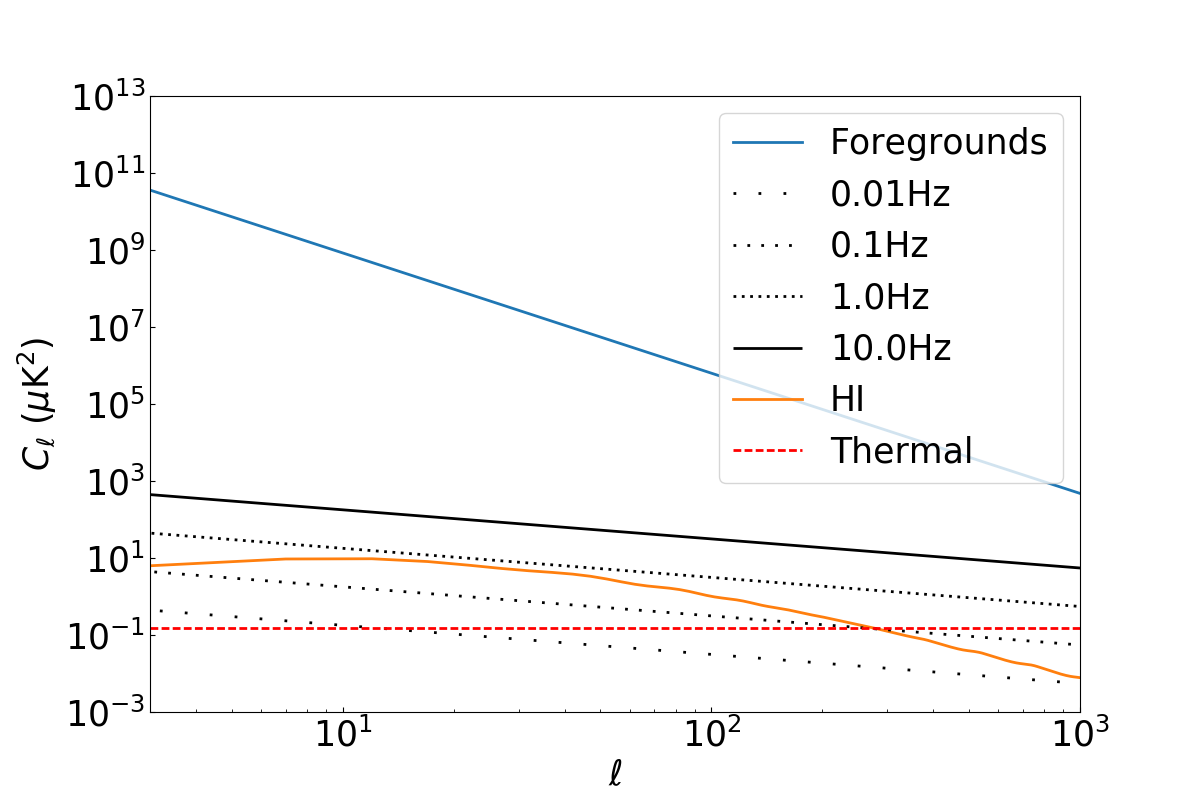}
\caption{Angular power spectra of the foregrounds signal (\textit{blue}), HI (\textit{orange}), thermal noise (\textit{red}) and $1/f$ noise with different knee frequencies (\textit{black} lines). These power spectra are  for 20\,MHz wide channels in the centre of SKA band 2 at 1190\,MHz. }\label{fig:sim1}
\end{figure}

To begin this Section will examine the properties of the simulated $1/f$ noise outputs. Fig.~\ref{fig:sim1} shows a comparison between the angular power spectra of the foregrounds, input HI and $1/f$ noise for different knee frequencies with a slew speed of 1\,deg\,s$^{-1}$ and $\beta = 1$. For knee frequencies less than 10\,mHz the thermal noise power is greater than the $1/f$ noise power at all temporal scales for a channel width of 20\,MHz. When the knee frequency approaches 1\,Hz then the amplitude of the $1/f$ noise at low-$\ell$ is comparable to the amplitude of the underlying HI signal. It is critical to be aware of the knee frequency at which the $1/f$ noise exceeds the HI signal because it will be shown in the results later in this Section that the impact of $1/f$ noise on the recovered HI angular power spectrum is most significant when the $1/f$ noise power is comparable to or exceeds the HI angular power. 

\begin{figure}
\centering 
\includegraphics[width=0.5\textwidth]{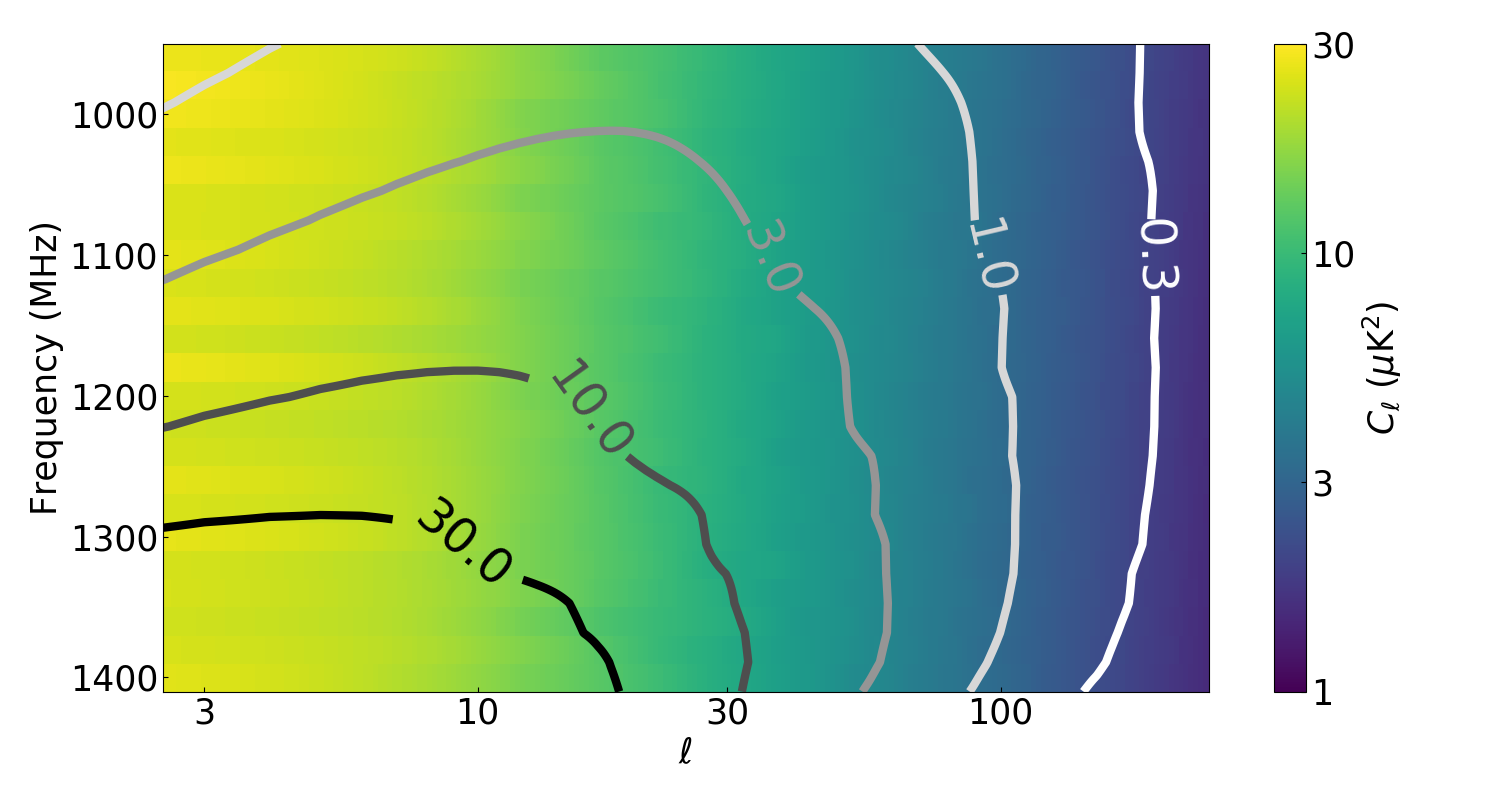}
\caption{Distribution of $1/f$ noise angular power spectra with frequency and $\ell$ for the baseline model with $f_k = 1$\,Hz. The colour scale shows the $1/f$ noise angular power, which is largely constant with frequency and a power law in $\ell$. The contours show the equivalent input HI angular power spectrum. Both colour scale and contours are in units of $\mu$K$^2$.}\label{fig:sim2}
\end{figure}

Fig.~\ref{fig:sim2} shows the same information as Fig.~\ref{fig:sim1} but in 2D for just 1\,Hz  knee frequency $1/f$ noise assuming the baseline model. As the $1/f$ noise is coupled to the foreground brightness, the $1/f$ noise angular power spectrum amplitude is slightly larger at low frequencies. Conversely, the amplitude of the HI variations decrease at low frequency. This results in the $1/f$ noise having significantly more impact on high redshift observations than low redshift observations. The impact of this on the low redshift simulations discussed here is small, but it could potentially become problematic for high redshift EoR HI IM surveys. Similarly, the $1/f$ noise impacts large-scales more than small-scales, leading to a decreased SNR at $\ell < 20$.

\begin{figure*}
\centering 
\includegraphics[width=0.95\textwidth]{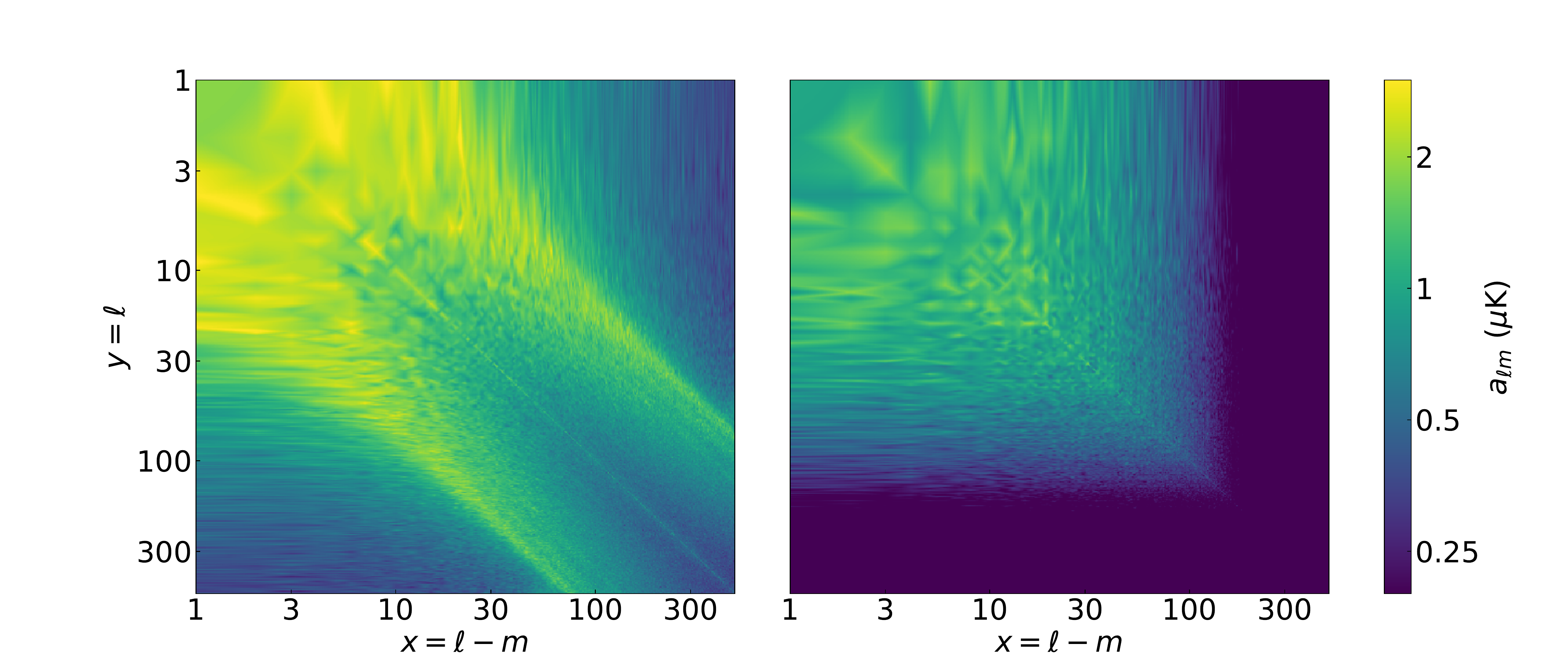}
\caption{Standard deviation of spherical harmonic coefficients projected into a 2D image plane for the simulated $1/f$ noise at 0.5\,deg\,s$^{-1}$ (\textit{left}), and the input HI signal (\textit{right}). The cut-off in HI power at high $\ell$ is due to the beam of the telescope. For $1/f$ noise the beam is not a factor to consider and therefore it still has a contribution at the smallest scales.}\label{fig:fnoise_alm}
\end{figure*}

Another informative way to view the $1/f$ noise is through the reprojection of the spherical harmonic coefficients into a 2D image. Fig.~\ref{fig:fnoise_alm} shows the standard deviation of the $a_{\ell m}$'s for 100 different $1/f$ noise and input HI realisations. The mapping between the $x-y$ Cartesian grid and the $\ell-m$ plane is $x = \ell - m$ and $y = \ell$ where it has been assumed that there is a diagonal symmetry such that $m = -m$. The images in Fig.~\ref{fig:fnoise_alm} reveal that the combination of the correlated $1/f$ noise with the observing strategy preferentially gives more power to some $a_{\ell m}$'s over others and has a  structure distinct from the underlying HI signal. The difference in the noise and the HI imply that there is a possibility of performing component separation in the harmonic space of the data, which may be complementary to performing component separation in image space. Though not explored further by this work, such differences in the harmonic structure does lead to the possibility of novel power spectrum analysis methods such as $m$-mode analysis \citep{Shaw2014,Shaw2015,Berger2016}. 

In the above cases, and for the rest of the results presented in this paper, the simulation survey is taken to last 30\,days. However, it would be interesting to know whether the results presented here can be simply scaled to different total observing times, number of telescopes or survey sky area. To determine this, SKA simulations for periods of 2 to 512\,days were generated that include only $1/f$ noise with $\alpha = 1$ and $\alpha = 2$. As may be expected, as the $1/f$ noise is simulated with Gaussian random variates and stationary values of $\alpha$ and $f_k$, the r.m.s. of the maps decrease as $1/\sqrt{T_\text{obs}}$. Exploration of how the r.m.s. of non-Gaussian and non-stationary $1/f$ noise is left to future work.

\subsection{Power Spectra Uncertainty}\label{sec:results3}

\begin{figure*}
\centering 
\includegraphics[width=0.98\textwidth]{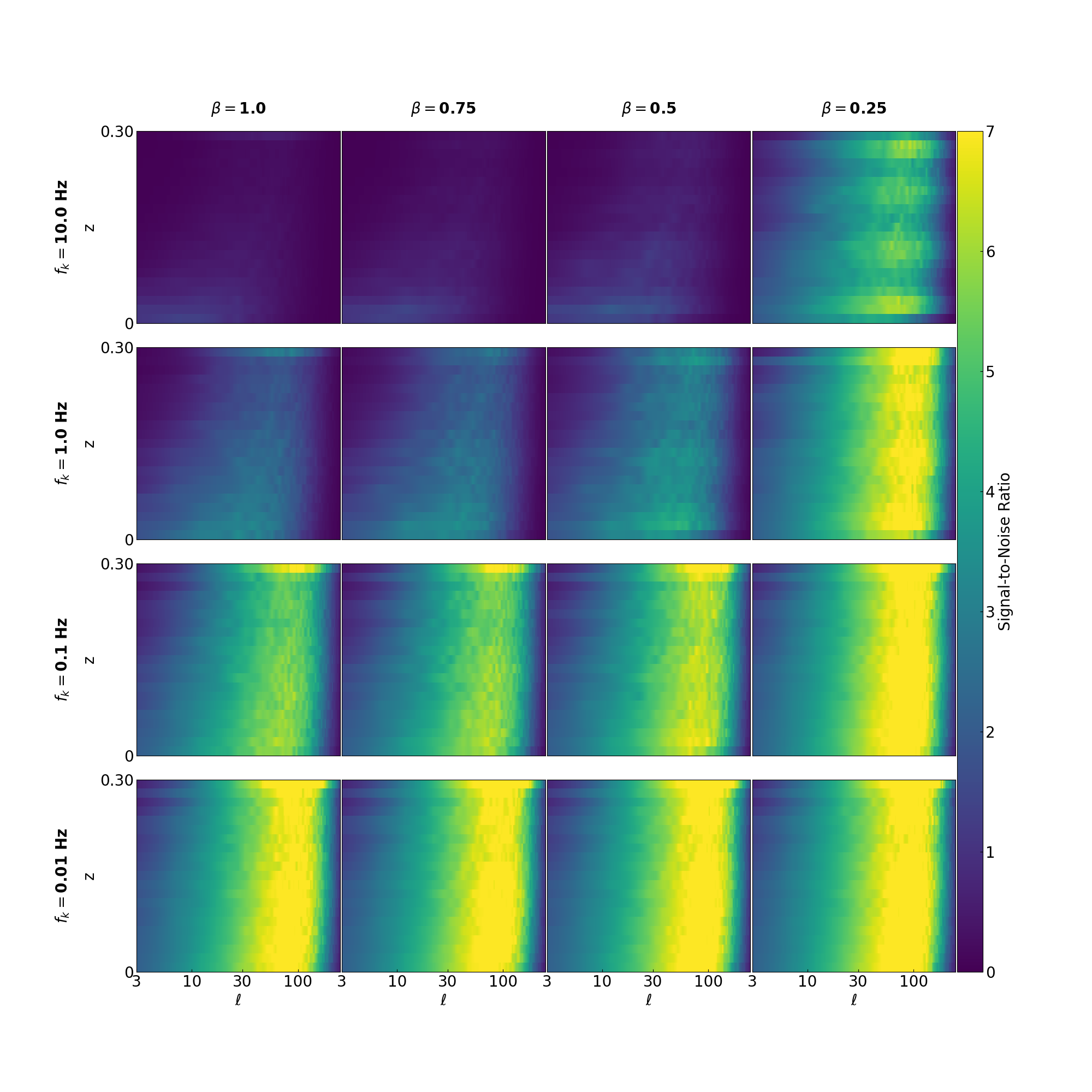}
\caption{The expected signal-to-noise ratio when considering only statistical fluctuations in the angular power spectrum due to cosmic, $1/f$, and white noise variances as described by Eqn.~\ref{eqn:snrnr1}. All plots are using the baseline simulation of $\alpha = 1$, $\vt = 1$\,deg\,s$^{-1}$. The statistical fluctuations in these plots are around 10\,per\,cent as expected for 100 Monte-Carlo realisations.}\label{fig:snrnr1}
\end{figure*}

The first question to ask of the $1/f$ noise in an SKA HI IM survey is how much additional statistical uncertainty does it contribute to the recovered angular power spectrum. One method of determining this is to measure the SNR of the input signal with the combined uncertainties due to noise and cosmic variance. The uncertainty in the HI angular power spectrum is defined as 
\begin{equation}
 \Delta C_\ell = \sqrt{\frac{2}{ \left(2 \ell + 1 \right) \Delta \ell}} C_\ell  + \Delta F_\ell
\end{equation}
where the first term on the \textit{right side} is the cosmic variance, calculated analytically from the input HI power spectrum, and the second term $\Delta F_\ell$ is the uncertainty due to both the thermal and $1/f$ noise contributions defined in Eqn.~\ref{eqn:fmodel}. The SNR ratio is then defined as 
\begin{equation}\label{eqn:snrnr1}
	\snr = \frac{C_\ell}{\Delta C_\ell} ,
\end{equation}
where, as before, $C_\ell$ denotes the angular power spectrum for the input realisation of the HI sky signal. Fig.~\ref{fig:snrnr1} shows the $\snr$ for the frequency range $950 < \nu < 1410$\,MHz for the baseline model of $\vt = 1$\,deg\,s$^{-1}$, and $\alpha = 1$. The plot reveals that when the $1/f$ noise is highly correlated ($\beta = 0.25$) in frequency then the contribution to the final HI statistical uncertainty is very small, irrespective of the input $1/f$ noise knee frequency as expected when using PCA for foreground subtraction and shown in Appendix~\ref{app:A}. The figure also shows that the survey will be more sensitive to low redshift HI emission, and to scales defined by the $\ell$ range $10 < \ell < 100$. The effective redshift of the survey can be calculated as 
\begin{equation}\label{eqn:weightedz}
	\bar{z} = \frac{\sum \snr^2 z }{\sum \snr^2}
\end{equation}
where $\bar{z}$ is the mean redshift, and $z$ is the redshift of each pixel in the $C_\ell - \nu$ plane. The mean redshift is found to be $\bar{z} \approx 0.2$. 

\begin{figure*}
\centering 
\includegraphics[width=0.98\textwidth,trim={2cm 1cm 2cm 2cm},clip]{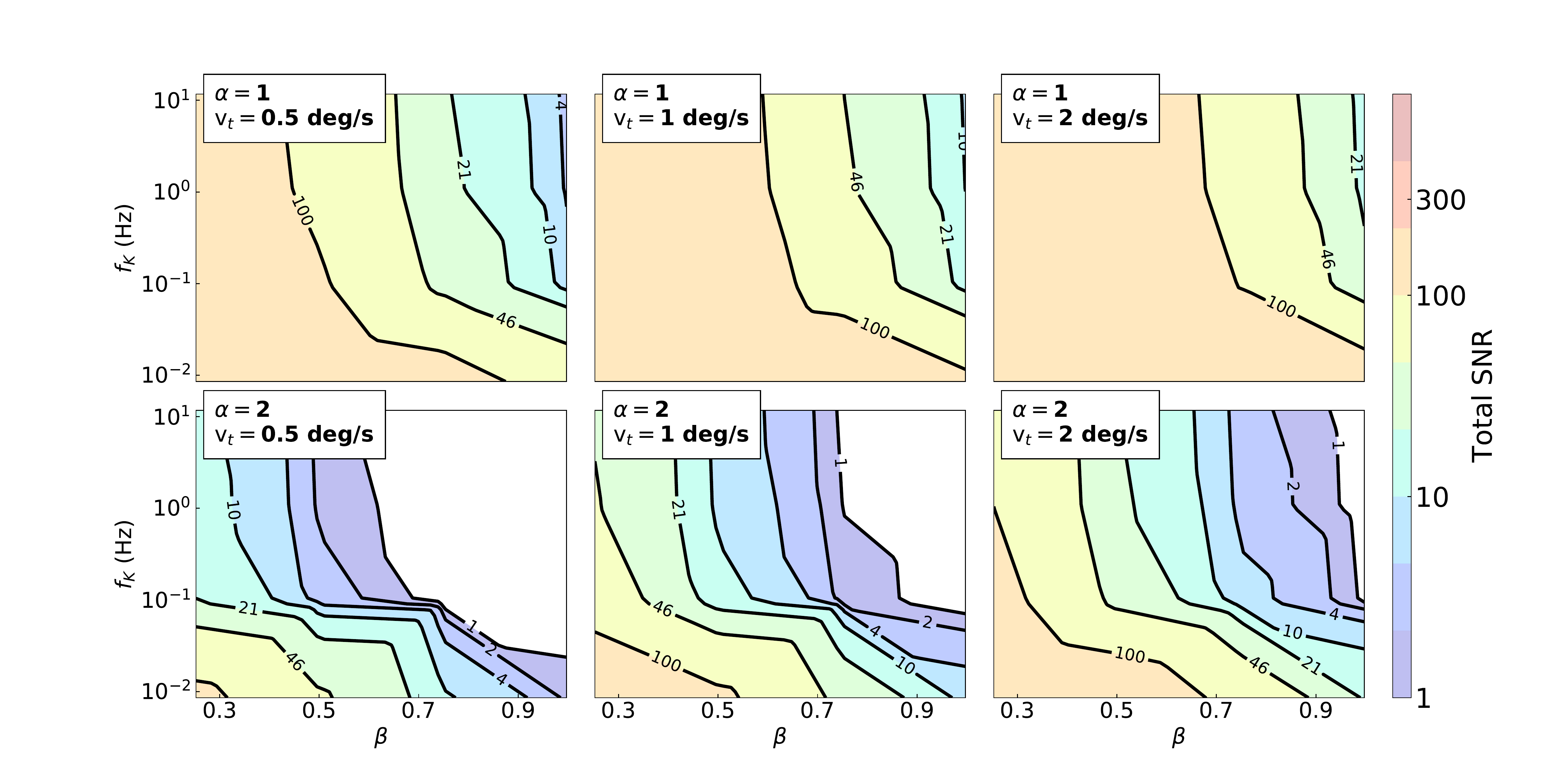}

\caption{The expected total signal-to-noise ratio when considering only statistical fluctuations in the angular power spectrum due to cosmic, $1/f$, and white noise variances as defined by Eqn.~\ref{eqn:snrnr1} and Eqn.~\ref{eqn:totalsnr}. The \textit{top} row is for $\alpha = 1$, and from \textit{left} to \textit{right} increasing slew speeds of $\vt = 0.5$, $\vt = 1$ and $\vt = 2$\,deg\,s$^{-1}$. The \textit{bottom} row is same again but with $1/f$ noise with $\alpha = 2$. The contours show the importance of $1/f$ noise correlations are when considering the statistical impact of $1/f$ noise on signal-to-noise.}\label{fig:snrnr1cont}
\end{figure*}

Summing over each $\snr$ contribution in Fig.~\ref{fig:snrnr1} results in a combined \textit{total} $\snr$
\begin{equation}\label{eqn:totalsnr}
	\snr_\text{T} = \sqrt{\sum\limits_\nu \sum\limits_\ell \snr^2} ,
\end{equation}
shown as contour plots in Fig.~\ref{fig:snrnr1cont}. This figure can be used to numerically determine the impact of the $1/f$ noise on the HI angular power spectrum recovery. The contour plots show the impact of the $1/f$ noise when $\beta = 0.25$ is comparable to the white noise only when $\alpha = 1$, with almost no detection when $\beta = 1$ and $f_k \geq 1$\,Hz. However, when $\alpha = 2$ the impact of the $1/f$ noise is significant even when the noise is highly correlated. This stresses the importance of measuring $\alpha$ of all future SKA receivers that will be used for a HI IM survey. Further the plot shows that small increases in $\beta$ can very quickly decrease the total SNR of the observation. The impact of $\alpha$ and $\beta$ on the SNR should be very carefully considered as the observing time is proportional to the square of the SNR, and very quickly the increase in observing time required to achieve the desire sensitivity may become untenable.

\begin{figure}
\centering 
\includegraphics[width=0.48\textwidth]{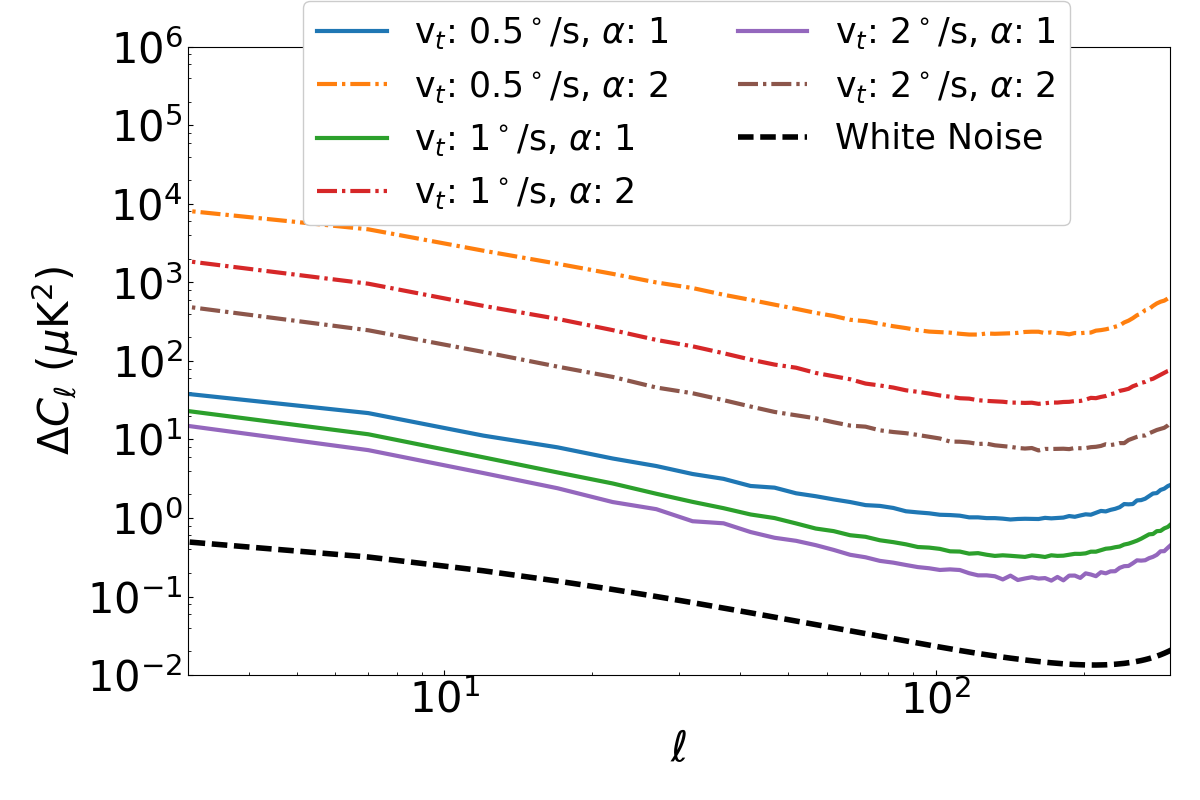}
\caption{Uncertainty in angular power spectrum  averaged over all simulated frequencies for $1/f$ noise with telescope slew speeds of 0.5, 1 and 2\,deg\,s$^{-1}$ and spectral indices of $\alpha = 1$ and 2. All curves are for a knee frequency of 1\,Hz and the black-dashed line shows the expected white noise level. There are two principal divisions of the results shown in this figure. The first division is between the \textit{solid} and \textit{dot-dashed} lines representing $1/f$ noise with $\alpha = 1$ and $\alpha = 2$ respectively. The difference in uncertainty within these two divisions is due to the three different scan speeds used, with 2\,deg\,s$^{-1}$ having the lowest uncertainty in both groups. The key point to take is that the spectral index of the $1/f$ noise is more important than telescope scan speed.}\label{fig:snrnr3}
\end{figure}

Fig.~\ref{fig:snrnr3} shows the uncertainty in $1/f$ noise angular power spectra for $\alpha = 1$ to 2, and $\vt = 0.5$, 1 and 2\,deg\,s$^{-1}$ for $f_k=1$\,Hz and $\beta = 1$. The figure shows how the $1/f$ noise $\alpha$ is a far larger contributor to the overall uncertainty than the slew speed of the telescope. This implies that stable receivers are an order-of-magnitude more important than choice of scan speed. However, the figure also shows how scan speed is coupled with the spectral index of the $1/f$ noise as the scan speed is seen to have greater impact for steep spectrum $1/f$ noise. This coupling between the scan speed and the spectral index is complex but will be strongly dependent on the choice of observing strategy.

\begin{figure*}
\centering 
\includegraphics[width=0.98\textwidth]{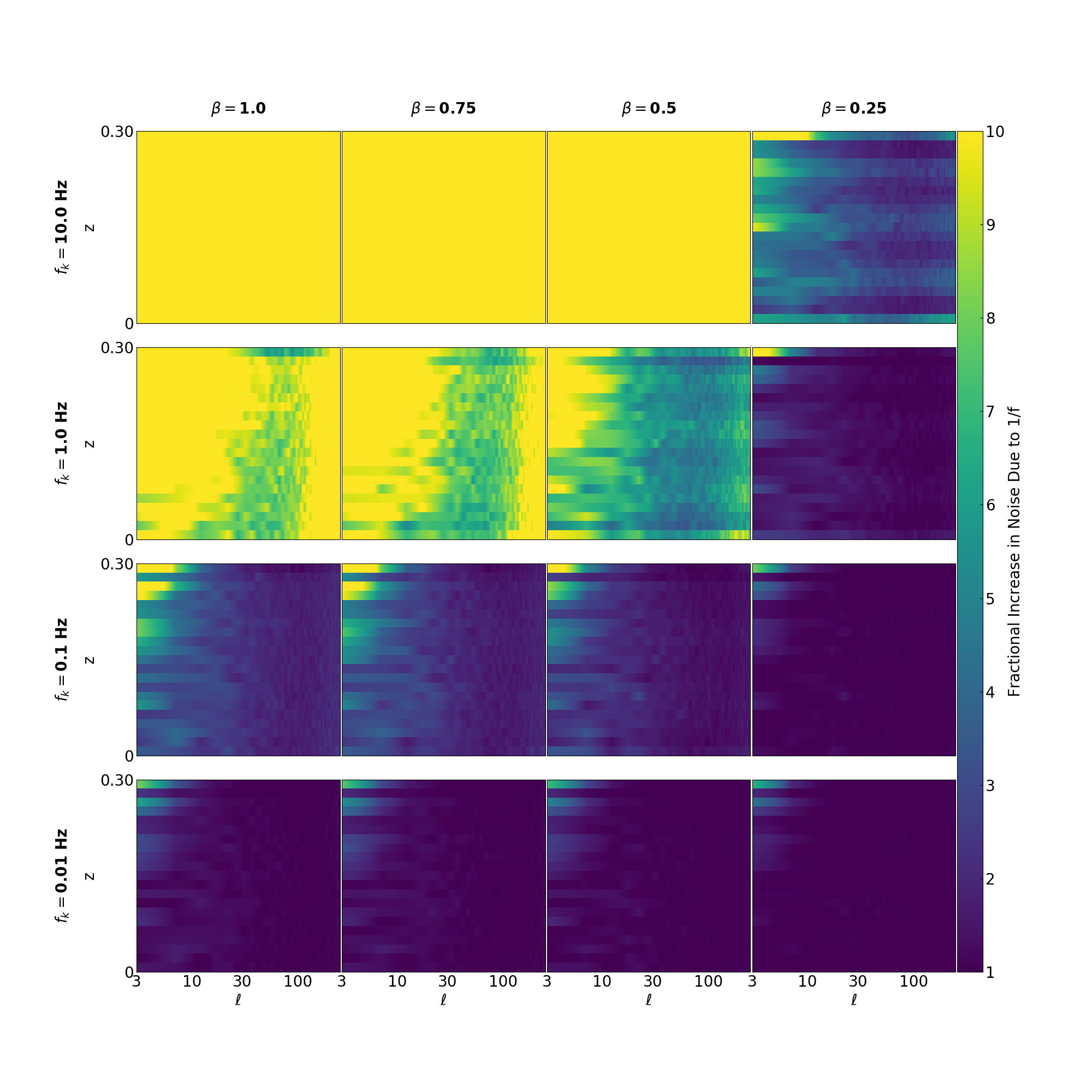}
\caption{The variation in ratio between the uncertainty in the angular power spectrum from $1/f$ noise over white noise calculated using Eqn.~\ref{eqn:f2w} a function of $\ell$ and $z$. All plots are using the baseline simulation of $\alpha = 1$, $\vt = 1$\,deg\,s$^{-1}$. The statistical fluctuations in these plots are at the 10\,per\,cent level expected for 100 Monte-Carlo realisations. }\label{fig:f2w}
\end{figure*}

As well as the $\snr$ it is also interesting to know how the ratio of the white noise to the $1/f$ noise is distributed in $\ell$ and frequency. The ratio of the $1/f$ noise to the thermal noise was defined as
\begin{equation}\label{eqn:f2w}
	r = \frac{\Delta F_\ell}{\Delta N_\ell} ,
\end{equation}
where $\Delta F_\ell$ is the total uncertainty measured from the simulations, and $\Delta N_\ell$ is the expected uncertainty due to the thermal noise (Eqn.~\ref{eqn:knoxnew}). Fig.~\ref{fig:f2w} shows how the ratio $r$ varies with $\ell$ and frequency for different knee frequencies and $\beta$ values for the baseline model. The figure shows how $1/f$ noise can contribute additional uncertainty on all scales, which is dependent only on the ratio of the HI and $1/f$ noise angular power spectra. At $f_k = 1$\,Hz, the residual $1/f$ noise power exceeds the recovered HI power at all scales and hence there is also additional uncertainty at all scales. The ripples that can be seen in the $\beta = 0.25$ column are residual foregrounds from the component separation step.

\subsection{Power Spectra Bias}\label{sec:results4}

As well as the noise uncertainty it is also important to quantify the magnitude of the systematics that are introduced by the $1/f$ noise. There are two methods for characterising the bias (not to be confused with cosmological bias), the first is the residual of the angular power spectrum between the input and recovered HI signals defined as
\begin{equation}\label{eqn:bias1}
	\epsilon_\ell = \frac{\left<\hat{C}_\ell\right> - \left<\hat{C}_\ell (f_k = \mathrm{0\,Hz})\right>}{C_\ell} ,
\end{equation}
where $\left<\hat{C}_\ell\right> $ is average over the recovered HI realisations, and $\left<\hat{C}_\ell (f_k = \mathrm{0\,Hz})\right>$ is the average recovered HI power spectrum assuming white noise only. Here realisations of the white noise are used, instead of the analytically derived white noise level, to account for any bias originating from the component separation step.

\begin{figure*}
\centering
\includegraphics[width=0.98\textwidth]{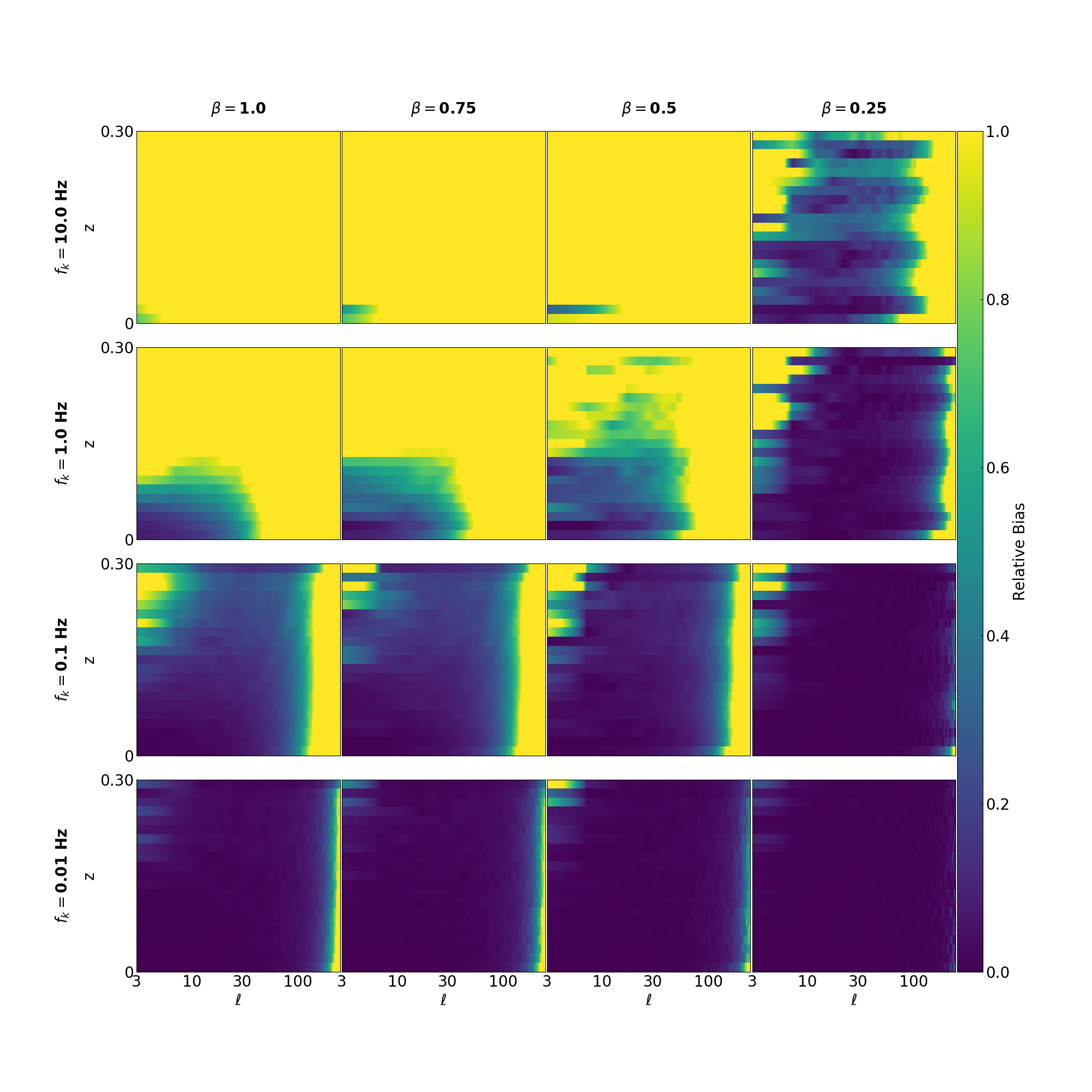}
\caption{The relative difference between the mean estimated HI and the input HI angular power spectra, normalised by the input HI spectrum as described by Eqn.~\ref{eqn:bias1}. All plots are using the baseline simulation of $\alpha = 1$, $\vt = 1$\,deg\,s$^{-1}$.  The statistical fluctuations seen at all scales are consistent with 10\,per\,cent, as expected for  100 Monte-Carlo with realisations.}\label{fig:bias1}
\end{figure*}

In Fig.~\ref{fig:bias1} $\epsilon_\ell$ is presented as a function of frequency and $\ell$ for the baseline model and a range of $f_k$ and $\beta$ values. The interpretation of the bias in the power spectra is that anything greater than unity implies that the HI power spectrum is dominated by the residual $1/f$ noise. The bias is only seen to be significant when the $1/f$ noise power exceeds the HI angular power (i.e. when $f_k \geq 1$\,Hz), however some small scale contributions to the noise can be seen when $f_k = 0.1$\,Hz. Encouragingly, when the $1/f$ noise is highly correlated ($\beta = 0.25$), the bias is close to zero until the $1/f$ noise greatly exceeds the HI noise power (e.g., $f_k > 10$\,Hz). Also note that, similarly to Fig.~\ref{fig:f2w}, the ripples that can be seen in the $\beta = 0.25$ column are due to poor foreground subtraction. In this case however these ripples are more interesting as, from Eqn.~\ref{eqn:bias1}, any intrinsic bias due to residual Galactic foregrounds after the component separation step should be subtracted. Therefore these ripples must be due to residual $1/f$ noise.

One possible question to ask is whether it would be possible to model this $1/f$ noise bias and remove it. For these simulations that would be trivial as the input $1/f$ noise is known, however in real data that may exhibit $1/f$ noise that is non-Gaussian, non-stationary and coupled with other systematics, accurate modelling may prove challenging. However, should the bias in the real data prove to be a significant difficulty one alternative possibility is to produce cross-correlation HI power spectra from subsets of dishes within the SKA array, but this will come at a cost of increasing the power spectrum uncertainty by at least a factor of 4.

\subsection{Combined Uncertainties}\label{sec:results5}

An estimate of the overall uncertainty in the recovered HI $\hat{C}_\ell$ is 
\begin{equation}\label{eqn:snr1}
 \Delta C_\ell = \sqrt{\frac{2}{(2 \ell + 1 ) \Delta \ell}} C_\ell + \Delta F_\ell + \left< \epsilon_\ell \right> C_\ell,
\end{equation}
where $\Delta F_\ell$ is the total uncertainty as measured in the simulations discussed in Section \ref{sec:power},  and $\left< \epsilon_\ell \right>$ is the mean bias in the recovered power spectra as discussed in Section \ref{sec:results4}. The first term in Eqn.~\ref{eqn:snr1} is the cosmic variance term, which has been added to account for the lack of cosmic variance in $\Delta F_\ell$. The bias ($\epsilon_\ell$) is not a true uncertainty however it is added here to give a tentative value to the impact $1/f$ noise bias may have on a real observation. 

\begin{figure*}
\centering 
\includegraphics[width=0.98\textwidth]{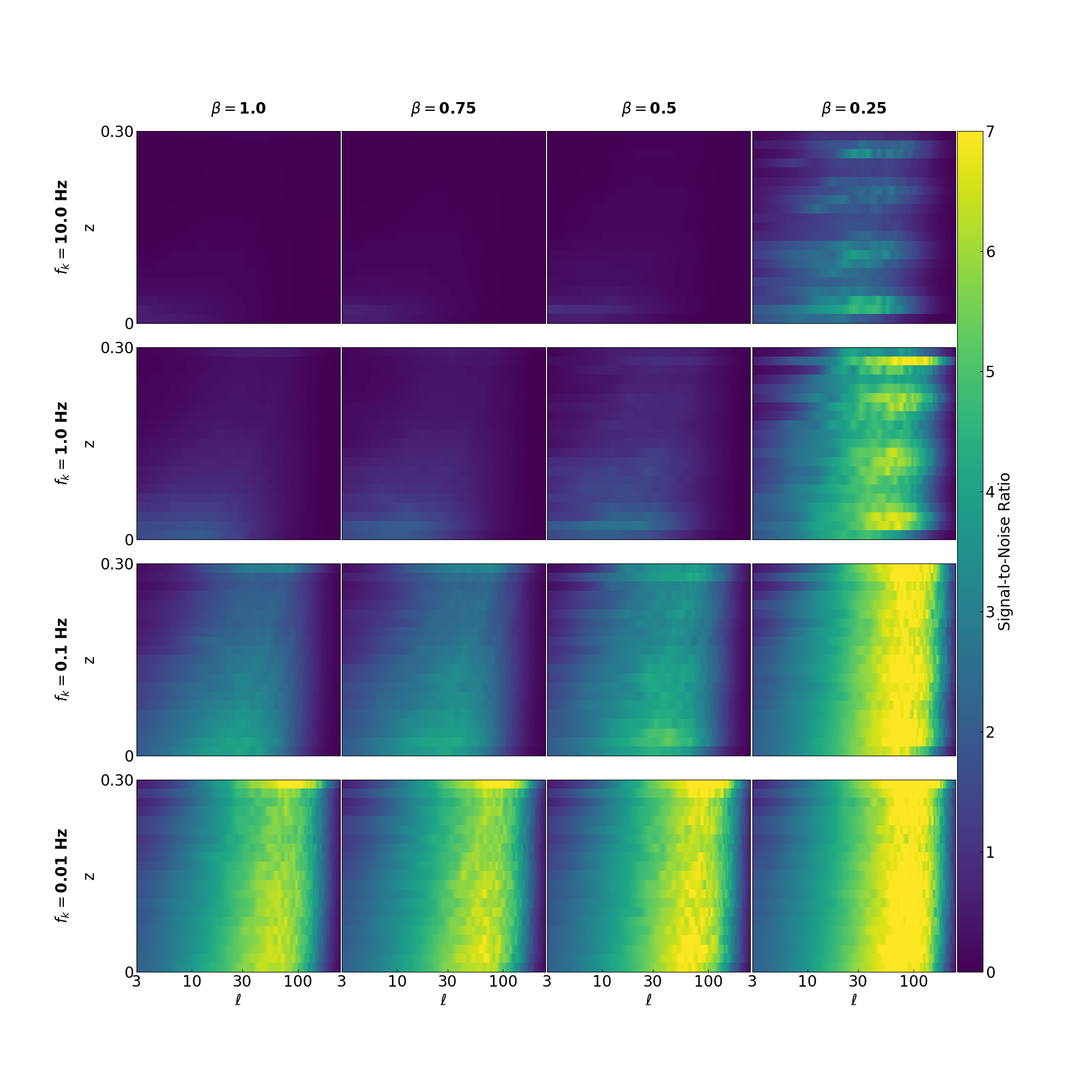}
\caption{The signal-to-noise ratio when including not just statistical variations from Eqn.~\ref{eqn:snrnr1} but also includes the systematic residual power from  \ref{eqn:bias1}, as described by Eqn.~\ref{eqn:snr1}. The overall distribution of the highest signal-to-noise regions in this plot are dominated by the residuals induced by poor foreground subtraction and masking. The statistical variations in this plot are consistent with the 10\,per\,cent level expected for 100 Monte-Carlo noise realisations.}\label{fig:snr1}
\end{figure*}

When the bias is added into the uncertainty Fig.~\ref{fig:snr1} shows how the $\snr$ changes. The figure shows that there is a dramatic decrease in the $\snr$ for uncorrelated $1/f$ noise with $\beta \geq 0.5$. However, even at 10\,Hz when $\beta = 0.25$ some bins do exhibit a $> 3\sigma$ detection on the HI angular power spectrum, which likely could be improved using a more advanced component separation method. 

\begin{figure*}
\centering 
\includegraphics[width=0.98\textwidth,trim={2cm 1cm 2cm 2cm},clip]{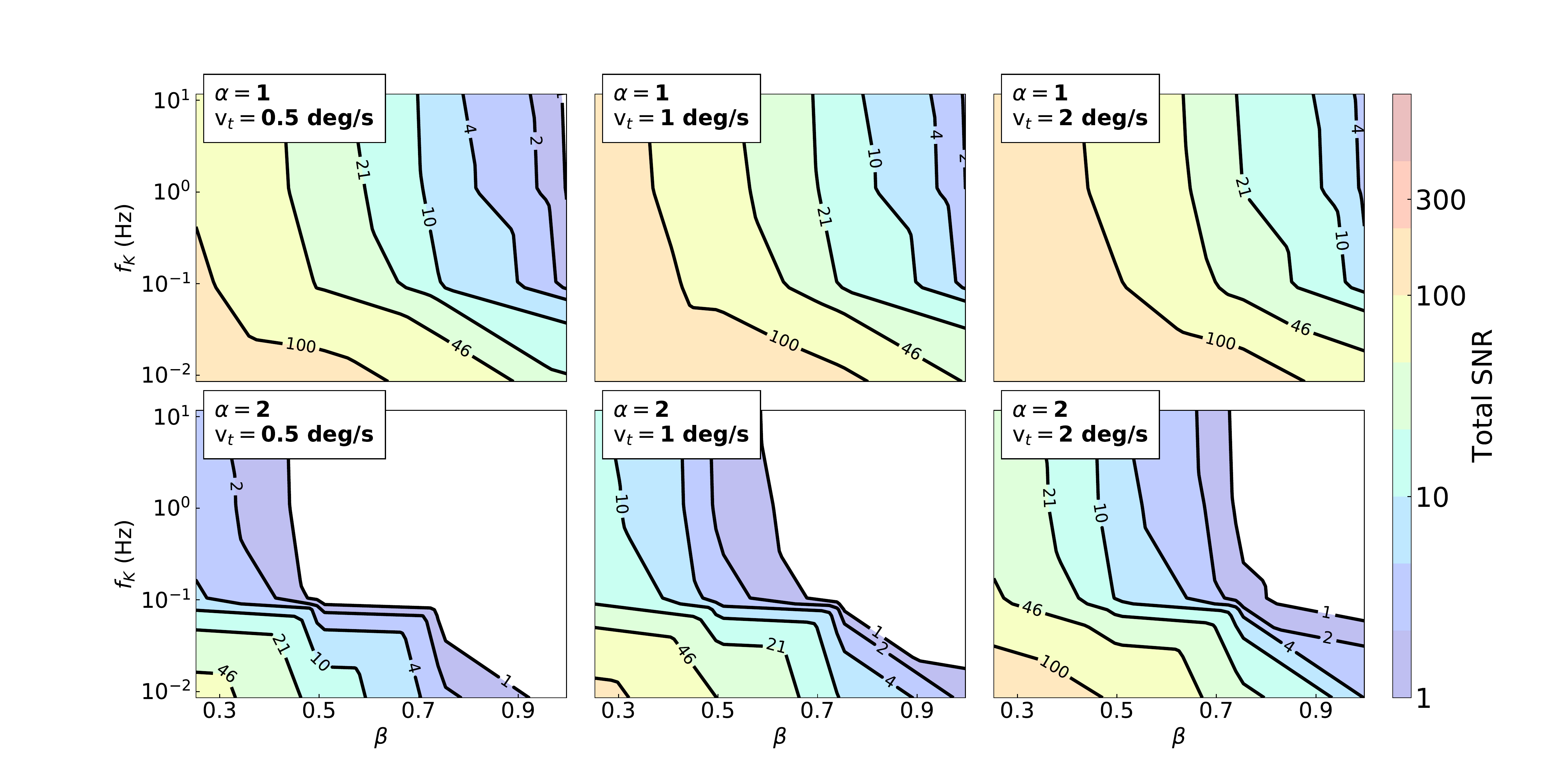}

\caption{The sum in quadrature over all the signal-to-noise in each plot of Fig.~\ref{fig:snr1}, using Eqn.~\ref{eqn:totalsnr}. These figures are similar to those shown in Fig.~\ref{fig:snrnr1cont} but include the mean residual power as an uncertainty contribution. The \textit{top} row is for $\alpha = 1$, and from \textit{left} to \textit{right} increasing slew speeds of $\vt = 0.5$, $\vt = 1$ and $\vt = 2$\,deg\,s$^{-1}$. The \textit{bottom} row is same again but with $1/f$ noise with $\alpha = 2$.}\label{fig:totalsnr} 
\end{figure*}

The six plots in Fig.~\ref{fig:totalsnr} shows the total $\snr$ ratio for the entire parameter space. Comparing this figure to Fig.~\ref{fig:snrnr1cont} it can be seen that the residual $1/f$ noise bias has significant impact for high values of $\beta$, decreasing the SNR by factors of 2 to 5 depending on $\alpha$ and $\vt$. As has previously been discussed, this plot further emphasizes the importance of $\alpha$ on the SNR of the recovered HI angular power spectra. For $\vt = 0.5$\,deg\,s$^{-1}$ and $\alpha=2$, nearly the entire $1/f$ noise parameter space has a $\snr < 1$, while for $\vt = 0.5$\,deg\,s$^{-1}$ and $\alpha=1$, there is a reasonable detection of the HI spectra at all but the highest $\beta$ and $f_k$ values.

\section{Discussion}\label{sec:discussion}

This Section will present some discussion, informed from the results in Section~\ref{sec:results}, about $1/f$ noise in the context of HI IM surveys. To begin in subsection~\ref{sec:dis1} an empirical model for the $1/f$ noise angular power spectra measured in these simulations is presented. In subsection~\ref{sec:dis2} some general comments about how the impact of $1/f$ noise in an SKA HI IM survey may differ from the results presented here due to interactions with the chosen observing strategy, data analysis methods or other systematics. Finally, subsection~\ref{sec:dis3} attempts to provide some practical intuitions about $1/f$ noise that can be inferred from Section~\ref{sec:results}.

\subsection{An Empirical 1/f Noise Model from Power Spectrum Reconstruction}\label{sec:dis1}

By combining the $1/f$ noise angular power spectra and the discussions of the previous section an approximate empirical model of the real spectra can be derived. The first step is to bring together all the constituent parts of the model that are known already. To simplify the analysis it will be assumed that the $1/f$ noise has no dependence on frequency (i.e. $T_{\text{rx}} + T_{\text{cmb}} \gg T_{\text{sky}}$), the noise is Gaussian and therefore integrates down as $1/\sqrt{T_\text{obs}}$ as discussed in subsection~\ref{sec:results2}, and that there is no correlation in frequency such that $\beta = 1$ (however an estimate for $\beta < 1$ is given later). The $\ell$-independent amplitude of the empirical $1/f$ noise model is defined as 
\begin{equation}\label{eqn:emp1}
	\begin{aligned}
	A = 
    	\left( \frac{\tsys}{21\,K} \right)^2
        	\left(\frac{f_k}{1\,\text{Hz}}\right)^\alpha
            	\left(\frac{\delta \nu}{20\,\text{MHz}}\right)^{-1} 
                	\left(\frac{N_t}{200}\right)^{-1} \\
                    	\left(\frac{T_\text{obs}}{30\,\text{days}}\right)^{-1} 
    \left(\frac{\Omega_\text{s}}{20500\text{\,deg}^2}\right)
    	10^{2}  \mu\text{K}^2,
    \end{aligned}
\end{equation}
where $\tsys$ is system temperature normalised to 21\,K to account for a mean $\approx 1$\,K sky contribution within the simulation, the knee frequency $f_k$, the channel width $\delta \nu$, $N_t$ the number of telescopes, $T_\text{obs}$ the number of observing days and $\Omega_s$ the survey sky coverage.

Eqn.~\ref{eqn:emp1} describes the amplitude of the $1/f$ noise angular power spectrum, but the slope of the $1/f$ noise angular power spectra shown in Fig.~\ref{fig:sim1} is dependent on a combination of the $1/f$ noise $\alpha$ and the choice of observing strategy. The impact of the scan speed on the amplitude of the $1/f$ noise power spectrum is shown in Fig.~\ref{fig:snrnr3} to be also dependent on the $1/f$ noise $\alpha$. The best fit model that describes the relationship between $\alpha$, $\vt$ and the slope of the $1/f$ noise angular power spectra for the range of parameters explored in this work was determined to be
\begin{equation}\label{eqn:emp2}
	\begin{aligned}
	\log_{10}\left(\frac{F_l}{\mu\mathrm{K}^2}\right) = \log_{10}\left(\frac{A}{\mu\mathrm{K}^2}\right)  +
    	a\left[\alpha - 1\right]  \\ + 
        	b \sqrt{\alpha} \log_{10}\left(\frac{\vt}{\mathrm{deg\,s}^{-1}}\right) -
            	\sqrt{c \alpha} \log_{10}(\ell) ,
    \end{aligned}
\end{equation}
where $A$ is the parameter described by Eqn.~\ref{eqn:emp1}, $\alpha$ is the $1/f$ spectral index, $\vt$ is the telescope speed in deg\,s$^{-1}$, and $a$, $b$ and $c$ are fitted constants. The best fit values for the three parameters are: $a = 1.5$, $b=-1.5$, $c=0.5$. This model is accurate to the 10\,per\,cent level over the parameter space explored in this work for spherical harmonics of $\ell < 100$. The $\sqrt{\alpha}$ dependence on the slope of the angular power spectrum is due to geometric considerations of the TOD as it is projected on the sphere.

There are several limitations of this empirical $1/f$ noise model. The principle limitation of the Eqn.~\ref{eqn:emp2} model is determining how to account for the $1/f$ noise frequency correlations described by $\beta$. How $\beta$ impacts the amplitude of the $1/f$ noise power in the recovered HI angular power spectrum will depend on many aspects of the observations such as the observing strategy, how the noise averages, the choice of component separation, $1/f$ noise filtering on long time-scales and more. Further, $\beta$ will also have an impact on the residual $1/f$ noise angular power spectrum \textit{bias} that will add additional systematic uncertainty not accounted for here. Considering these provisos it is possible to determine the fractional decrease in $F_\ell$ predicted by these particular simulations by measuring the mean fractional change in the angular power spectrum \textit{bias} with $\beta$. Through inspection of the data the following two parameter model was determined to be a reasonable fit to the fractional change in bias
\begin{equation}\label{eqn:emp3}
 \frac{F_\ell(\beta)}{F_\ell(\beta = 1)} = a \sin\left(2 \pi \beta \right) + \beta
\end{equation}
where the fitted parameter $a = -0.16$ for $0 < \beta < 1$. Again, to reiterate, care should be taken when using this model in combination with Eqn.~\ref{eqn:emp2} to extrapolate the impact of $\beta$ on to different HI IM experimental designs.

The forms of Eqn.~\ref{eqn:emp2} and Eqn.~\ref{eqn:emp3} are not physically motivated, are limited largely to the description of the observations simulated within this work, and give limited insight into the correlations between spectral index and slew speed. Still, one particular use of Eqn.~\ref{eqn:emp2} and Eqn.~\ref{eqn:emp3} could be to improve forecasts of future SKA HI IM surveys by adding predictions for the $1/f$ noise power spectrum into Fisher matrix analyses such as those presented in \citet{Bull2015} and \citet{Pourtsidou2016}.

\begin{figure}
\centering 
\includegraphics[width=0.48\textwidth]{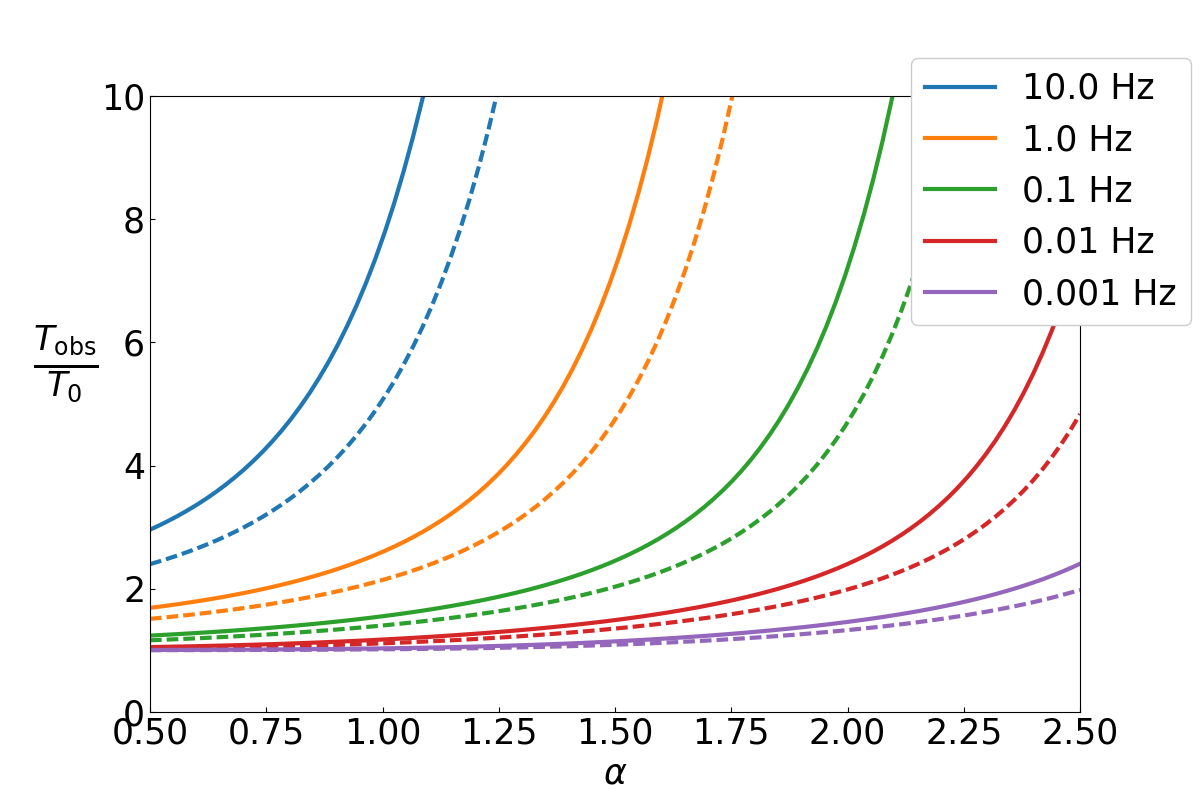}
\caption{Increase in observing time due to $1/f$ noise calculated using the empirical $1/f$ noise model of Eqn.~\ref{eqn:emp2} assuming $\beta = 1$ (\textit{solid lines}) and $\beta = 0.5$ (\textit{dashed lines}). The $T_0$ observing time assumes the same input parameters as the main simulations to calculate the white noise level.}\label{fig:tobs} 
\end{figure}

One practical example of using the Eqn.~\ref{eqn:emp2} is to predict how much additional observing time may be required for a range of receiver $1/f$ noise parameters. For simplicity only the worst case of $\beta = 1$ is considered. As SNR is proportional to the total observing time, then the following relationship is true
\begin{equation}
	T_\text{obs} = T_0 \frac{\snr_w}{\snr} ,
\end{equation}
where $T_0$ is the observing time to achieve the signal-to-noise ratio of $\snr_w$ given just white noise, $T_\text{obs}$ is the observing time required to achieve $\snr_w$ given that $\snr$ was achieved in time $T_0$ due to additional $1/f$ noise. Fig.~\ref{fig:tobs} shows difference in observing times calculated for $1/f$ noise with $1\,\text{mHz} < f_k < 10$\,Hz and $0.5 < \alpha < 2.5$. This analysis assumed the full bandwidth ($\Delta \nu = 450\,\text{MHz}$) of the instrument, the signal to be at the weighted mean redshift of $\bar{z} = 0.2$, and includes a sample variance contribution. The plot shows how important the spectral index of the $1/f$ noise is, as it can easily increase the effective statistical uncertainty in the HI angular power spectrum by an order-of-magnitude between $\alpha = 1$ and $\alpha = 2$. Consideration of the $1/f$ noise spectral index (and $\beta$) should be factored in when designing the instrumentation for a HI IM experiment. However, this may be challenging for multi-purpose instruments such as the SKA as competing factors may take precedence.

\subsection{Impact of Specific Data Analysis Methods and Systematics}\label{sec:dis2}

An interesting question to ask is could the impact of the $1/f$ noise presented in this work be reduced in a real HI IM survey? The answer is almost certainly yes, the most obvious improvement would be to use a more sophisticated component separation technique in place of PCA. For example the GNILC \citep{Olivari2016} method can perform spatially localised component separation for individual needlet scales, which could suppress the large-scale $1/f$ noise bias discussed in Section \ref{sec:results}. 

Another improvement would be to use a more advanced map-making method, such as Destriping \citep[e.g.,][]{Sutton2010} or optimal map-making \citep[e.g.,][]{Natoli2001}. However, such map-making methods at present are not optimised for HI IM data as they do not consider the spectral covariance of the noise or the sky signal. Without modifying existing map-making codes to consider spectral correlations there is the possibility that the map-making procedure itself could induce spectral variations in to the $1/f$ noise spectrum that is effectively similar to increasing $\beta$. 

Further improvements could also be achieved by using a more carefully considered observing strategy. For example, this analysis does not utilise of any of instrument specific advantages of the SKA. The observing strategy presented in this work assumes that all telescopes are all continuously slewing at the same constant elevation of 55\,deg. However, each dish or subsets of dishes could be given individual scanning strategies that are designed to maximise cross-linking of scanning tracks, or equalise integration times across the sky. Similarly, no attempt is made to use cross-correlation information between dishes or take advantage of any potential interferometric observations that are taken in parallel.

$1/f$ noise can also be suppressed during the calibration process. This can be achieved by injecting a signal from a calibration diode  into the receiver timestream at regular intervals. The use of a diode for calibration purposes can present its own challenges. For example, accurate calibration using a diode requires that over the interval between two diode injections the stability of the diode must be significantly better than the receiver system being calibrated. However, to calibrate the $1/f$ noise on shorter time scales would require a brighter diode signal, which can rapidly become impractical. As an example, a simple estimate of the required calibration diode stability implies that calibrating an SKA receiver with a channel width of 50\,MHz on relatively long time scales of 100\,seconds will require the diode stabilities of $\frac{\Delta T_\mathrm{cal}}{T_\mathrm{cal}} \approx 10^{-4}$ and a diode brightness as 25\,K. \citet{Roychowdhury2017} intends to explore this problem in greater detail. 

Unfortunately, for the real observations suppression of $1/f$ noise will be far more challenging due to the presence of other systematics in the data. Some of these systematics will be intrinsic to the $1/f$ noise such as different functional forms for the $\psd$ of the frequency correlations, the non-Gaussianity in the distribution of the $1/f$ noise amplitudes, and non-stationary properties for $\alpha$ or $f_k$. Other systematics will be intrinsic to the instrument or the observations such as the combined interaction of thousands of standing waves unique to each line-of-sight and each telescope, which in large numbers may exhibit $1/f$ noise-like properties in frequency and time.  Also, how the data is processed, calibrated and binned into maps from which HI power spectra are extracted are all important, as each step can unintentionally introduce additional frequency structure into the $1/f$ noise spectrum. In the context of this work, such effects could be seen as increase the effective $\beta$ value of the $1/f$ noise, and as such will make it more challenging for component separation methods to recover the underlying HI intensity field. 

A final point to consider is the impact of the $1/f$ noise and other systematics on the recovery of the frequency space correlations in the HI signal. Though not the focus of this paper, it should be pointed out that such correlations in the HI intensity field, such as redshift space distortions, will be lost when using the data processing method outlined in this work. Therefore it may be necessary to have separate analysis pipelines, or perhaps different observations altogether, for spatial and radial science objectives.

\subsection{Intuitions for $1/f$ Noise in HI IM Surveys}\label{sec:dis3}

As the results and discussions in this work have shown, there is no simple way of determining the exact impact of $1/f$ noise on any given future HI IM survey. However, several important guidelines can be inferred. First, and most importantly, if $1/f$ noise is completely correlated across the bandpass and is not interacting with any other systematics then it can be perfectly subtracted from the data. This ideal scenario is not likely to be the case in real data, but the existence of this limit is encouraging as it sets an ultimate goal in terms of receiver frequency stability. Therefore in the instance that $1/f$ noise is not fully correlated across the band the following guidelines should be considered:
\begin{enumerate}
	\item First, and most importantly, attempts to measure the $\alpha$ and $\beta$ of SD HI IM receivers should be made in order to inform the planning of any future survey.
	\item The frequency correlations as described by $\beta$ (or any other functional form) are critical to determine. Instruments with highly uncorrelated $1/f$ noise in frequency will find HI IM significantly challenging.
    \item Care should taken to preserve the statistical properties of the $1/f$ noise frequency spectrum to avoid inadvertently increasing the effective $\beta$ of the $1/f$ noise.
	\item The spatial power of $1/f$ noise integrates down as $1/\sqrt{T_\mathrm{obs}}$ (assuming Gaussainity and stationarity), but the initial r.m.s. of the $1/f$ noise fluctuations depends strongly on $\alpha$, which in turn can significantly impact the observing time required to achieve a desired SNR.
    \item Scan speeds should be as fast as feasibly possible, with the aim of at least achieving a speed that matches the period of a slew to the knee frequency of the $1/f$ noise integrated across the channel width to be used in the final data analysis.
    \item The observing time of the experiment should be long enough such that the integrated angular power of the $1/f$ noise in the map is less than the HI angular power spectrum at all scales of interest. 
\end{enumerate}

\section{Conclusion}\label{sec:conclusion}

This work has presented for the first time a set of end-to-end simulations that describe the impact of $1/f$ noise on a potential future SKA HI IM survey. As HI IM surveys use spectroscopic measurements of the extragalactic HI signal it was necessary to expand the existing models of $1/f$ noise to include correlated spectral fluctuations, which were parametrised in this work by $\beta$. It was found that in the ideal case that $1/f$ noise is completely correlated across the bandpass of spectroscopic receivers then it is trivial for existing component separations methods to remove it. However, should the $1/f$ noise not be perfectly correlated and have a $\beta > 0$ then $1/f$ noise can have a significant impact on the recovery of the HI angular power spectrum through both increased statistical uncertainty and an intrinsic bias. Exactly how much impact the $1/f$ noise has depends on the degree of $1/f$ noise correlation, and a combination of the intrinsic statistical properties of the $1/f$ noise such as the knee frequency and $\alpha$, and the choice of telescope observing strategy.

As this is the first detailed exploration of $1/f$ noise in HI IM, numerous assumptions have been made to simplify the $1/f$ noise signal. However, in real data some of these assumptions are unlikely to be valid and therefore $1/f$ noise should be expected to be more challenging to remove than implied by these simulations. First the intrinsic properties of the $1/f$ noise are assumed to have Gaussian distributed amplitudes, stationary values of $\alpha$ and knee frequency, and a simple power law functional form for the frequency correlations. None of these assumptions are necessarily true, and should be investigated during the pre-commissioning stages of single-dish IM experiments. Further, the interactions of the $1/f$ noise with additional systematics is not explored, however in real data there will be standing waves, radio-frequency interference, calibration errors, beam deconvolution errors and many more potential instrumental systematics. 

Finally, given the above assumptions, this paper does provide some rough numerical guidelines for which a future SKA HI IM survey should attempt to achieve. First, a design consideration for receivers should be to make the 1/f noise as correlated as possible across the band (e.g., $\beta \rightarrow 0$). Following this, the data analysis pipeline should be carefully designed as to not corrupt the receiver $1/f$ noise properties in order to avoid inadvertent increases in the effective value of $\beta$. It was found $1/f$ noise $\alpha$ can have a major impact on the recovery of the HI angular power spectrum, and ideally $\alpha$ would be minimised during the design of the receivers. However, this may not always be practical due to other design constraints, therefore careful consideration of how to mitigate $\alpha$ should be implemented into any HI IM data analysis pipeline. Finally, the impact of the knee frequency on an HI IM experiment is to increase the potential observing time required to achieve a desired SNR, for a 30\,day SKA HI IM survey the knee frequency ideally would be $f_k < 10$\,mHz per 20\,MHz channel. In terms of scanning strategy a slew speed of 1\,deg\,s$^{-1}$ should be sufficient. In conclusion, the analysis presented here finds that $1/f$ noise could be significantly detrimental to future HI IM surveys, but careful choices of observing strategy, data analysis and component separation method should be sufficient to ensure successful recovery of the cosmological HI signal.

\section{Acknowledgements}

SH, CD and RAB acknowledge support from an STFC Consolidated Grant (ST/L000768/1). SH and CD acknowledges support from an ERC Starting (Consolidator) Grant (no.$\sim$307209) under FP7. LCO acknowledges funding from CNPq, Conselho Nacional de Desenvolvimento Cient\'{i}fico e Tecnol\'{o}gico $–$ Brazil. YZM acknowledges the support from the National Research Foundation of South Africa with Grant no.~105925 and no.~104800. The authors would also like to acknowledge Mario Santos for providing useful feedback during the development of this work.

\bibliography{references,FlickerNoise,fullpapers}

\newcommand{\noop}[1]{}
\begin{thebibliography}{}
\makeatletter
\relax
\def\mn@urlcharsother{\let\do\@makeother \do\$\do\&\do\#\do\^\do\_\do\%\do\~}
\def\mn@doi{\begingroup\mn@urlcharsother \@ifnextchar [ {\mn@doi@}
  {\mn@doi@[]}}
\def\mn@doi@[#1]#2{\def\@tempa{#1}\ifx\@tempa\@empty \href
  {http://dx.doi.org/#2} {doi:#2}\else \href {http://dx.doi.org/#2} {#1}\fi
  \endgroup}
\def\mn@eprint#1#2{\mn@eprint@#1:#2::\@nil}
\def\mn@eprint@arXiv#1{\href {http://arxiv.org/abs/#1} {{\tt arXiv:#1}}}
\def\mn@eprint@dblp#1{\href {http://dblp.uni-trier.de/rec/bibtex/#1.xml}
  {dblp:#1}}
\def\mn@eprint@#1:#2:#3:#4\@nil{\def\@tempa {#1}\def\@tempb {#2}\def\@tempc
  {#3}\ifx \@tempc \@empty \let \@tempc \@tempb \let \@tempb \@tempa \fi \ifx
  \@tempb \@empty \def\@tempb {arXiv}\fi \@ifundefined
  {mn@eprint@\@tempb}{\@tempb:\@tempc}{\expandafter \expandafter \csname
  mn@eprint@\@tempb\endcsname \expandafter{\@tempc}}}

\bibitem[\protect\citeauthoryear{{Alam} et~al.,}{{Alam}
  et~al.}{2015}]{Alam2015}
{Alam} S.,  et~al., 2015, \mn@doi [\apjs] {10.1088/0067-0049/219/1/12}, \href
  {http://adsabs.harvard.edu/abs/2015ApJS..219...12A} {219, 12}

\bibitem[\protect\citeauthoryear{{Albrecht} et~al.,}{{Albrecht}
  et~al.}{2006}]{Albrecht2006}
{Albrecht} A.,  et~al., 2006, ArXiv Astrophysics e-prints, \href
  {http://adsabs.harvard.edu/abs/2006astro.ph..9591A} {}

\bibitem[\protect\citeauthoryear{{Alonso}, {Bull}, {Ferreira}  \&
  {Santos}}{{Alonso} et~al.}{2015}]{Alonso2015}
{Alonso} D.,  {Bull} P.,  {Ferreira} P.~G.,   {Santos} M.~G.,  2015, \mn@doi
  [\mnras] {10.1093/mnras/stu2474}, \href
  {http://adsabs.harvard.edu/abs/2015MNRAS.447..400A} {447, 400}

\bibitem[\protect\citeauthoryear{{Alves}, {Davies}, {Dickinson}, {Calabretta},
  {Davis}  \& {Staveley-Smith}}{{Alves} et~al.}{2012}]{Alves2012}
{Alves} M.~I.~R.,  {Davies} R.~D.,  {Dickinson} C.,  {Calabretta} M.,  {Davis}
  R.,   {Staveley-Smith} L.,  2012, \mn@doi [\mnras]
  {10.1111/j.1365-2966.2012.20796.x}, \href
  {http://adsabs.harvard.edu/abs/2012MNRAS.422.2429A} {422, 2429}

\bibitem[\protect\citeauthoryear{{Anderson} et~al.,}{{Anderson}
  et~al.}{2017}]{Anderson2017}
{Anderson} C.~J.,  et~al., 2017, preprint, \href
  {http://adsabs.harvard.edu/abs/2017arXiv171000424A} {} (\mn@eprint {arXiv}
  {1710.00424})

\bibitem[\protect\citeauthoryear{{Aubourg} et~al.,}{{Aubourg}
  et~al.}{2015}]{Aubourg2015}
{Aubourg} {\'E}.,  et~al., 2015, \mn@doi [\prd] {10.1103/PhysRevD.92.123516},
  \href {http://adsabs.harvard.edu/abs/2015PhRvD..92l3516A} {92, 123516}

\bibitem[\protect\citeauthoryear{{Bandura} et~al.,}{{Bandura}
  et~al.}{2014}]{Bandura2014}
{Bandura} K.,  et~al., 2014, in Ground-based and Airborne Telescopes V. p.
  914522 (\mn@eprint {arXiv} {1406.2288}), \mn@doi{10.1117/12.2054950}

\bibitem[\protect\citeauthoryear{{Battye}, {Davies}  \& {Weller}}{{Battye}
  et~al.}{2004}]{Battye2004}
{Battye} R.~A.,  {Davies} R.~D.,   {Weller} J.,  2004, \mn@doi [\mnras]
  {10.1111/j.1365-2966.2004.08416.x}, \href
  {http://adsabs.harvard.edu/abs/2004MNRAS.355.1339B} {355, 1339}

\bibitem[\protect\citeauthoryear{{Battye}, {Browne}, {Dickinson}, {Heron},
  {Maffei}  \& {Pourtsidou}}{{Battye} et~al.}{2013}]{Battye2013}
{Battye} R.~A.,  {Browne} I.~W.~A.,  {Dickinson} C.,  {Heron} G.,  {Maffei} B.,
    {Pourtsidou} A.,  2013, \mn@doi [\mnras] {10.1093/mnras/stt1082}, \href
  {http://adsabs.harvard.edu/abs/2013MNRAS.434.1239B} {434, 1239}

\bibitem[\protect\citeauthoryear{{Berger}, {Oppermann}, {Pen}  \&
  {Shaw}}{{Berger} et~al.}{2017}]{Berger2016}
{Berger} P.,  {Oppermann} N.,  {Pen} U.-L.,   {Shaw} J.~R.,  2017, \mn@doi
  [\mnras] {10.1093/mnras/stx2329}, \href
  {http://adsabs.harvard.edu/abs/2017MNRAS.472.4928B} {472, 4928}

\bibitem[\protect\citeauthoryear{{Betoule} et~al.,}{{Betoule}
  et~al.}{2014}]{Betoule2014}
{Betoule} M.,  et~al., 2014, \mn@doi [\aap] {10.1051/0004-6361/201423413},
  \href {http://adsabs.harvard.edu/abs/2014A%26A...568A..22B} {568, A22}

\bibitem[\protect\citeauthoryear{{Bigot-Sazy} et~al.,}{{Bigot-Sazy}
  et~al.}{2015}]{BigotSazy2015}
{Bigot-Sazy} M.-A.,  et~al., 2015, \mn@doi [\mnras] {10.1093/mnras/stv2153},
  \href {http://adsabs.harvard.edu/abs/2015MNRAS.454.3240B} {454, 3240}

\bibitem[\protect\citeauthoryear{{Bigot-Sazy} et~al.,}{{Bigot-Sazy}
  et~al.}{2016}]{Bigot2016}
{Bigot-Sazy} M.-A.,  et~al., 2016, in {Qain} L.,  {Li} D.,  eds,  Astronomical
  Society of the Pacific Conference Series Vol. 502, Frontiers in Radio
  Astronomy and FAST Early Sciences Symposium 2015. p.~41 (\mn@eprint {arXiv}
  {1511.03006})

\bibitem[\protect\citeauthoryear{{Bonaldi} \& {Brown}}{{Bonaldi} \&
  {Brown}}{2015}]{Bonaldi2015}
{Bonaldi} A.,  {Brown} M.~L.,  2015, \mn@doi [\mnras] {10.1093/mnras/stu2601},
  \href {http://adsabs.harvard.edu/abs/2015MNRAS.447.1973B} {447, 1973}

\bibitem[\protect\citeauthoryear{{Bowman} et~al.,}{{Bowman}
  et~al.}{2013}]{Bowman2013}
{Bowman} J.~D.,  et~al., 2013, \mn@doi [\pasa] {10.1017/pas.2013.009}, \href
  {http://adsabs.harvard.edu/abs/2013PASA...30...31B} {30, e031}

\bibitem[\protect\citeauthoryear{{Bull}, {Ferreira}, {Patel}  \&
  {Santos}}{{Bull} et~al.}{2015}]{Bull2015}
{Bull} P.,  {Ferreira} P.~G.,  {Patel} P.,   {Santos} M.~G.,  2015, \mn@doi
  [\apj] {10.1088/0004-637X/803/1/21}, \href
  {http://adsabs.harvard.edu/abs/2015ApJ...803...21B} {803, 21}

\bibitem[\protect\citeauthoryear{Caloyannides}{Caloyannides}{1974}]{Caloyannides1974}
Caloyannides M.~A.,  1974, \mn@doi [Journal of Applied Physics]
  {10.1063/1.1662977}, 45, 307

\bibitem[\protect\citeauthoryear{{Catinella} \& {Cortese}}{{Catinella} \&
  {Cortese}}{2015}]{Catinella2015}
{Catinella} B.,  {Cortese} L.,  2015, \mn@doi [\mnras] {10.1093/mnras/stu2241},
  \href {http://adsabs.harvard.edu/abs/2015MNRAS.446.3526C} {446, 3526}

\bibitem[\protect\citeauthoryear{{Chang}, {Pen}, {Bandura}  \&
  {Peterson}}{{Chang} et~al.}{2010}]{Chang2010}
{Chang} T.-C.,  {Pen} U.-L.,  {Bandura} K.,   {Peterson} J.~B.,  2010, \mn@doi
  [\nat] {10.1038/nature09187}, \href
  {http://adsabs.harvard.edu/abs/2010Natur.466..463C} {466, 463}

\bibitem[\protect\citeauthoryear{{Chapman} et~al.,}{{Chapman}
  et~al.}{2012}]{Chapman2012}
{Chapman} E.,  et~al., 2012, \mn@doi [\mnras]
  {10.1111/j.1365-2966.2012.21065.x}, \href
  {http://adsabs.harvard.edu/abs/2012MNRAS.423.2518C} {423, 2518}

\bibitem[\protect\citeauthoryear{{Chen}}{{Chen}}{2012}]{Chen2012}
{Chen} X.,  2012, in International Journal of Modern Physics Conference Series.
  pp 256--263 (\mn@eprint {arXiv} {1212.6278}),
  \mn@doi{10.1142/S2010194512006459}

\bibitem[\protect\citeauthoryear{{Chon}, {Challinor}, {Prunet}, {Hivon}  \&
  {Szapudi}}{{Chon} et~al.}{2004}]{Chon2004}
{Chon} G.,  {Challinor} A.,  {Prunet} S.,  {Hivon} E.,   {Szapudi} I.,  2004,
  \mn@doi [\mnras] {10.1111/j.1365-2966.2004.07737.x}, \href
  {http://adsabs.harvard.edu/abs/2004MNRAS.350..914C} {350, 914}

\bibitem[\protect\citeauthoryear{{Condon}}{{Condon}}{1992}]{Condon1992}
{Condon} J.~J.,  1992, \mn@doi [\araa] {10.1146/annurev.aa.30.090192.003043},
  \href {http://adsabs.harvard.edu/abs/1992ARA%26A..30..575C} {30, 575}

\bibitem[\protect\citeauthoryear{{Crites} et~al.,}{{Crites}
  et~al.}{2014}]{Crites2014}
{Crites} A.~T.,  et~al., 2014, in Millimeter, Submillimeter, and Far-Infrared
  Detectors and Instrumentation for Astronomy VII. p. 91531W,
  \mn@doi{10.1117/12.2057207}

\bibitem[\protect\citeauthoryear{{Croft} et~al.,}{{Croft}
  et~al.}{2016}]{Croft2016}
{Croft} R.~A.~C.,  et~al., 2016, \mn@doi [\mnras] {10.1093/mnras/stw204}, \href
  {http://adsabs.harvard.edu/abs/2016MNRAS.457.3541C} {457, 3541}

\bibitem[\protect\citeauthoryear{Dalcin, Paz, Kler  \& Cosimo}{Dalcin
  et~al.}{2011}]{Dalcin2011}
Dalcin L.~D.,  Paz R.~R.,  Kler P.~A.,   Cosimo A.,  2011, \mn@doi [Advances in
  Water Resources] {http://dx.doi.org/10.1016/j.advwatres.2011.04.013}, 34,
  1124

\bibitem[\protect\citeauthoryear{{Datta}, {Choudhury}  \& {Bharadwaj}}{{Datta}
  et~al.}{2007}]{Datta2007}
{Datta} K.~K.,  {Choudhury} T.~R.,   {Bharadwaj} S.,  2007, \mn@doi [\mnras]
  {10.1111/j.1365-2966.2007.11747.x}, \href
  {http://adsabs.harvard.edu/abs/2007MNRAS.378..119D} {378, 119}

\bibitem[\protect\citeauthoryear{{DeBoer} et~al.,}{{DeBoer}
  et~al.}{2017}]{DeBoer2017}
{DeBoer} D.~R.,  et~al., 2017, \mn@doi [\pasp]
  {10.1088/1538-3873/129/974/045001}, \href
  {http://adsabs.harvard.edu/abs/2017PASP..129d5001D} {129, 045001}

\bibitem[\protect\citeauthoryear{{Dickinson}, {Davies}  \& {Davis}}{{Dickinson}
  et~al.}{2003}]{Dickinson2003}
{Dickinson} C.,  {Davies} R.~D.,   {Davis} R.~J.,  2003, \mn@doi [\mnras]
  {10.1046/j.1365-8711.2003.06439.x}, \href
  {http://adsabs.harvard.edu/abs/2003MNRAS.341..369D} {341, 369}

\bibitem[\protect\citeauthoryear{{Draine}}{{Draine}}{2011}]{Draine2011}
{Draine} B.~T.,  2011, {Physics of the Interstellar and Intergalactic Medium}

\bibitem[\protect\citeauthoryear{{Eisenstein} et~al.,}{{Eisenstein}
  et~al.}{2005}]{Eisenstein2005}
{Eisenstein} D.~J.,  et~al., 2005, \mn@doi [\apj] {10.1086/466512}, \href
  {http://adsabs.harvard.edu/abs/2005ApJ...633..560E} {633, 560}

\bibitem[\protect\citeauthoryear{{Emerson} \& {Graeve}}{{Emerson} \&
  {Graeve}}{1988}]{Emerson1988}
{Emerson} D.~T.,  {Graeve} R.,  1988, \aap, \href
  {http://adsabs.harvard.edu/abs/1988A%26A...190..353E} {190, 353}

\bibitem[\protect\citeauthoryear{{Fan}, {Carilli}  \& {Keating}}{{Fan}
  et~al.}{2006}]{Fan2006}
{Fan} X.,  {Carilli} C.~L.,   {Keating} B.,  2006, \mn@doi [\araa]
  {10.1146/annurev.astro.44.051905.092514}, \href
  {http://adsabs.harvard.edu/abs/2006ARA%26A..44..415F} {44, 415}

\bibitem[\protect\citeauthoryear{{Fern{\'a}ndez} et~al.,}{{Fern{\'a}ndez}
  et~al.}{2016}]{Fernandez2016}
{Fern{\'a}ndez} X.,  et~al., 2016, \mn@doi [\apjl]
  {10.3847/2041-8205/824/1/L1}, \href
  {http://adsabs.harvard.edu/abs/2016ApJ...824L...1F} {824, L1}

\bibitem[\protect\citeauthoryear{Gilden, Thornton  \& Mallon}{Gilden
  et~al.}{1995}]{gilden1995}
Gilden D.~L.,  Thornton T.,   Mallon M.~W.,  1995, Science, 267, 1837

\bibitem[\protect\citeauthoryear{{Giovanelli} et~al.,}{{Giovanelli}
  et~al.}{2005}]{Giovanelli2005}
{Giovanelli} R.,  et~al., 2005, \mn@doi [\aj] {10.1086/497431}, \href
  {http://adsabs.harvard.edu/abs/2005AJ....130.2598G} {130, 2598}

\bibitem[\protect\citeauthoryear{{G{\'o}rski}, {Hivon}, {Banday}, {Wandelt},
  {Hansen}, {Reinecke}  \& {Bartelmann}}{{G{\'o}rski}
  et~al.}{2005}]{Gorski2005}
{G{\'o}rski} K.~M.,  {Hivon} E.,  {Banday} A.~J.,  {Wandelt} B.~D.,  {Hansen}
  F.~K.,  {Reinecke} M.,   {Bartelmann} M.,  2005, \mn@doi [\apj]
  {10.1086/427976}, \href {http://adsabs.harvard.edu/abs/2005ApJ...622..759G}
  {622, 759}

\bibitem[\protect\citeauthoryear{{Haslam}, {Salter}, {Stoffel}  \&
  {Wilson}}{{Haslam} et~al.}{1982}]{Haslam1982}
{Haslam} C.~G.~T.,  {Salter} C.~J.,  {Stoffel} H.,   {Wilson} W.~E.,  1982,
  \aaps, \href {http://adsabs.harvard.edu/abs/1982A%26AS...47....1H} {47, 1}

\bibitem[\protect\citeauthoryear{{Hauser} \& {Peebles}}{{Hauser} \&
  {Peebles}}{1973}]{Hauser1973}
{Hauser} M.~G.,  {Peebles} P.~J.~E.,  1973, \mn@doi [\apj] {10.1086/152453},
  \href {http://adsabs.harvard.edu/abs/1973ApJ...185..757H} {185, 757}

\bibitem[\protect\citeauthoryear{{Hellbourg}, {Chippendale}, {Kesteven}  \&
  {Jeffs}}{{Hellbourg} et~al.}{2014}]{Hellbourg2014}
{Hellbourg} G.,  {Chippendale} A.~P.,  {Kesteven} M.~J.,   {Jeffs} B.~D.,
  2014, in Signal and Information Processing (GlobalSIP), 2014 IEEE Global
  Conference on, p. 1286-1290. pp 1286--1290,
  \mn@doi{10.1109/GlobalSIP.2014.7032330}

\bibitem[\protect\citeauthoryear{{Hinton} et~al.,}{{Hinton}
  et~al.}{2017}]{Hinton2017}
{Hinton} S.~R.,  et~al., 2017, \mn@doi [\mnras] {10.1093/mnras/stw2725}, \href
  {http://adsabs.harvard.edu/abs/2017MNRAS.464.4807H} {464, 4807}

\bibitem[\protect\citeauthoryear{{Hunt}, {Pisano}  \& {Edel}}{{Hunt}
  et~al.}{2016}]{Hunt2016}
{Hunt} L.~R.,  {Pisano} D.~J.,   {Edel} S.,  2016, \mn@doi [\aj]
  {10.3847/0004-6256/152/2/30}, \href
  {http://adsabs.harvard.edu/abs/2016AJ....152...30H} {152, 30}

\bibitem[\protect\citeauthoryear{{Irfan} et~al.,}{{Irfan}
  et~al.}{2015}]{Irfan2015}
{Irfan} M.~O.,  et~al., 2015, \mn@doi [\mnras] {10.1093/mnras/stv212}, \href
  {http://adsabs.harvard.edu/abs/2015MNRAS.448.3572I} {448, 3572}

\bibitem[\protect\citeauthoryear{Johnson}{Johnson}{1925}]{johnson1925}
Johnson J.~B.,  1925, Physical review, 26, 71

\bibitem[\protect\citeauthoryear{{Keating} et~al.,}{{Keating}
  et~al.}{2015}]{Keating2015}
{Keating} G.~K.,  et~al., 2015, \mn@doi [\apj] {10.1088/0004-637X/814/2/140},
  \href {http://adsabs.harvard.edu/abs/2015ApJ...814..140K} {814, 140}

\bibitem[\protect\citeauthoryear{{Keating}, {Marrone}, {Bower}, {Leitch},
  {Carlstrom}  \& {DeBoer}}{{Keating} et~al.}{2016}]{Keating2016}
{Keating} G.~K.,  {Marrone} D.~P.,  {Bower} G.~C.,  {Leitch} E.,  {Carlstrom}
  J.~E.,   {DeBoer} D.~R.,  2016, \mn@doi [\apj] {10.3847/0004-637X/830/1/34},
  \href {http://adsabs.harvard.edu/abs/2016ApJ...830...34K} {830, 34}

\bibitem[\protect\citeauthoryear{{Knox}}{{Knox}}{1995}]{Knox1995}
{Knox} L.,  1995, \mn@doi [\prd] {10.1103/PhysRevD.52.4307}, \href
  {http://adsabs.harvard.edu/abs/1995PhRvD..52.4307K} {52, 4307}

\bibitem[\protect\citeauthoryear{Kobayashi \& Musha}{Kobayashi \&
  Musha}{1982}]{kobayashi1982}
Kobayashi M.,  Musha T.,  1982, IEEE transactions on Biomedical Engineering, pp
  456--457

\bibitem[\protect\citeauthoryear{{Lah} et~al.,}{{Lah} et~al.}{2009}]{Lah2009}
{Lah} P.,  et~al., 2009, \mn@doi [\mnras] {10.1111/j.1365-2966.2009.15368.x},
  \href {http://adsabs.harvard.edu/abs/2009MNRAS.399.1447L} {399, 1447}

\bibitem[\protect\citeauthoryear{Lewis \& Bridle}{Lewis \&
  Bridle}{2002}]{Lewis2002}
Lewis A.,  Bridle S.,  2002, \mn@doi [Phys. Rev.] {10.1103/PhysRevD.66.103511},
  D66, 103511

\bibitem[\protect\citeauthoryear{{Li}, {Wechsler}, {Devaraj}  \& {Church}}{{Li}
  et~al.}{2016}]{Li2016}
{Li} T.~Y.,  {Wechsler} R.~H.,  {Devaraj} K.,   {Church} S.~E.,  2016, \mn@doi
  [\apj] {10.3847/0004-637X/817/2/169}, \href
  {http://adsabs.harvard.edu/abs/2016ApJ...817..169L} {817, 169}

\bibitem[\protect\citeauthoryear{{Lidz}, {Furlanetto}, {Oh}, {Aguirre},
  {Chang}, {Dor{\'e}}  \& {Pritchard}}{{Lidz} et~al.}{2011}]{Lidz2011}
{Lidz} A.,  {Furlanetto} S.~R.,  {Oh} S.~P.,  {Aguirre} J.,  {Chang} T.-C.,
  {Dor{\'e}} O.,   {Pritchard} J.~R.,  2011, \mn@doi [\apj]
  {10.1088/0004-637X/741/2/70}, \href
  {http://adsabs.harvard.edu/abs/2011ApJ...741...70L} {741, 70}

\bibitem[\protect\citeauthoryear{{Loeb} \& {Wyithe}}{{Loeb} \&
  {Wyithe}}{2008}]{Loeb2008}
{Loeb} A.,  {Wyithe} J.~S.~B.,  2008, \mn@doi [Physical Review Letters]
  {10.1103/PhysRevLett.100.161301}, \href
  {http://adsabs.harvard.edu/abs/2008PhRvL.100p1301L} {100, 161301}

\bibitem[\protect\citeauthoryear{{Loeb} \& {Zaldarriaga}}{{Loeb} \&
  {Zaldarriaga}}{2004}]{Loeb2004}
{Loeb} A.,  {Zaldarriaga} M.,  2004, \mn@doi [Physical Review Letters]
  {10.1103/PhysRevLett.92.211301}, \href
  {http://adsabs.harvard.edu/abs/2004PhRvL..92u1301L} {92, 211301}

\bibitem[\protect\citeauthoryear{{Ma} \& {Scott}}{{Ma} \&
  {Scott}}{2016}]{Ma2016}
{Ma} Y.-Z.,  {Scott} D.,  2016, \mn@doi [\prd] {10.1103/PhysRevD.93.083510},
  \href {http://adsabs.harvard.edu/abs/2016PhRvD..93h3510M} {93, 083510}

\bibitem[\protect\citeauthoryear{Mandal, Arfin  \& Sarpeshkar}{Mandal
  et~al.}{2009}]{Mandal2009}
Mandal S.,  Arfin S.~K.,   Sarpeshkar R.,  2009, \mn@doi [Electronics Letters]
  {10.1049/el:20092638}, 45, 81

\bibitem[\protect\citeauthoryear{Mandelbrot \& Wallis}{Mandelbrot \&
  Wallis}{1969}]{Mandelbrot1969}
Mandelbrot B.~B.,  Wallis J.~R.,  1969, \mn@doi [Water Resources Research]
  {10.1029/WR005i002p00321}, 5, 321

\bibitem[\protect\citeauthoryear{{Masui}, {McDonald}  \& {Pen}}{{Masui}
  et~al.}{2010}]{Masui2010}
{Masui} K.~W.,  {McDonald} P.,   {Pen} U.-L.,  2010, \mn@doi [\prd]
  {10.1103/PhysRevD.81.103527}, \href
  {http://adsabs.harvard.edu/abs/2010PhRvD..81j3527M} {81, 103527}

\bibitem[\protect\citeauthoryear{{Masui} et~al.,}{{Masui}
  et~al.}{2013}]{Masui2013}
{Masui} K.~W.,  et~al., 2013, \mn@doi [\apjl] {10.1088/2041-8205/763/1/L20},
  \href {http://adsabs.harvard.edu/abs/2013ApJ...763L..20M} {763, L20}

\bibitem[\protect\citeauthoryear{{McQuinn}, {Lidz}, {Zaldarriaga}, {Hernquist}
  \& {Dutta}}{{McQuinn} et~al.}{2008}]{McQuinn2008}
{McQuinn} M.,  {Lidz} A.,  {Zaldarriaga} M.,  {Hernquist} L.,   {Dutta} S.,
  2008, \mn@doi [\mnras] {10.1111/j.1365-2966.2008.13271.x}, \href
  {http://adsabs.harvard.edu/abs/2008MNRAS.388.1101M} {388, 1101}

\bibitem[\protect\citeauthoryear{{Natoli}, {de Gasperis}, {Gheller}  \&
  {Vittorio}}{{Natoli} et~al.}{2001}]{Natoli2001}
{Natoli} P.,  {de Gasperis} G.,  {Gheller} C.,   {Vittorio} N.,  2001, \mn@doi
  [\aap] {10.1051/0004-6361:20010393}, \href
  {http://adsabs.harvard.edu/abs/2001A%26A...372..346N} {372, 346}

\bibitem[\protect\citeauthoryear{{Neeleman}, {Prochaska}, {Ribaudo}, {Lehner},
  {Howk}, {Rafelski}  \& {Kanekar}}{{Neeleman} et~al.}{2016}]{Neeleman2016}
{Neeleman} M.,  {Prochaska} J.~X.,  {Ribaudo} J.,  {Lehner} N.,  {Howk} J.~C.,
  {Rafelski} M.,   {Kanekar} N.,  2016, \mn@doi [\apj]
  {10.3847/0004-637X/818/2/113}, \href
  {http://adsabs.harvard.edu/abs/2016ApJ...818..113N} {818, 113}

\bibitem[\protect\citeauthoryear{{Newburgh} et~al.,}{{Newburgh}
  et~al.}{2016}]{Newburgh2016}
{Newburgh} L.~B.,  et~al., 2016, in Society of Photo-Optical Instrumentation
  Engineers (SPIE) Conference Series. p. 99065X (\mn@eprint {arXiv}
  {1607.02059}), \mn@doi{10.1117/12.2234286}

\bibitem[\protect\citeauthoryear{{Noterdaeme} et~al.,}{{Noterdaeme}
  et~al.}{2012}]{Noterdaeme2012}
{Noterdaeme} P.,  et~al., 2012, \mn@doi [\aap] {10.1051/0004-6361/201220259},
  \href {http://adsabs.harvard.edu/abs/2012A%26A...547L...1N} {547, L1}

\bibitem[\protect\citeauthoryear{Nyquist}{Nyquist}{1928}]{Nyquist1928}
Nyquist H.,  1928, Physical Review, 32, 110

\bibitem[\protect\citeauthoryear{{Olivari}, {Remazeilles}  \&
  {Dickinson}}{{Olivari} et~al.}{2016}]{Olivari2016}
{Olivari} L.~C.,  {Remazeilles} M.,   {Dickinson} C.,  2016, \mn@doi [\mnras]
  {10.1093/mnras/stv2884}, \href
  {http://adsabs.harvard.edu/abs/2016MNRAS.456.2749O} {456, 2749}

\bibitem[\protect\citeauthoryear{{Paciga} et~al.,}{{Paciga}
  et~al.}{2011}]{Paciga2011}
{Paciga} G.,  et~al., 2011, \mn@doi [\mnras]
  {10.1111/j.1365-2966.2011.18208.x}, \href
  {http://adsabs.harvard.edu/abs/2011MNRAS.413.1174P} {413, 1174}

\bibitem[\protect\citeauthoryear{{Padmanabhan}, {Choudhury}  \&
  {Refregier}}{{Padmanabhan} et~al.}{2015}]{Padmanabhan2015}
{Padmanabhan} H.,  {Choudhury} T.~R.,   {Refregier} A.,  2015, \mn@doi [\mnras]
  {10.1093/mnras/stu2702}, \href
  {http://adsabs.harvard.edu/abs/2015MNRAS.447.3745P} {447, 3745}

\bibitem[\protect\citeauthoryear{{Parsons} et~al.,}{{Parsons}
  et~al.}{2010}]{Parsons2010}
{Parsons} A.~R.,  et~al., 2010, \mn@doi [\aj] {10.1088/0004-6256/139/4/1468},
  \href {http://adsabs.harvard.edu/abs/2010AJ....139.1468P} {139, 1468}

\bibitem[\protect\citeauthoryear{{Peebles}}{{Peebles}}{1973}]{Peebles1973}
{Peebles} P.~J.~E.,  1973, \mn@doi [\apj] {10.1086/152431}, \href
  {http://adsabs.harvard.edu/abs/1973ApJ...185..413P} {185, 413}

\bibitem[\protect\citeauthoryear{{Pen}, {Staveley-Smith}, {Peterson}  \&
  {Chang}}{{Pen} et~al.}{2009}]{Pen2009}
{Pen} U.-L.,  {Staveley-Smith} L.,  {Peterson} J.~B.,   {Chang} T.-C.,  2009,
  \mn@doi [\mnras] {10.1111/j.1745-3933.2008.00581.x}, \href
  {http://adsabs.harvard.edu/abs/2009MNRAS.394L...6P} {394, L6}

\bibitem[\protect\citeauthoryear{{Peterson} \& {Suarez}}{{Peterson} \&
  {Suarez}}{2012}]{Peterson2012}
{Peterson} J.~B.,  {Suarez} E.,  2012, preprint, \href
  {http://adsabs.harvard.edu/abs/2012arXiv1206.0143P} {} (\mn@eprint {arXiv}
  {1206.0143})

\bibitem[\protect\citeauthoryear{{Planck Collaboration} et~al.,}{{Planck
  Collaboration} et~al.}{2016a}]{PlanckX2016}
{Planck Collaboration} et~al., 2016a, \mn@doi [\aap]
  {10.1051/0004-6361/201525967}, \href
  {http://adsabs.harvard.edu/abs/2016A%26A...594A..10P} {594, A10}

\bibitem[\protect\citeauthoryear{{Planck Collaboration} et~al.,}{{Planck
  Collaboration} et~al.}{2016b}]{PlanckXXV2016}
{Planck Collaboration} et~al., 2016b, \mn@doi [\aap]
  {10.1051/0004-6361/201526803}, \href
  {http://adsabs.harvard.edu/abs/2016A%26A...594A..25P} {594, A25}

\bibitem[\protect\citeauthoryear{{Platania}, {Burigana}, {Maino}, {Caserini},
  {Bersanelli}, {Cappellini}  \& {Mennella}}{{Platania}
  et~al.}{2003}]{Platania2003}
{Platania} P.,  {Burigana} C.,  {Maino} D.,  {Caserini} E.,  {Bersanelli} M.,
  {Cappellini} B.,   {Mennella} A.,  2003, \mn@doi [\aap]
  {10.1051/0004-6361:20031125}, \href
  {http://adsabs.harvard.edu/abs/2003A%26A...410..847P} {410, 847}

\bibitem[\protect\citeauthoryear{{Pourtsidou}}{{Pourtsidou}}{2016}]{Pourtsidou2016}
{Pourtsidou} A.,  2016, \mn@doi [\mnras] {10.1093/mnras/stw1406}, \href
  {http://adsabs.harvard.edu/abs/2016MNRAS.461.1457P} {461, 1457}

\bibitem[\protect\citeauthoryear{{Press}}{{Press}}{1978}]{Press1978}
{Press} W.~H.,  1978, Comments on Astrophysics, \href
  {http://adsabs.harvard.edu/abs/1978ComAp...7..103P} {7, 103}

\bibitem[\protect\citeauthoryear{{Pritchard} \& {Loeb}}{{Pritchard} \&
  {Loeb}}{2008}]{Pritchard2008}
{Pritchard} J.~R.,  {Loeb} A.,  2008, \mn@doi [\prd]
  {10.1103/PhysRevD.78.103511}, \href
  {http://adsabs.harvard.edu/abs/2008PhRvD..78j3511P} {78, 103511}

\bibitem[\protect\citeauthoryear{{Pritchard} \& {Loeb}}{{Pritchard} \&
  {Loeb}}{2012}]{Pritchard2012}
{Pritchard} J.~R.,  {Loeb} A.,  2012, \mn@doi [Reports on Progress in Physics]
  {10.1088/0034-4885/75/8/086901}, \href
  {http://adsabs.harvard.edu/abs/2012RPPh...75h6901P} {75, 086901}

\bibitem[\protect\citeauthoryear{{Remazeilles}, {Dickinson}, {Banday},
  {Bigot-Sazy}  \& {Ghosh}}{{Remazeilles} et~al.}{2015}]{Remazeilles2015}
{Remazeilles} M.,  {Dickinson} C.,  {Banday} A.~J.,  {Bigot-Sazy} M.-A.,
  {Ghosh} T.,  2015, \mn@doi [\mnras] {10.1093/mnras/stv1274}, \href
  {http://adsabs.harvard.edu/abs/2015MNRAS.451.4311R} {451, 4311}

\bibitem[\protect\citeauthoryear{{Rhee}, {Lah}, {Chengalur}, {Briggs}  \&
  {Colless}}{{Rhee} et~al.}{2016}]{Rhee2016}
{Rhee} J.,  {Lah} P.,  {Chengalur} J.~N.,  {Briggs} F.~H.,   {Colless} M.,
  2016, \mn@doi [\mnras] {10.1093/mnras/stw1097}, \href
  {http://adsabs.harvard.edu/abs/2016MNRAS.460.2675R} {460, 2675}

\bibitem[\protect\citeauthoryear{{Riess} et~al.,}{{Riess}
  et~al.}{1998}]{Riess1998}
{Riess} A.~G.,  et~al., 1998, \mn@doi [\aj] {10.1086/300499}, \href
  {http://adsabs.harvard.edu/abs/1998AJ....116.1009R} {116, 1009}

\bibitem[\protect\citeauthoryear{{Roychowdhury}}{{Roychowdhury}}{prep}]{Roychowdhury2017}
{Roychowdhury} S.,  \noop{3001}in prep.

\bibitem[\protect\citeauthoryear{{Rubi{\~n}o-Mart{\'{\i}}n}
  et~al.,}{{Rubi{\~n}o-Mart{\'{\i}}n} et~al.}{2010}]{Rubino2010}
{Rubi{\~n}o-Mart{\'{\i}}n} J.~A.,  et~al., 2010, \mn@doi [Astrophysics and
  Space Science Proceedings] {10.1007/978-3-642-11250-8_12}, \href
  {http://adsabs.harvard.edu/abs/2010ASSP...14..127R} {14, 127}

\bibitem[\protect\citeauthoryear{{SKA Collaboration}, {Turner}, {Stringhetti},
  {Braun}  \& {McPherson}}{{SKA Collaboration} et~al.}{2016}]{SKABaseline2016}
{SKA Collaboration} {Turner} W.,  {Stringhetti} L.,  {Braun} R.,   {McPherson}
  A.,  2016, Technical Report SKA-TEL-SKO-0000008, SKA Phase 1 System
  Requirements Specification.
SKA

\bibitem[\protect\citeauthoryear{{Santos}, {Cooray}  \& {Knox}}{{Santos}
  et~al.}{2005}]{Santos2005}
{Santos} M.~G.,  {Cooray} A.,   {Knox} L.,  2005, \mn@doi [\apj]
  {10.1086/429857}, \href {http://adsabs.harvard.edu/abs/2005ApJ...625..575S}
  {625, 575}

\bibitem[\protect\citeauthoryear{{Santos}, {Alonso}, {Bull}, {Camera}  \&
  {Ferreira}}{{Santos} et~al.}{2014}]{Santos2014}
{Santos} M.~G.,  {Alonso} D.,  {Bull} P.,  {Camera} S.,   {Ferreira} P.~G.,
  2014, in {Heavens} A.,  {Starck} J.-L.,   {Krone-Martins} A.,  eds,  IAU
  Symposium Vol. 306, Statistical Challenges in 21st Century Cosmology. pp
  165--176, \mn@doi{10.1017/S174392131401388X}

\bibitem[\protect\citeauthoryear{{Santos} et~al.,}{{Santos}
  et~al.}{2015}]{Santos2015}
{Santos} M.,  et~al., 2015, Advancing Astrophysics with the Square Kilometre
  Array (AASKA14), \href {http://adsabs.harvard.edu/abs/2015aska.confE..19S}
  {p.~19}

\bibitem[\protect\citeauthoryear{{Santos} et~al.,}{{Santos}
  et~al.}{2017}]{Santos2017}
{Santos} M.~G.,  et~al., 2017, preprint, \href
  {http://adsabs.harvard.edu/abs/2017arXiv170906099S} {} (\mn@eprint {arXiv}
  {1709.06099})

\bibitem[\protect\citeauthoryear{{Scheuer} \& {Williams}}{{Scheuer} \&
  {Williams}}{1968}]{Scheuer1968}
{Scheuer} P.~A.~G.,  {Williams} P.~J.~S.,  1968, \mn@doi [\araa]
  {10.1146/annurev.aa.06.090168.001541}, \href
  {http://adsabs.harvard.edu/abs/1968ARA%26A...6..321S} {6, 321}

\bibitem[\protect\citeauthoryear{{Seiffert}, {Mennella}, {Burigana},
  {Mandolesi}, {Bersanelli}, {Meinhold}  \& {Lubin}}{{Seiffert}
  et~al.}{2002}]{Seiffert2002}
{Seiffert} M.,  {Mennella} A.,  {Burigana} C.,  {Mandolesi} N.,  {Bersanelli}
  M.,  {Meinhold} P.,   {Lubin} P.,  2002, \mn@doi [\aap]
  {10.1051/0004-6361:20020880}, \href
  {http://adsabs.harvard.edu/abs/2002A%26A...391.1185S} {391, 1185}

\bibitem[\protect\citeauthoryear{{Sethi}}{{Sethi}}{2005}]{Sethi2005}
{Sethi} S.~K.,  2005, \mn@doi [\mnras] {10.1111/j.1365-2966.2005.09485.x},
  \href {http://adsabs.harvard.edu/abs/2005MNRAS.363..818S} {363, 818}

\bibitem[\protect\citeauthoryear{{Shaw}, {Sigurdson}, {Pen}, {Stebbins}  \&
  {Sitwell}}{{Shaw} et~al.}{2014}]{Shaw2014}
{Shaw} J.~R.,  {Sigurdson} K.,  {Pen} U.-L.,  {Stebbins} A.,   {Sitwell} M.,
  2014, \mn@doi [\apj] {10.1088/0004-637X/781/2/57}, \href
  {http://adsabs.harvard.edu/abs/2014ApJ...781...57S} {781, 57}

\bibitem[\protect\citeauthoryear{{Shaw}, {Sigurdson}, {Sitwell}, {Stebbins}  \&
  {Pen}}{{Shaw} et~al.}{2015}]{Shaw2015}
{Shaw} J.~R.,  {Sigurdson} K.,  {Sitwell} M.,  {Stebbins} A.,   {Pen} U.-L.,
  2015, \mn@doi [\prd] {10.1103/PhysRevD.91.083514}, \href
  {http://adsabs.harvard.edu/abs/2015PhRvD..91h3514S} {91, 083514}

\bibitem[\protect\citeauthoryear{{Smoot} \& {Debono}}{{Smoot} \&
  {Debono}}{2017}]{Smoot2014}
{Smoot} G.~F.,  {Debono} I.,  2017, \mn@doi [\aap]
  {10.1051/0004-6361/201526794}, \href
  {http://adsabs.harvard.edu/abs/2017A%26A...597A.136S} {597, A136}

\bibitem[\protect\citeauthoryear{{Sutton} et~al.,}{{Sutton}
  et~al.}{2010}]{Sutton2010}
{Sutton} D.,  et~al., 2010, \mn@doi [\mnras]
  {10.1111/j.1365-2966.2010.16954.x}, \href
  {http://adsabs.harvard.edu/abs/2010MNRAS.407.1387S} {407, 1387}

\bibitem[\protect\citeauthoryear{{Switzer} et~al.,}{{Switzer}
  et~al.}{2013}]{Switzer2013}
{Switzer} E.~R.,  et~al., 2013, \mn@doi [\mnras] {10.1093/mnrasl/slt074}, \href
  {http://adsabs.harvard.edu/abs/2013MNRAS.434L..46S} {434, L46}

\bibitem[\protect\citeauthoryear{{Traficante} et~al.,}{{Traficante}
  et~al.}{2011}]{Traficante2011}
{Traficante} A.,  et~al., 2011, \mn@doi [\mnras]
  {10.1111/j.1365-2966.2011.19244.x}, \href
  {http://adsabs.harvard.edu/abs/2011MNRAS.416.2932T} {416, 2932}

\bibitem[\protect\citeauthoryear{{Verheijen} et~al.,}{{Verheijen}
  et~al.}{2010}]{Verheijen2010}
{Verheijen} M.~A.~W.,  et~al., 2010, in ISKAF2010 Science Meeting. p.~54

\bibitem[\protect\citeauthoryear{{Villaescusa-Navarro}, {Alonso}  \&
  {Viel}}{{Villaescusa-Navarro} et~al.}{2017}]{Villaescusa2017}
{Villaescusa-Navarro} F.,  {Alonso} D.,   {Viel} M.,  2017, \mn@doi [\mnras]
  {10.1093/mnras/stw3224}, \href
  {http://adsabs.harvard.edu/abs/2017MNRAS.466.2736V} {466, 2736}

\bibitem[\protect\citeauthoryear{Voss \& Clarke}{Voss \&
  Clarke}{1978}]{voss1978}
Voss R.~F.,  Clarke J.,  1978, The Journal of the Acoustical Society of
  America, 63, 258

\bibitem[\protect\citeauthoryear{{Wandelt}, {Hivon}  \& {G{\'o}rski}}{{Wandelt}
  et~al.}{2001}]{Wandelt2001}
{Wandelt} B.~D.,  {Hivon} E.,   {G{\'o}rski} K.~M.,  2001, \mn@doi [\prd]
  {10.1103/PhysRevD.64.083003}, \href
  {http://adsabs.harvard.edu/abs/2001PhRvD..64h3003W} {64, 083003}

\bibitem[\protect\citeauthoryear{{Wilkinson}}{{Wilkinson}}{1991}]{Wilkinson1991}
{Wilkinson} P.~N.,  1991, in {Cornwell} T.~J.,  {Perley} R.~A.,  eds,
  Astronomical Society of the Pacific Conference Series Vol. 19, IAU Colloq.
  131: Radio Interferometry. Theory, Techniques, and Applications. pp 428--432

\bibitem[\protect\citeauthoryear{{Wilson}, {Rohlfs}  \&
  {H{\"u}ttemeister}}{{Wilson} et~al.}{2009}]{Wilson2009}
{Wilson} T.~L.,  {Rohlfs} K.,   {H{\"u}ttemeister} S.,  2009, {Tools of Radio
  Astronomy}.
Springer-Verlag, \mn@doi{10.1007/978-3-540-85122-6}

\bibitem[\protect\citeauthoryear{{Wolz} et~al.,}{{Wolz}
  et~al.}{2017}]{Wolz2015}
{Wolz} L.,  et~al., 2017, \mn@doi [\mnras] {10.1093/mnras/stw2556}, \href
  {http://adsabs.harvard.edu/abs/2017MNRAS.464.4938W} {464, 4938}

\bibitem[\protect\citeauthoryear{{Wyithe} \& {Loeb}}{{Wyithe} \&
  {Loeb}}{2008}]{Wyithe2008b}
{Wyithe} J.~S.~B.,  {Loeb} A.,  2008, \mn@doi [\mnras]
  {10.1111/j.1365-2966.2007.12568.x}, \href
  {http://adsabs.harvard.edu/abs/2008MNRAS.383..606W} {383, 606}

\bibitem[\protect\citeauthoryear{{Zwaan}, {van Dokkum}  \& {Verheijen}}{{Zwaan}
  et~al.}{2001}]{Zwaan2001}
{Zwaan} M.~A.,  {van Dokkum} P.~G.,   {Verheijen} M.~A.~W.,  2001, \mn@doi
  [Science] {10.1126/science.1063034}, \href
  {http://adsabs.harvard.edu/abs/2001Sci...293.1800Z} {293, 1800}

\bibitem[\protect\citeauthoryear{{Zwaan} et~al.,}{{Zwaan}
  et~al.}{2004}]{Zwaan2004}
{Zwaan} M.~A.,  et~al., 2004, \mn@doi [\mnras]
  {10.1111/j.1365-2966.2004.07782.x}, \href
  {http://adsabs.harvard.edu/abs/2004MNRAS.350.1210Z} {350, 1210}

\bibitem[\protect\citeauthoryear{{de Oliveira-Costa}, {Tegmark}, {Gaensler},
  {Jonas}, {Landecker}  \& {Reich}}{{de Oliveira-Costa}
  et~al.}{2008}]{OliveiraCosta2008}
{de Oliveira-Costa} A.,  {Tegmark} M.,  {Gaensler} B.~M.,  {Jonas} J.,
  {Landecker} T.~L.,   {Reich} P.,  2008, \mn@doi [\mnras]
  {10.1111/j.1365-2966.2008.13376.x}, \href
  {http://adsabs.harvard.edu/abs/2008MNRAS.388..247D} {388, 247}

\bibitem[\protect\citeauthoryear{{van Haarlem} et~al.,}{{van Haarlem}
  et~al.}{2013}]{vanHaarlem2013}
{van Haarlem} M.~P.,  et~al., 2013, \mn@doi [\aap]
  {10.1051/0004-6361/201220873}, \href
  {http://adsabs.harvard.edu/abs/2013A%26A...556A...2V} {556, A2}

\makeatother
\end{thebibliography}

\appendix

\section{PCA Subtraction of Perfectly Correlated Signals}\label{app:A}

In the main text it was stated that $1/f$ noise that is perfectly correlated in frequency ($\beta = 0$) was not included in the suite of simulations because in that instance PCA can perfectly subtract all the $1/f$ noise fluctuations. Recall from Eqn.~\ref{eqn:freqcorr} that the model $1/f$ noise frequency correlations are described by a power law. The model is designed such that if if $\beta = 0$ there is zero power in the 2D PSD for all Fourier modes apart from the zeroth-mode that describes the temporal fluctuations. This implies that the frequency fluctuations can be described by a single zero-order Fourier mode or constant offset in the TOD. Removing such a constant is trivial. However, the simulations use a receiver model that multiplies the $1/f$ noise gain fluctuations with the measured $\tsys$, which imprints a power law spectral curvature into the $1/f$ noise due to the sky signal. This implies that the $1/f$ noise is not perfectly correlated, but critically is described by only one functional form.

The following demonstrates a simple example of how PCA can easily remove $1/f$ noise described by a single functional form. Imagine a simple \textit{map} with $N_{\mathrm{pix}}$, two frequency channels, and a single sky component of arbitrary complexity in the spatial domain but can be described in frequency by a single function $f(\nu)$. Such a map can be written as
\begin{equation}
	\textbf{m}(\Omega, \nu) = A(\Omega) \begin{pmatrix}
    	f(\nu_1) \\
       	f(\nu_2) 
        \end{pmatrix} ,
\end{equation}
where $\textbf{m}(\Omega,\nu)$ is a function of direction $\Omega$ and frequency $\nu$, where the amplitude of the spatial variations are described by $A$ and the two frequency channels are presented as a vector containing $f(\nu_1)$ and $f(\nu_2)$. 

Next generate the covariance matrix of the map by taking the outer-product
\begin{equation}
	\mathbf{C} = \mathbf{m}\mathbf{m}^{\mathrm{T}} = A^2(\Omega)
    	\begin{pmatrix}
    	f^2(\nu_1) & f(\nu_1)f(\nu_2) \\
        f(\nu_1)f(\nu_2) &  f^2(\nu_2) \\
        \end{pmatrix} = \mathbf{U}\Lambda\mathbf{U}^\mathrm{T} ,
\end{equation}
where the covariance matrix has been decomposed into the orthogonal matrix $\mathbf{U}$ of eigenvectors and diagonal matrix $\Lambda$ that contains the corresponding eigenvalues. 

Solving for the eigenvalues $\lambda$ can then be done by solving for the condition $\mathrm{det}(\mathbf{C} - \mathbf{I}\lambda) = 0$. The resulting determinant is then
\begin{equation}
	\mathrm{det}(\mathbf{C} - \mathbf{I}\lambda) = \lambda \left(\lambda - f^2(\nu_1) - f^2(\nu_2) \right)
\end{equation}
where the two possible eigenvalues are
\begin{equation}
\lambda = \left\{\sum_i^{\mathrm{N}_\nu} f^2(\nu_i), 0\right\} ,
\end{equation}
or simply there is only one non-zero eigenmode that is the sum of squares of each frequency channel $\lambda = \sum_i^{\mathrm{N}_\nu} f^2(\nu_i)$. This statement holds true regardless of the number of frequency channels.

Assuming then that the $1/f$ noise is described by a single power law in all pixels, then it will only need one eigenmode to describe it entirely. Since then the HI signal of interest is \textit{uncorrelated} in frequency (assuming sufficiently wide redshift bins on the scales of interest) it requires $N_{\mathrm{chan}}$ modes to be fully described, and so can be easily separated from the $1/f$ noise signal. The effect then of increasing $\beta$ is to increase the information content of the $1/f$ noise, with higher values of $\beta$ requiring more eigenmodes to be fully described.

Of course there are further complications such as each pixel may have different spectral indices, or in the real data the $1/f$ noise may have a complex relationship with other systematics meaning more than one functional form is required to describe $1/f$ noise even with $\beta = 0$. In this case $1/f$ noise will not be perfectly separated from the signal of interest, and the resulting residual $1/f$ noise power will depend on this \textit{effective} value of $\beta$.

\section{Derivation of White Noise Uncertainty}\label{app:B}

This section will show how to derive Eqn.~\ref{eqn:knoxnew}. It is assumed in this derivation that there are two components of interest, one is the signal of interest and the other is the noise. The spherical harmonic coefficients ($a_{\ell m}$) are a linear combination of the signal and noise such that
\begin{equation}\label{eqn:cl1}
	a_{\ell m} = s_{\ell m} + n_{\ell m} 
\end{equation}
where $ s_{\ell m}$ describes the signal coefficients and $n_{\ell m}$ the noise coefficients.  Taking the absolute square and the fourth power of Eqn.~\ref{eqn:cl1} results in
\begin{equation}\label{eqn:cl2}
	\left| a_{\ell m} \right|^2 = \left| s_{\ell m} \right|^2 + \left| n_{\ell m} \right|^2 + 2\left| s_{\ell m} \right| \left| n_{\ell m} \right|  ,
\end{equation}
and
\begin{equation}\label{eqn:cl3}
		| a_{\ell m} |^4 = | s_{\ell m} |^4 + | n_{\ell m} |^4 + 6| s_{\ell m} |^2 | n_{\ell m} |^2 + 4 \left[ |s_{\ell m}|^3 |n_{\ell m}|  + |s_{\ell m}| |n_{\ell m}|^3 \right] .
\end{equation}
Both the signal and noise are Gaussian random variables sampled from distributions with variances $S_\ell$ and $N_\ell$ respectively. Gaussianity implies that $\left< | a_{\ell m} |^2 \right> = C_\ell$ and $\left< | a_{\ell m} |^4 \right> = 3 C_\ell^2$. Therefore the average over Eqn.~\ref{eqn:cl2} and Eqn.~\ref{eqn:cl3} results in
\begin{equation}\label{eqn:cl4}
	\left< | a_{\ell m} |^2 \right> = S_\ell + N_\ell ,
\end{equation}
and 
\begin{equation}\label{eqn:cl5}
	\left< | a_{\ell m} |^4 \right> = 3S_\ell^2 + 3N_\ell^2 + 6 S_\ell N_\ell ,
\end{equation}
where $S_\ell$ and $N_\ell$ are the angular power spectra of the signal and noise respectively. Taking the square of Eqn.~\ref{eqn:cl3} from Eqn.~\ref{eqn:cl4} gives the uncertainty in the square of the angular coefficients
\begin{equation}\label{eqn:cl6}
	\Delta C_\ell^2 = \frac{2}{2 \ell + 1} ( S_\ell + N_\ell)^2,
\end{equation}
where $C_\ell = S_\ell + N_\ell$ and the factor of $2 \ell + 1$ comes from remembering that each $C_\ell$ has $2\ell + 1$ $m$-modes that are independently drawn from the same random distribution. Eqn.~\ref{eqn:cl6} is the familiar form of power spectrum uncertainty as stated in \citet{Knox1995}.

To derive Eqn.~\ref{eqn:knoxnew} it must be assumed that there is only one sky realisation (as per the simulations described in this work) such that
\begin{equation}\label{eqn:cl7}
	\left< | a_{\ell m} |^2 \right> = \sqrt{\left< | a_{\ell m} |^4 \right>} = C_\ell
\end{equation}
where $C_\ell$ is the measured variance of signal described by $a_{\ell m}$. Substituting Eqn.~\ref{eqn:cl7} for the signal in Eqn.~\ref{eqn:cl2} and Eqn.~\ref{eqn:cl3} results in 
\begin{equation}\label{eqn:cl8}
	\left< | a_{\ell m} |^2 \right> = S_\ell + N_\ell ,
\end{equation}
and 
\begin{equation}\label{eqn:cl9}
	\left< | a_{\ell m} |^4 \right> = S_\ell^2 + 3N_\ell^2 + 6 S_\ell N_\ell .
\end{equation}
Differencing the square of Eqn.~\ref{eqn:cl8} from Eqn.~\ref{eqn:cl9} as before, again assuming $2 \ell + 1$ independent $m$-modes, results in
\begin{equation}
	\Delta C_\ell^2 = \frac{2}{2 \ell + 1} N_\ell( 1 + 2\frac{N_\ell}{S_\ell})^2,
\end{equation}
which is Eqn.~\ref{eqn:knoxnew}.

\label{lastpage}

\end{document}